\documentclass[usenatbib]{mn2e}
\usepackage{amsmath}
\usepackage{graphicx}
\usepackage{txfonts}
\usepackage{natbib}
\usepackage{array}
\usepackage{longtable}
\usepackage[usenames]{color}

\bibpunct{(}{)}{;}{a}{}{,}

\DeclareMathAlphabet{\mymath}{U}{eus}{m}{n}

\def\la{\mathrel{\hbox{\rlap{\hbox{\lower4pt\hbox{$\sim$}}}\hbox{$<$}}}}
\def\ga{\mathrel{\hbox{\rlap{\hbox{\lower4pt\hbox{$\sim$}}}\hbox{$>$}}}}

\newcommand{\be}{\begin{equation}}
\newcommand{\ee}{\end{equation}}

\newcommand{\bi}{\begin{itemize}}
\newcommand{\ei}{\end{itemize}}

\newcommand{\ben}{\begin{enumerate}}
\newcommand{\een}{\end{enumerate}}

\newcommand{\bfig}{\begin{figure}\begin{minipage}{140mm}}
\newcommand{\efig}{\end{minipage}\end{figure}}

\newcommand{\btab}{\begin{table}\begin{minipage}{140mm}}
\newcommand{\etab}{\end{minipage}\end{table}}

\newcommand{\bfigMore}{\begin{figure}\begin{minipage}{160mm}}
\newcommand{\efigMore}{\end{minipage}\end{figure}}

\newcommand{\btabMore}{\begin{table}\begin{minipage}{160mm}}
\newcommand{\etabMore}{\end{minipage}\end{table}}

\newcommand{\bea}{\begin{eqnarray}}
\newcommand{\eea}{\end{eqnarray}}

\newcommand{\bega}{\begin{gather}}
\newcommand{\eega}{\end{gather}}

\newcommand{\bc}{\begin{center}}
\newcommand{\ec}{\end{center}}

\newcommand{\dif}{{\rm d}}

\newcommand{\ave}[1]{\langle #1 \rangle}

\newcommand{\bm}{\boldsymbol}

\def\mof{MACSJ0417.5$-$1154}

\def\pz{$P\left ( z\right )~$}
\def\photoz{photo-$z$~}
\def\BVRiz{{\it B}$_{\rm J}${\it V}$_{\rm J}${\it R}$_{\rm C}${\it I}$_{\rm C}${\it z}$^{+}$}
\def\BVriz{{\it B}$_{\rm J}${\it V}$_{\rm J}${\it R}$_{\rm C}${\it i}$^{+}${\it z}$^{+}$}

\def\avebeta{\langle \beta_s \rangle}

\title [Weighing the Giants - III. Methods \& Measurements of Accurate 
  Lensing Cluster Masses] {Weighing The Giants -  III. Methods and
  Measurements of Accurate Galaxy Cluster Weak-Lensing Masses}

\author[D. E. Applegate et al.]
{\parbox[t]{\textwidth}{
Douglas E. Applegate$^{1,2,3,4}$
\thanks{E-mail:dapple@stanford.edu}, 
Anja von der Linden$^{1,2,5}$,
Patrick L. Kelly$^{1,2,3}$,
Mark T. Allen$^{1,2}$,
Steven W. Allen$^{1,2,3}$,
Patricia R. Burchat$^{1,2}$,
David L. Burke$^{1,3}$,
Harald Ebeling$^{6}$,
Adam Mantz$^{7,8}$,
R. Glenn Morris$^{1,3}$}\\
      \vspace*{3pt}
\\
$^{1}$Kavli Institute for Particle Astrophysics and Cosmology,
Stanford University,
452 Lomita Mall,
Stanford, CA  94305-4085, USA\\
$^{2}$Department of Physics,
Stanford University,
382 Via Pueblo Mall, 
Stanford, CA  94305-4060, USA\\
$^{3}$SLAC National Accelerator Laboratory, 
2575 Sand Hill Road, 
Menlo Park, CA 94025, USA\\
$^{4}$Argelander-Institut f\"ur Astronomie,
Universit\"at Bonn, 
Auf dem H\"ugel 71, 
53121 Bonn, Germany\\
$^{5}$Dark Cosmology Centre , Niels Bohr Institute, 
University of Copenhagen, 
Juliane Maries Vej 30, 
2100 Copenhagen Ø, Denmark\\
$^{6}$Institute for Astronomy, 
2680 Woodlawn Drive, 
Honolulu, HI 96822, USA\\
$^{7}$Kavli Institute for Cosmological Physics,
University of Chicago,
5640 South Ellis Avenue,
Chicago, IL 60637-1433, USA\\
$^{8}$Department of Astronomy and Astrophysics, 
University of Chicago,
5640 South Ellis Avenue,
Chicago, IL 60637-1433, USA\\
}

\begin{document}

\date{}

\pagerange{\pageref{firstpage}--\pageref{lastpage}} \pubyear{2010}

\maketitle

\label{firstpage}

\begin{abstract}
  We report weak-lensing masses for 51 of the most X-ray luminous
  galaxy clusters known. This cluster sample, introduced earlier in
  this series of papers, spans redshifts $0.15 \lesssim z_{\rm cl}
  \lesssim 0.7$, and is well suited to calibrate mass proxies for
  current cluster cosmology experiments. Cluster masses are measured
  with a standard `color-cut' lensing method from three-filter
  photometry of each field. Additionally, for 27 cluster fields with
  at least five-filter photometry, we measure high-accuracy masses
  using a new method that exploits all information available in the
  photometric redshift posterior probability distributions of
  individual galaxies.  Using simulations based on the COSMOS-30
  catalog, we demonstrate control of systematic biases in the mean
  mass of the sample with this method, from photometric redshift
  biases and associated uncertainties, to better than 3\%. In
  contrast, we show that the use of single-point estimators in place
  of the full photometric redshift posterior distributions can lead to
  significant redshift-dependent biases on cluster masses. The
  performance of our new photometric redshift-based method allows us
  to calibrate `color-cut' masses for all 51 clusters in the present
  sample to a total systematic uncertainty of $\approx7\%$ on the mean
  mass, a level sufficient to significantly improve current cosmology
  constraints from galaxy clusters. Our results bode well for future
  cosmological studies of clusters, potentially reducing the need for
  exhaustive spectroscopic calibration surveys as compared to other
  techniques, when deep, multi-filter optical and near-IR imaging
  surveys are coupled with robust photometric redshift methods.

\end{abstract}


\section{Introduction}

Galaxy clusters have become a cornerstone of the experimental evidence
supporting the standard $\Lambda$CDM cosmological model.  Recent
studies of statistical samples of clusters have placed precise and
robust constraints on fundamental parameters, including the amplitude
of the matter power spectrum, the dark energy equation of state, and
departures from General Relativity on large scales.  For a review of
recent progress and future prospects, see \citet*{aem11}.

Typical galaxy cluster number count experiments require a mass-observable scaling relation
to infer cluster masses from survey data, which in turn requires calibration of the mass-proxy bias and scatter.
Weak lensing follow-up of clusters can be used, and to some extent
has already been used, to set the absolute calibrations for the
mass-observable relations employed in current X-ray and optical
cluster count surveys \citep[e.g.][]{mae08, mantz10a, vikhlinin09,
  rwr10}.  However, targeted weak lensing follow-up efforts of cluster surveys
have not yet studied a sufficient number of clusters nor have
demonstrated a sufficient control over systematic uncertainties to
meaningfully impact on cosmological constraints.

For the current generation of X-ray cluster surveys drawn from ROSAT observations
\citep[e.g.][]{bcs98,reflex,fourhundred, brightmacs}, the uncertainty
in the absolute mass calibration of the survey proxy, which is of the
order $\approx 15\%$, dominates the systematic uncertainty on the
matter power spectrum normalization $\sigma_8$ \citep{mantz10a, vikhlinin09}. For
\citet{mantz10a}, the current limits on this systematic uncertainty
are derived from simulations of non-thermal pressure support in relaxed
clusters \citep[e.g.,][]{nvk07} and uncertainties in the
\textit{Chandra} calibration, whereas for \citet{vbe09} the
limits are derived from weak lensing calibrations \citep{hoekstra07,
  zhang08}, quoted as a 9\% uncertainty but neglecting an additional
systematic uncertainty on the lensing masses known to be at least 10\%
\citep{mahdavi08}. The absolute mass
calibration from weak lensing follow-up therefore needs to be accurate
to better than 15\% to impact significantly on current work. Future surveys will face even more stringent
systematics requirements on the absolute calibration of
multiwavelength mass proxies if they are to utilize fully their
statistical potential. For example, the Dark Energy Survey will
require an absolute mass calibration at the 5\% level for the dark
energy constraints to be within 10\% of their maximum potential
sensitivity \citep{heidi10}, requiring a combination of weak lensing and high-precision mass proxies, i.e. X-ray
observations. Similar arguments apply to cluster surveys across the
electromagnetic spectrum, e.g. the South Pole Telescope
\citep[SPT,][]{wbh11} and eRosita \citep{pab10}. 

To achieve such calibration with weak lensing, one needs to follow up a large sample of
clusters. For individual clusters weak lensing typically offers mass measurements with a precision
of $\approx30\%$ \citep{becker11, okabe_masses,
  hoekstra07}, driven approximately equally by a limited number of
well measured galaxies and line of sight structure.  However, simulations
show that weak-lensing measurements can in principle provide
accurate, approximately unbiased, estimates of the \emph{mean} mass
for statistical samples of galaxy clusters \citep{becker11,
  cok07}. Small systematic biases in the mean mass can still arise
from, e.g., the details of the assumed mass model, shear
calibration, and the lensed-galaxy redshift distribution. Such sources
of uncertainty, in particular the lensed-galaxy redshift distribution,
have not yet been sufficiently understood for upcoming, or even
current, surveys, as we show in this work.

In the \emph{Weighing the Giants} project, we aim to provide absolute
mass-calibration for galaxy cluster mass proxies, including
specifically X-ray mass proxies, to better than 10\% accuracy. We have
gathered extensive optical imaging of 51 clusters in at least three
wide photometric filters, where clusters are mostly drawn from the
X-ray selected cosmological cluster sample of
\citet{mantz10a} and the relaxed cluster sample from \citet{ars08}. Of
these, 27 were observed in at least five filters. The clusters span a
redshift range of $0.15 < z < 0.7$. To ensure an accurate
mass-calibration, we have pursued a `blind' analysis where we have
deliberately delayed comparing our lensing masses to X-ray masses
 and the lensing masses of others reported in the
literature. Such a simple procedure prevents us from introducing
observer's bias into our results. Given the redshift range, data
quality, filter coverage, and blind analysis, our study represents the
most extensive analysis of its type to date, and should be considered a
pathfinder for the challenges facing upcoming optical, submillimetre, and X-ray
cluster surveys.

Here, we report weak-lensing masses for the 51 clusters in
our sample, and show that the total systematic uncertainty on the mean
mass of the sample is controlled to $\approx$7\%. In particular, we
focus on controlling systematic uncertainties associated with the redshift
distribution of lensed galaxies. We approach this problem
in two ways. For the entire sample, we employ a
standard analysis technique \citep[the ``color-cut''
  method;][]{hoekstra07}, albeit with some improvements, where the lensed
redshift distribution for each cluster field is estimated from
separate, deep field photometric redshift (\photoz) measurements. We show that this
method alone does not sufficiently control systematic
uncertainties to the accuracy required for current surveys. Alternatively, the redshift distribution of background galaxies may be
measured using photometric redshift estimates in fields with
at least five filter coverage. While previous large photometric surveys
\citep{wmk04, ilbert09} have shown that high fidelity photometric
redshift point estimators are possible through the use of many (e.g., greater than 15) broad,
medium, and narrow band filters for objects down to $i^{+}<25$ magnitude,
observations of cluster fields usually lack coverage with such a
comprehensive array of photometric filters and future optical surveys
will typically have only six broad filters. We show that with such
limited photometric coverage, \photoz point estimates are insufficient
to recover unbiased cluster masses. We therefore develop a method that
uses the full \photoz posterior probability distribution \pz for
individual galaxies in each cluster field, referred to as the
``$P\left ( z\right )$'' method, and show that it can be used to measure robust cluster
weak-lensing masses. Using the COSMOS-30 \photoz catalog
\citep{ilbert09}, we create a series of simulations to test the
sensitivity of our reconstructed masses to \photoz errors. We show
that $P\left ( z\right )$ distributions from current photometric
redshift codes, with \BVRiz photometry, enable control of systematic
uncertainties on the mean mass for the sample to better than 2\%
accuracy for clusters at $0.15 < z < 0.7$ -- a result that provides
significant encouragement for future cluster-cosmology work.

\label{contents}

This is the third in a series of papers describing the project.
Paper I describes our cluster sample, data reduction
procedures, and shear measurements \citep{paper1}. Paper II  details
our photometric redshift measurements, including the development of a
scattered-light correction for SuprimeCam, and an improved relative
photometric calibration procedure based on fitting the stellar locus \citep{paper2}.
This paper reports our lensing masses and estimates of
systematic uncertainties in the sample mean mass. Forthcoming papers
will use these accurate cluster masses to
calibrate X-ray mass proxies and determine
improved cosmological constraints.

The structure of this paper is as follows. In
Section~\ref{sec:lensing_theory}, we review cluster mass measurements
with weak lensing. We describe our data set and analysis procedures in
Section~\ref{sec:data_reduction}. In Section~\ref{sec:color_cuts}, we
develop and apply our implementation of the color-cut method to all
clusters in the sample. In Sections~\ref{sec:maxlike_method} \&
\ref{sec:shape_distribution}, we introduce our \photoz lensing
framework that incorporates \photoz posterior probability
distributions for each galaxy observed.  In
Section~\ref{sec:testing_framework}, we investigate the expected
systematic errors present in mass measurements given the empirical
performance of \photoz estimators. In Section~\ref{sec:final_masses},
we report measured masses using both the \pz and color-cut methods,
and cross-calibrate the color-cut results. In
Section~\ref{sec:systematics}, we perform checks of other potential
systematic uncertainties. In Section~\ref{sec:aveconcentration}, we
briefly digress to measure the average concentration of massive
clusters. We compare our lensing mass measurements to
other efforts in the literature, based on overlapping samples, in
Section~\ref{sec:lit_comp}.  Finally, we provide concluding remarks in
Section~\ref{sec:conclusions}.

Unless otherwise noted, all mass measurements assume a flat
$\Lambda CDM$ reference cosmology with $\Omega_{\rm m} = 0.3$,
$\Omega_{\Lambda} = 0.7$ and $H_0 = 100 \,h\, \mbox{km/s/Mpc}$, where
$h=0.7$.


\section{Weak-Lensing Mass Measurements}
\label{sec:lensing_theory}

The mass of a gravitational lens, in this case a massive galaxy
cluster, may be inferred from the systematic distortion of images of
background galaxies as measured by the reduced shear. For a review of
weak lensing, see \citet{bas01} and \citet{saasfee}. Here, we review the redshift
dependence of the reduced shear and how it relates to the cluster mass
profile.

\begin{figure}
\includegraphics[width=0.95\columnwidth]{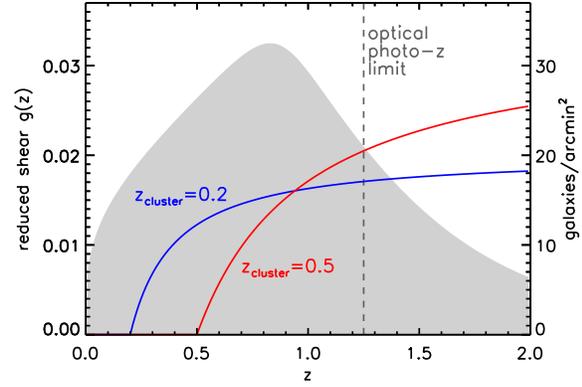}
\caption{The reduced shear $g$, as a function of source galaxy redshift
  $z$, for cluster lenses at redshifts $z_{\rm cluster} = $0.2 and 0.5. The function $\beta_s$, a
  ratio of angular diameter distances, controls the shape of the
  curve (see Eq.~\ref{eq:beta}). $\beta_s$ is zero for sources at redshifts less than $z_{\rm cluster}$ and rises
  steeply above $z_{\rm cluster}$, eventually flattening off
  at high redshift. The shape of the function is cosmology
  dependent. A typical galaxy redshift distribution for a typical ground-based $i^{+} < 25$ mag survey is shown in light gray,
  peaking at $z\approx0.8$.}
\label{fig:beta_s}
\end{figure}

The ellipticity of a galaxy, corrected for PSF effects, provides a
noisy estimate of the reduced shear at the galaxy position. Assuming a
single lens plane, the theoretical expectation for the reduced shear $\bm{g}\left(\vec \theta\right)$ is given by
\begin{equation}
\bm{g}\left(\vec \theta\right) = \frac{\beta_s(z_b) \bm{\gamma}_{\infty}\left(\vec \theta\right)}{1 - \beta_s(z_b)\kappa_{\infty}\left(\vec \theta\right)}\quad,
\label{eq:reducedshear}
\end{equation} %
where the shear $\bm{\gamma}_{\infty}\left(\vec \theta\right)$ and convergence $\kappa_{\infty}\left(\vec \theta\right)$ are set by the mass distribution of the lens, evaluated at
the source position $\vec \theta$, assuming a lensed source at
infinite redshift. For an axisymmetric lens,
Eq.~\ref{eq:reducedshear} reduces to a scalar equation, as the only
shear will be tangential to the lens. The dependence of the distortion on the background-galaxy redshift is set by $\beta_s(z_b)$:
\begin{equation}
\beta_s = \frac{D_{LS}}{D_S}\frac{D_{\infty}}{D_{L,\infty}}\quad.
\label{eq:beta}
\end{equation} %
$\beta_s$ is a ratio of angular diameter distances, where $D_{LS}$ is
the distance between the lens and the source, $D_S$ is the distance to
the source, and $D_{L,\infty}$ and $D_{\infty}$ are the corresponding
distances from the lens and the observer to a source at infinite redshift, respectively. Figure~\ref{fig:beta_s}
shows how the reduced shear $g$ scales as a function of background-galaxy redshift for
lenses at two redshifts. For reference, a typical analytical approximation to a
ground-based $i^{+} < 25$ magnitude redshift distribution, peaking at
$z\approx0.8$, is shown as the shaded region \citep{shj10}. $\beta_s$ rises
rapidly from zero for redshifts just beyond a lens, and approaches a constant
value at high redshift.

To facilitate comparisons to other mass proxies, especially X-ray
proxies, we measure the total mass enclosed within a sphere of fixed
radius. While a general 2D mass distribution can in principle be
recovered \citep{swunited}, this approach would be limited by the
depth of our images and an inability to break the mass sheet
degeneracy from weak-lensing data alone \citep{bls04}, especially for
low redshift clusters that fill the SuprimeCam field of view.  Another
alternative is to measure the mass within a 2D aperture, which
determines the total projected mass within a cylinder. Operationally,
aperture mass measurements would require us to deproject an
ill-constrained, noisy, 2D mass profile to make the needed comparison
to X-ray mass measurements, and requires an assumed profile at large
radius to break the mass sheet degeneracy.

We instead fit the estimated reduced shear at each galaxy position to
the lensing signal predicted by a spherical Navarro-Frenk-White (NFW)
halo \citep{nfw97} profile. The parametrized mass profile, known to
be a reasonable description of dark matter halos, automatically breaks
the mass-sheet degeneracy. The NFW profile has two free parameters,
the scale radius $r_s$ and the concentration $c_{200} \equiv
r_{200}/r_s$ (where overdensity is defined with respect to the critical density), or alternatively
the mass within a particular radius. We implement the detailed radial,
lens redshift, and cosmology dependence of $\gamma_{\infty}$ and
$\kappa_{\infty}$ for a spherical NFW profile found in
\citet{wright00}. Extensive simulation work in the literature shows
that fitting such a profile to the reduced shear, averaged over a
sample of clusters, can in principle return an unbiased mass,
depending on details in the analysis. Triaxiality, nearby correlated
structure, and uncorrelated structure along the line of sight
contribute 20-25\% scatter to individual mass measurements, not
including the statistical uncertainty due to the finite number of lensed sources \citep{hoe03, cok07, becker11, bahe2011,
  hhh11}. Efforts are underway to verify this result for the mass
range spanned by clusters in our sample ($M_{500} > 10^{15}
M_{\odot}$).


\section{Data \& Processing}
\label{sec:data_reduction}

In this section, we describe the data set, data processing, and sample
selection used as input to the mass measurement algorithms. We
analyze a sample of 51 X-ray selected, luminous galaxy clusters imaged with SuprimeCam \citep{suprimecam} at the Subaru Telescope
and Megaprime at the Canada-France-Hawaii Telescope. Paper I contains a
detailed description of the clusters observed, filters, and
processing details. All clusters in the sample were imaged with at
least three broad optical filters, and 27 were imaged with at least
five broad optical filters. Raw CCD exposures were processed using a
modified {\sc GaBoDS/Theli} pipeline \citep{esd05, ehl09, schirmer13}. We detect objects
using {\sc SExtractor} \citep{bea96}. Shape measurements were made
with the code {\sc analyseldac\ }\citep{erben01}, based on the KSB algorithm
\citep{ksb95}, to produce shear catalogs.  Shape measurements were
calibrated using the STEP2 simulations \citep{step2}.

The heterogeneous nature of our dataset requires us to adopt two
different strategies to measure the redshift distributions of galaxies in each
cluster field. For the 27 cluster fields where we have five or more filters, we
compute photometric redshifts for each galaxy in our shear
catalogs. Photometric redshifts require strict control of the relative
photometric calibration between filters. Paper II describes the
position-dependent zeropoint corrections (a ``star flat'')
and a custom implementation of the stellar locus regression 
technique that we use to calibrate colors. For \photoz calculations, we use the BPZ code
\citep{bpz00} with templates optimized by \citet{caa07}. Paper II includes
detailed quality checks on the \photoz calculations. For the remaining 24 cluster fields
where we have observations in less than five filters, we use the COSMOS-30 photometric redshift catalog
\citep{ilbert09} as a reference deep field for a traditional
``color-cut'' analysis.

\begin{figure}
\includegraphics[width=\hsize]{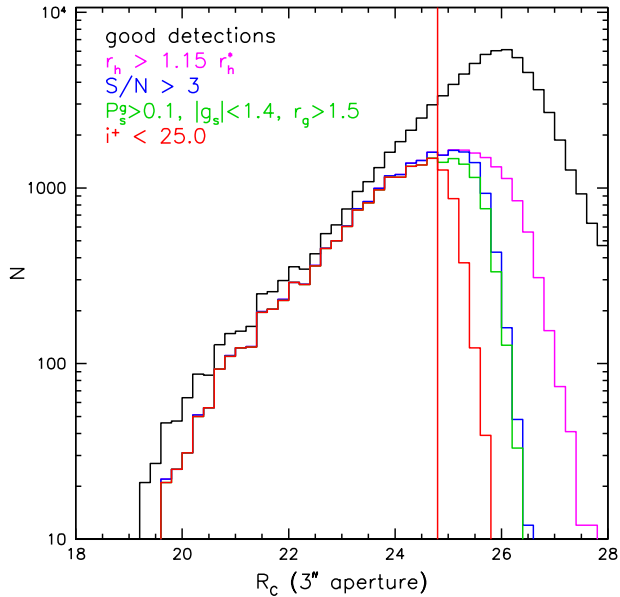}
\caption{The number of background galaxies for the \mof\ field(as a function of {\it
    R}$_{\rm C}$ magnitude), after lensing quality cuts. The
  black histogram shows the distribution of unsaturated objects that
  are detected in an image region with an exposure weight of at least
  half of the maximum value. Because the initial object detection is highly
  complete, most of these detections are smaller than the minimum size
  requirement for lensing ($r_h\ge 1.15 r_{\rm h}^{*}$); the magenta
  histogram shows objects that survive this size cut. Our minimum
  signal-to-noise requirement ($S/N\ge 3$) removes further objects at
  the faint end (blue histogram).  We furthermore reject objects with
  exceptionally large shear estimates, small values of $P^{\rm g}_s$,
  or a small KSB filter size; these cuts remove only a few objects at the
  faint end (green histogram). To ensure robust photometric redshift
  estimates, we remove objects with $i^{+}>25$, which
  is the COSMOS-30 completeness limit (red histogram). These cuts (and those applied in Fig.~\ref{fig:cmd}) are
  applied to the catalogs used for both the \pz and the color-cut
  methods. For the color-cut method, we estimate the detection
  completeness magnitude from this histogram, shown here as the
  vertical red line. }
\label{fig:mag_histo}
\end{figure}

We apply a series of cuts to the galaxy catalogs, based on the shape
and photometry measurements, to minimize bias while maximizing
sensitivity. We do not use measurements based on tangential shear,
measured mass of the cluster, or X-ray derived mass when establishing
these cuts. The color-cut method and the \pz method, use a similar set
of cuts which are described below. Differences, where they exist, are noted explicitly. In
the color-cut analysis, we keep between 1500 to 15,000 galaxies per
cluster field. For the \photoz based analysis, roughly 500 to 5,000
galaxies remain, where the main difference is a redshift cut that only
selects galaxies behind the cluster. Figure~\ref{fig:mag_histo}
illustrates how some of the major cuts, detailed below, affect the
object number counts in an example field.

\paragraph*{Lensing quality cuts:} For the lensing analysis we require
that an object ellipticity is measured with $S/N\ge 3$, as defined by
{\sc analyseldac}. We also require that the objects are 15\% larger
than the point spread function (PSF) of the observation as measured
by the median half-light radius $r_{\rm h}$ reported by {\sc
  analyseldac} of stars in the image ($r_{\rm h}\ge 1.15 r_{\rm
  h}^{*}$; see Paper 1 and Section~\ref{sec:shape_distribution} for
motivation of these cuts). These criteria remove a large fraction of
the initially detected objects (Fig.~\ref{fig:mag_histo}). In
addition, we guard against failures in the shape measurement code by
accepting only galaxies with a minimum KSB filter size $r_g >1.5$
pixels, measured shear $|\hat g| < 1.4$, and shear susceptibility
$P^{g} > 0.1$. We do not explicitly cut
on objects that are close to one another (a ``nearest neighbor''
cut); we have verified that the explicit removal of objects who have nearest
companions within a radius $3r_g$ does not induce a systematic shift in our
mass measurements. We also remove large objects from the catalog, as these
objects are unlikely to be background galaxies and the success rate of
{\sc analyseldac} drops for large objects, mostly because the centroid
may vary with isophote radius. We choose the upper galaxy size limit
to be the $r_g$ radius where the success rate drops below 75\%. This
removes (2-10)\% of the objects that otherwise pass all lensing
criteria, with higher rates at the cluster center.

\paragraph*{Bright magnitude cut:} The brightest galaxies in each field are
unlikely to lie behind the cluster. We therefore remove galaxies with
a magnitude brighter than 22 in the detection band.

\paragraph*{COSMOS completeness limit cut:} The COSMOS-30
photometric catalog is publicly available to {\it i}$^{+} < 25.0$
\citep{ilbert09}. For objects fainter than this, the uncertainty in
the COSMOS-30 photometric redshifts is large \citep{ilbert09}, and the
outlier fraction may be significant \citep{shj10}. Therefore, we limit
our catalogs to the same depth, even when our data are substantially
deeper. For fields not observed in {\it i}$^{+}$, we interpolate the
{\it i}$^{+}$ magnitude from the best-fit {\sc BPZ} template (for all cluster fields, regardless of filter coverage). This
faint cut applies to both the color-cut analysis and the \photoz based
analysis, as we use COSMOS-30 to verify the performance of \photoz
measurements.

\begin{figure}
\includegraphics[width=\hsize]{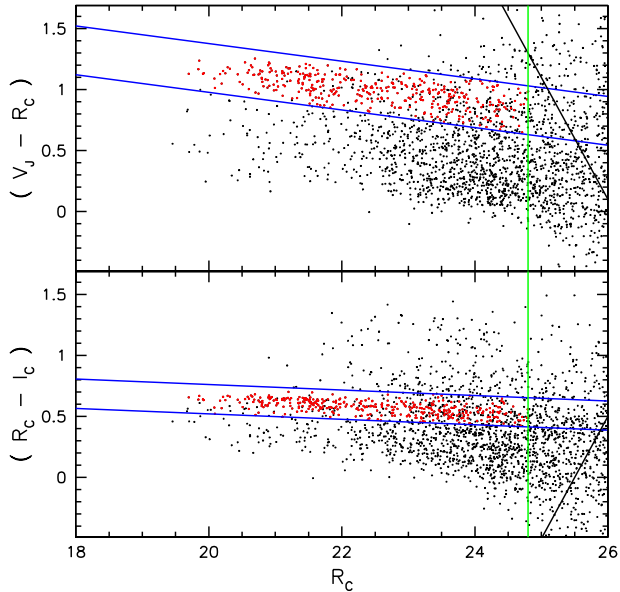}
\caption{({\it V}$_{\rm J}$ - {\it R}$_{\rm C}$) vs. {\it R}$_{\rm C}$
  (top panel) and ({\it R}$_{\rm C}$ - {\it I}$_{\rm C}$) vs. {\it
    R}$_{\rm C}$ (bottom panel) color-magnitude diagrams for the \mof\
  field. The band within which we select the red sequence is shown by
  the blue lines. Galaxies are classified as being in the red sequence
  if they fall within the band in both diagrams (shown as red
  symbols). These galaxies are excluded from both the \pz and
  color-cut methods. The green line shows the completeness limit of
  the lensing band (see Fig.~\ref{fig:mag_histo}). The black lines
  illustrate the completeness limits of the other two filters in these
  diagrams.}
\label{fig:cmd}
\end{figure}

\paragraph*{Red sequence cut:} The measured ellipticities of galaxies lying
on the cluster red sequence are not sensitive to the mass of the cluster,
and will dilute the average measured shear if not removed. The use of
a color-magnitude diagram (CMD) is an efficient way to identify
and remove these galaxies, independent of \photoz's. However, a single
color is not a monotonic function of redshift, and so a generic red
sequence band on a CMD can contain non-cluster  galaxies. This is
particularly true at faint magnitudes, where the majority of our
lensing sample lies. Furthermore, with increasing cluster redshift, the faint
end of the cluster red sequence is not well populated
\citep[e.g.][]{dpa07}. We identify the red sequence simultaneously in
two color-magnitude diagrams (Fig.~\ref{fig:cmd}). Galaxies are only
removed from the catalog if they lie on the red sequence in both
diagrams.

\paragraph*{Radial distance cut:} The NFW halo model does not provide an
adequate description of the mass distribution beyond the virial radius
of galaxy clusters \citep[e.g.,][]{becker11}. Also, near the cluster
center, we expect increased cluster galaxy contamination, and shears
departing from the weak-lensing approximation. In addition, we are
only able to calibrate our shear measurements over the narrow range of shears
probed by the STEP2 program \citep{step2}. Therefore, we accept
galaxies within a projected range $750 \textrm{kpc} < r < 3
\textrm{Mpc}$, which is approximately equivalent to the X-ray measured
$0.5r_{500} < r < 2.0r_{500}$ for these massive clusters
\citep{mae10}. By removing the centers of clusters, we are also less
sensitive to profile miscentering \citep{msb10, paper1}.

\paragraph*{Photo-$z$ cuts:} For the \pz analysis only, we make cuts on the measured \photoz's. We remove galaxies with a
BPZ single point redshift estimate $z_b > 1.25$. This is due to
degraded performance in the \photoz's as the $4000\AA$ break moves out
of the $z^{+}$ band. We also remove galaxies with $z_b <
z_{cluster} + 0.1$ to remove foreground and cluster
contamination. Finally, we only include objects where the difference between
the 2.5 and 97.5 redshift percentiles, measured by the \pz, is less than 2.5.  This excludes
objects with effectively no redshift constraint, for which \pz is
dominated by the BPZ prior. Note that we do not enforce a cut on the
BPZ $ODDS$ parameter, which is itself a measure of the redshift posterior
probability concentration around the most likely value. We
find no statistically significant systematic shift in our measured
masses when we enforce such a cut at $ODDS > 0.7$ or at $ODDS > 0.9$.


\section{Lensing Masses with the ``Color-Cut'' Method}

\label{sec:color_cuts}

In this section, we present a traditional ``color-cut'' analysis,
employing three-filter observations, for all clusters in the
sample. In the color-cut method, the average redshift of the lensed
galaxy population is measured from a statistically matched subset of
galaxies in a reference deep field where spectroscopic or high-quality
photometric redshifts are available. The color-cut method has the
advantage that it can be applied to fields with a modest filter coverage. However, the relative lack of color information leads to a shear dilution
from cluster galaxies, for which we need to correct.

Color-cut methods have been used extensively in previous studies
\citep{hoekstra07,okabe_masses,hhh11}, although the details of the
implementations differ in some regards from what we present here.


\subsection{Defining and Applying Color-Cuts}
\label{sect:color_cuts_cuts}

For our reference deep field, we use the COSMOS-30 photometry catalog
with photometric redshifts determined from all available bands, as described in \citet{ilbert09}. Although
the COSMOS field is a statistically limited sample with respect to cosmic variance effects,
it is the best-suited reference field for our study because of the
depth to which photometric redshifts are complete ($i^{+} < 25$), and
the overlap in filter coverage and data quality with our observations.

To avoid a statistical mismatch between the galaxy populations
selected from our cluster fields and the sample presented in
\citet{ilbert09}, we must first apply all of the cuts from
Section~\ref{sec:data_reduction} to the COSMOS-30 catalog. Below, we describe additional photometry cuts that are also required for the color-cut analysis, and proxies for cuts based on measurements
from {\sc analyseldac}, which are not available for the COSMOS field.

\paragraph*{Completeness cut in the detection band:} To emulate the effects of the {\sc analyseldac} $S/N > 3$ cut, we reject all galaxies in
both our catalogs and the COSMOS-30 catalog fainter than the limiting
magnitude in our detection band for a given cluster. The limiting magnitude of the galaxy
sample in a cluster field is defined as the magnitude where the number
counts turn over (see Fig~\ref{fig:mag_histo}). We define the limiting
magnitude after the {\it i}$^{+} < 25$ cut has already been applied.

\paragraph*{Completeness in colors:} Because we use three filters for
the identification and removal of red sequence galaxies, we must
match the detection limits in these filters with the COSMOS catalog
(Fig.~\ref{fig:cmd}). Since there are no lensing quality constraints
on these filters from our observations, the completeness limit is typically considerably
deeper than the lensing limit, and this cut removes only a few objects.

\paragraph*{Size cut proxy:}   We emulate the effects of the size
cut employed in each cluster field by determining the median {\sc SExtractor} half-light radius (the {\sc FLUX\_RADIUS} parameter) of
the objects flagged as stars in the COSMOS Subaru photometry, and
rejecting objects with {\sc FLUX\_RADIUS}~$<1.15$~{\sc
  FLUX\_RADIUS}$^{\star}$. There is significant scatter between the {\sc analyseldac}
measured galaxy half-light radii and the one reported in the COSMOS-30
catalog. The
inability to match the size cut precisely is a source for systematic error
in the mean mass of the galaxy cluster sample. We return to this issue
in Section~\ref{sec:color_cut_sys_errs}.


\subsection{Contamination Correction}
\label{sec:contamination_correction}

The remaining cluster field catalogs still contain some cluster galaxies,
predominantly galaxies bluer than the red sequence. These galaxies
dilute the lensing signal because they are not lensed by the cluster,
and are not accounted for in the redshift distribution from the COSMOS-30 deep
field. We follow the method of \citet{hoekstra07} to estimate the fraction
of contaminating cluster galaxies by examining the number density
profile of objects in the lensing catalog. The assumption here is that
the number density of background (and foreground) objects not associated with the cluster is uniform across the field. The number density of cluster-member 
galaxies, on the other hand, increases towards the cluster center.

\begin{figure*}
\includegraphics[width=0.48\hsize]{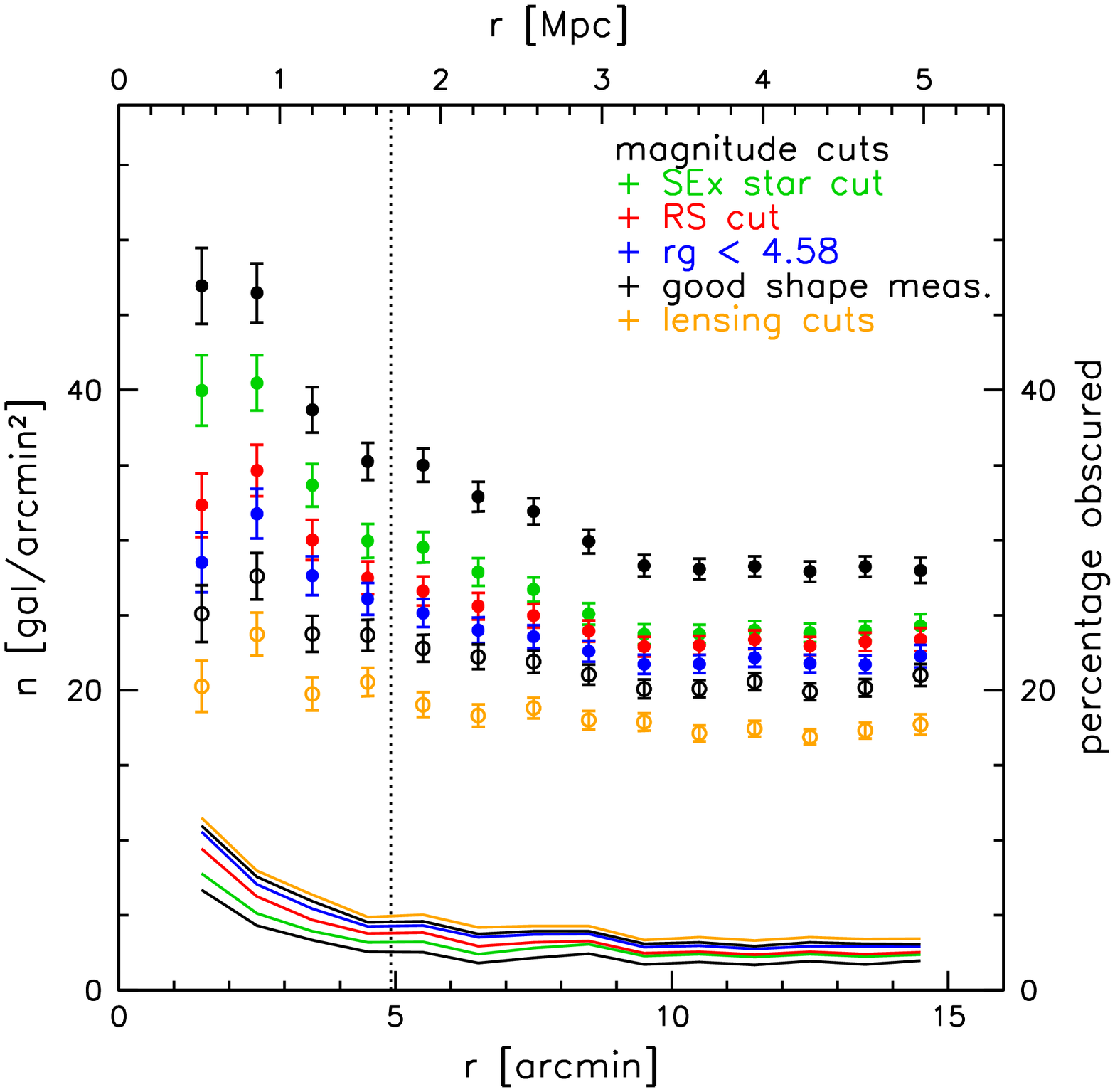}\hspace{0.03\hsize}
\includegraphics[width=0.48\hsize]{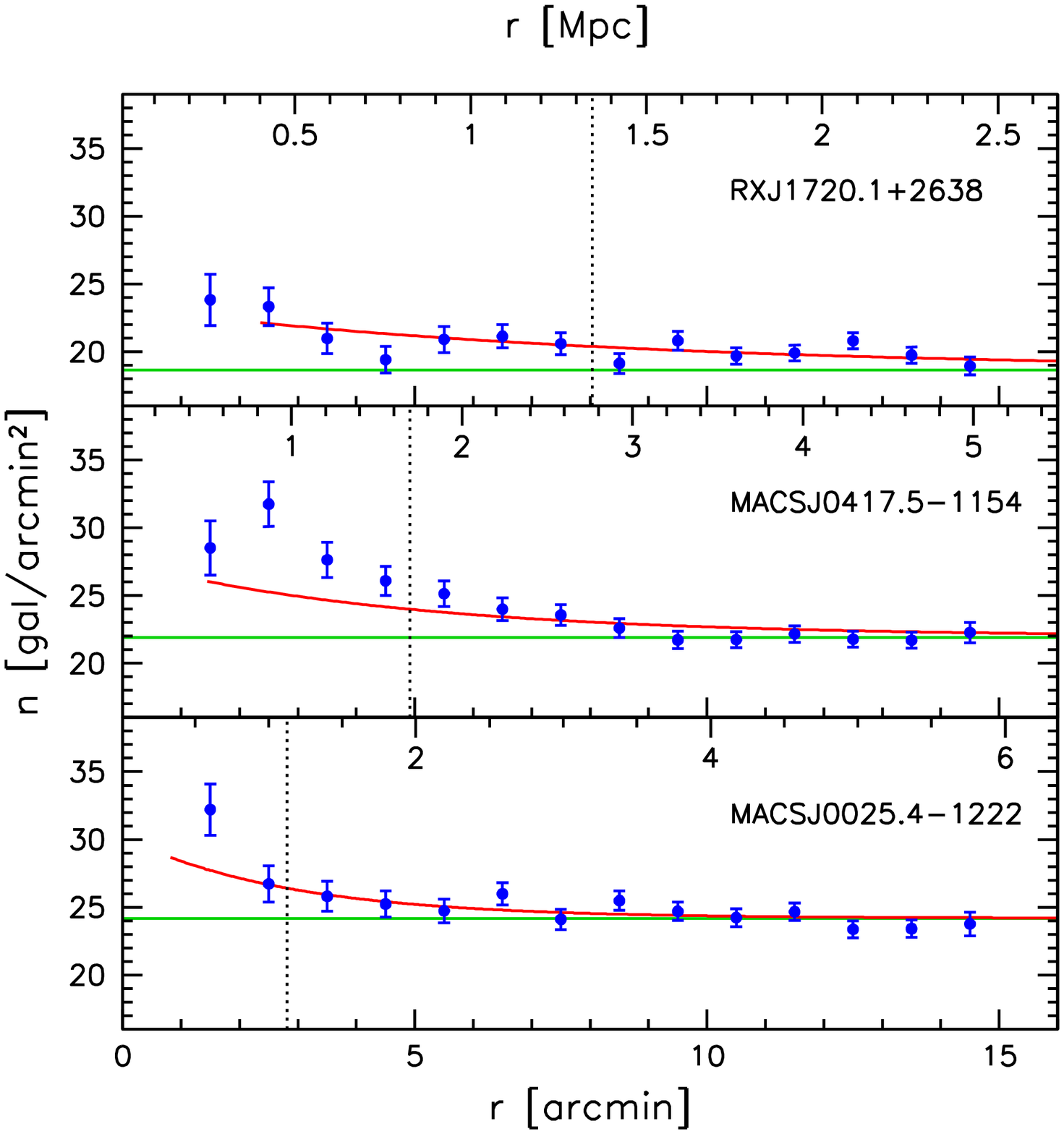}
\caption{Left: The number density profile of objects in the \mof\ field as a function of distance from the cluster, for several
  lensing catalog selection criteria. The black filled points show the
  number density of objects with $22<m_{R_{\rm C}}<24.8$, the
  completeness limit of the lensing catalog in this field (see
  Fig.~\ref{fig:mag_histo}).  For the green points, objects smaller
  than the PSF size are removed using only criteria based on {\sc
    SExtractor} output parameters - note how this removes a roughly
  constant number density of objects. The red points show the number
  counts after applying the red sequence cut; this predominantly
  removes objects close to the cluster center. For the blue points,
  objects with large half-light radii are removed, since these are
  likely to be foreground or cluster galaxies, and the shape
  measurement is often compromised, e.g., by centroid shifts. The blue
  number density profile is the basis of the contamination correction,
  as all the previous selection criteria are largely independent of local
  galaxy density. The black open points show the number density of
  objects for which the shape measurement is robust. Because this can
  be compromised by close neighbors, the shape measurement fails more
  frequently for objects near the cluster center. The orange points
  show the number density profile after the lensing cuts ($S/N>3$,
  $r_h>1.15r_h^{\star}$) have been applied - note how this profile is
  much flatter than the one based only on {\sc SExtractor} criteria
  (blue), underestimating the contamination of cluster galaxies. The
  lines at the bottom of the figure indicate for each number density
  profile the fraction of pixels obscured by masks or bright objects
  excluded from the sample. The number densities have been corrected for
  this obscuration.  Right: The number density profiles used for the
  contamination correction (blue points) for three clusters, along
  with the best-fit joint contamination profile (red line). The
  fraction of cluster galaxies is constrained to be the same at
  $r_{500,\rm X}$ across the sample (indicated by the vertical dotted
  lines), whereas the background number density (shown as green
  horizontal line) is a free parameter for each cluster.}
\label{fig:numdens}
\end{figure*}

For this measurement, one has to be careful to take into account
effects that mimic a decrease/increase in the number density of
measured objects as a function of cluster radius, which are at least as
large as the density effects induced by cluster magnification. For
instance, cluster galaxies obscure a fraction of the background
galaxies, with the fraction increasing towards the cluster center. Left
unaccounted for, this would lead to an underestimate of the cluster
galaxy contamination, or even an apparent depletion in the number density of background objects. When deriving the
number density profiles, it is therefore essential to track the areas masked by image
artifacts and objects rejected from the
background catalog.   (For each of these objects, the masked area is taken as the
{\sc ISOAREA\_IMAGE} SExtractor output parameter.) The area masked by
other objects, mostly bright objects and red sequence galaxies
(Fig.~\ref{fig:numdens}) is typically $\lesssim 5\%$ at large radii,
but $\sim 10\%$ at $0.5 R_{500}$.

The `lensing quality cuts' have an additional effect on the number density profiles of background objects, in that objects with close neighbors are less likely to have shape measurements of acceptable quality. This lowers the observed number of background galaxies in the lensing sample near the clear center,  below the density extrapolated from the cluster outskirts. In order to account for this, we derive the contamination correction from galaxy catalogs prior to applying the lensing cuts.  However, the galaxy
catalog from which the contamination correction cut is determined
should have statistical properties as close as possible, in terms of brightness,
color, and size distribution, to the final lensing catalog, so that
the contamination fraction is the same for both -- analogous to the
need for proxy cuts for the COSMOS-30 catalog in the previous
section. Here, however, we have considerably more information about
each object, and can refine the proxy cuts developed previously to be
more accurate. 

As before, the effects of the $S/N$ cut can be  mimicked by applying a limiting magnitude
cut. The size cut is mimicked as follows.  For those objects with $r_h < 1.15r_h^{\star}$, we determine the 33\% percentile in {\sc
  FLUX\_RADIUS}, {\sc FWHM\_IMAGE}, and major ($A$) and minor ($B$)
axis lengths.  We then remove objects in the full catalog which are
smaller than the 33\% percentile in any of these four variables. In
addition, we remove objects with ${\sc CLASS\_STAR}\ge0.99$.  These
proxies remove (50-60)\% of the objects caught by the $r_h < 1.15r_h^{\star}$
cut; however, only $\sim 2\%$ of objects with $r_h>1.15 r_h^{\star}$
are removed.

Although the expected increase in galaxy number density towards the cluster
core due to contamination is evident in most fields, the number counts are generally too
noisy to reliably estimate the contamination fraction in each cluster
independently, i.e. the counts are affected by correlated structures in the field.
We follow \citet{hoekstra07} and determine an average
contamination fraction for the cluster sample. Unlike
\citet{hoekstra07} however, we do not estimate the background number density
simply from the outer annuli. Nearby clusters will fill
most of the SuprimeCam field, preventing reliable measurements of the
background population density.  Instead, we fit the number density
profile of each cluster with a function of the form %
\bea 
f(r) &=& \frac{n_{\rm cluster}(r)}{n_{\rm cluster}(r) + n_{\rm background}} = f_{500} \; {\rm e}^{1-r/r_{500,\rm X}} \quad. \label{eq:contcorr}  
\eea

All clusters are fitted simultaneously. The fractional contamination at
$r_{500,\rm X}$, $f_{500}$, is linked across clusters, whereas $n_{\rm
  background}$ is free for each field.  We assume Gaussian errors
and fit the model with $\chi^2$ minimization. Since the cluster core
is not fit in the lensing analysis, we restrict the fit to the
projected radius $R>0.3 R_{500,\rm X}$. For this purpose only, we use a provisional value for
$r_{500,\rm X}$ from \citet{mantz10a}. By scaling with $R_{500,\rm
  X}$, we account for the mass range of the clusters. The noise
induced by cosmic variance from field to field prevent us from fitting
any functional dependence of $f_{500}$ (e.g., redshift, (scaled)
cluster mass, observing filter, limiting magnitude).


The best-fit contamination fraction for the full cluster sample is
$f_{500} = (8.6 \pm 0.9) \%$, where the uncertainties quoted here and below
are based on bootstrapping the cluster sample. Because of the presence
of systematic, unmodeled scatter (since for individual clusters, the
adopted model is not necessarily a good description), none of the
fits are formally acceptable. We have
chosen to fit an exponential, rather than $1/r$ profile -- the former
provides a slightly better fit than the latter \footnote{A
  projected $1/r$ profile implies a 3D $1/r^2$ profile. Although the total
  number of galaxies roughly follows a $1/r^2$ profile, the
  fraction of those galaxies that are not on the red sequence
  declines sharply towards the cluster core \citep[e.g.][]{lwk10}. The exponential profile is shallower and provides a better match to expectations.}.
The baseline contamination correction of $f_{500} = (8.6 \pm 0.9) \%$ boosts cluster masses measured within 1.5 Mpc by $\sim 8\%$.

To test for a possible dependence of $f_{500}$ on cluster
redshift, we split the sample in half at $z=0.38$. The
best-fit contamination fractions are then $f_{500}^{z<0.38} = (7.3 \pm 1.3) \%$ and
$f_{500}^{z>0.38} = (9.9 \pm 1.2) \%$.  To estimate the significance
of this dependence, we repeatedly split the sample into two random, equally-sized sets
and measure the difference in $f_{500}$, $\Delta f_{500}$. In 17\% of
the samples, the observed $\Delta f_{500}$ is larger than when the
sample is split into low- and high-redshift halves. Another way of
evaluating the significance of a redshift dependence is to fit $f_{500}$
for each cluster individually and test the correlation with
redshift. We bootstrap the sample and measure the Pearson, Spearman,
and Kendall correlation coefficients. The probability to find a
correlation coefficient randomly greater than zero is $\sim 83$\% for
all three correlation measures. We conclude that $f_{500}$ does not
have a significant dependence with redshift, to the limits of our data.

As described above, the cluster galaxy contamination fractions are
estimated from catalogs that approximate the final lensing sample
without applying criteria depending on the ``success rate'' of the
shape measurement algorithm (as this depends on local density and thus
distance from the cluster center). For comparison, the contamination
fraction inferred directly from the final lensing catalogs is $f_{500}^{\prime}
= (4.8 \pm 1.6)\%$. This serves as a lower bound to the true
contamination fraction.

\subsection{Color-Cut Mass Estimates}
\label{sec:cc_mass_est_proc}

Mass measurements with the color-cut method use bootstrapped galaxy samples drawn from the individual cluster fields.
For each bootstrap
realization, we calculate the weighted average tangential shear
$\ave{\hat{g}_t}_i$ in radial bins $i$ spanning the range 750kpc to 3.0Mpc. These bins are
chosen to contain an approximately equal number of galaxies, with at least 300
galaxies in each bin. For fields with less than 1800
galaxies available, we fix the number of bins to six, again with an
equal number of galaxies in each. Only two clusters have fewer than 1800
galaxies available.  Each galaxy is weighted by the inverse of the
variance of the distribution $p(\hat{g}|g)$ for the galaxy's $S/N$
from {\sc analyseldac} (which is otherwise marginalized over in
the \pz method, Section~\ref{sec:shape_distribution}).

The measured tangential shear at radius $r_i$ of the $i$th bin is then
corrected for cluster galaxy contamination according to
Eq.~\ref{eq:contcorr}:
\be
\ave{\hat{g}_t}_i \; \rightarrow \; \ave{\tilde{g}_t}_{i} = \frac{\ave{\hat{g}_t}_{i}}{1-f_{500}\,e^{(1-r_i/r_{500,X})}} \quad .
\ee
At this stage we also incorporate the statistical uncertainty on the
contamination correction by sampling $f_{500}$ from its posterior distribution for each bootstrap realization.

The corrected tangential shear is azimuthally averaged in each radial bin. The average tangential shear measures the quantity (similar to Eq.~\ref{eq:reducedshear}) %
\begin{equation}
\langle \hat g_t(r) \rangle = \left\langle
\frac{\beta_s\gamma_{t,\infty}(r)}{1 - \beta_s\kappa_{\infty}(r)}
\right\rangle \; .
\label{eq:aveshear}
\end{equation}%
Without individual redshift estimates, Eq.~\ref{eq:aveshear} cannot be computed. However, with
knowledge of the expected galaxy redshift
distribution for the cluster field, the right-hand side of Eq.~\ref{eq:aveshear} can be approximated by \citep{ss97}\footnote{Note that there is a typo in Eq.~(4.14) of
  \citet{ss97}, suggesting that the correction factor is
  $\frac{\ave{\beta_s^2}}{\ave{\beta_s}^2}$ instead of
  $\frac{\ave{\beta_s^2}}{\ave{\beta_s}}$.

We measure a bias of
  0.0070$\pm$0.0022 when using the correct version shown in Eq.~\ref{eq:color-cut-gamma-approx},
  which is not redshift dependent. While this small bias may be
  mitigated even further by including higher order moments of the $z$ distribution in
  Eq.~\ref{eq:color-cut-gamma-approx} \citep{ss97}, other sources of systematic uncertainty remain that are more difficult to characterize.
\citet{hfk00} advocate a
  similar, though not identical, approximation also based on
  $\frac{\ave{\beta_s^2}}{\ave{\beta_s}^2}$.
This approximation results in a nearly indistinguishable correction to the one we adopt.
 }:

\be%
g_{t,i}^{\rm model}\approx \frac{\ave{\beta_s} \gamma_{t,\infty}^{\rm model}(r_i)}{1-\frac{\ave{\beta_s^2}}{\ave{\beta_s}}\kappa_{\infty}^{\rm model}(r_i)} \quad.%
\label{eq:color-cut-gamma-approx}
\ee%
In the color-cut method, $\ave{\beta_s}$ and $\ave{\beta_s^2}$,
are
calculated from the redshifts of the galaxies in the reference field
(in our case the COSMOS field) and assumed to be the same in the
cluster fields.
For each bootstrap realization we draw a random pair of
$\ave{\beta_s}$ and $\ave{\beta_s^2}$ values measured on the COSMOS
field in an annulus of the same angular size as the radial fit range for each cluster (Sect.~\ref{sec:color_cuts}). This partially accounts for
sample variance associated with the limited area entering the fit,
but can only be a lower limit due to the limited size of the COSMOS
field.

To find the best-fit mass, we minimize for each bootstrap realization%
\be \chi^2 = \sum_{i} \frac{\left(\ave{\hat{g}_t}_i - g_{t}^{\rm
      model}(r_s)\right)^2}{\sigma_i^2} \ee%
with respect to the scale radius $r_s$ of the NFW profile,
keeping the concentration fixed at $c_{200}=4$. The parameter
$\sigma_i^2$ is the variance of the weighted mean shear in each bin. From
the best-fit profile, we calculate the mass within 1.5~Mpc.

The distributions of bootstrapped masses for individual clusters are very close to Gaussian. We quote the medians of these distributions
as our `best-fit' masses from the color-cut method, with the 16\% and 84\% percentile limits
as the upper and lower error bars. The statistical uncertainties on
the mass estimates are entirely dominated by shot noise from the galaxy
ellipticities; incorporating the statistical uncertainties on
$\ave{\beta}_s$, $\ave{\beta_s^2}$ and $f_{500}$ as we have done here
only marginally increases the error budget.


\subsection{Systematic Uncertainties on the Mean Sample Mass in the Color-Cut Method}
\label{sec:color_cut_sys_errs}

The ``color-cut'' analysis has systematic uncertainties associated
with calculating $\langle \beta_s \rangle$ and the contamination
correction, in addition to the uncertainties from shear estimation and
the assumed mass model (see
Section~\ref{sec:systematics}). The overall level of systematic
uncertainty in the color-cut method can be difficult to quantify, and
has generally not been quantified in previous lensing efforts.

\paragraph*{Cosmic Variance:} Cluster masses are roughly linearly
proportional to $\langle \beta_s \rangle$, measured from a common deep
field. Deep fields that are unrepresentative of the average cluster
field will lead to a bias in the average
cluster mass. (For studies of individual clusters, any deep field will always
lead to a biased mass.) In practice, \citet{mahdavi08} reported a
$\sim10\%$ shift in their cluster masses from \citet{hoekstra07},
based on the same data, simply due to using the larger CFHT Deep Legacy
Survey \citep{cfhtlsdeep} in place of the Hubble Deep Fields
\citep{hdf} as their reference. In principle, the
systematic uncertainty expected on using the 2-sq. degree COSMOS field may be
estimated from simulations\footnote{The cosmic variance for the average redshift of
  deep field galaxies has been estimated to be $\approx 3\%$, by
  \citet{wwh06}. However, the variance in $\left<z\right>$ cannot be
  substituted for the variance in $\left<\beta\right>$.}.

\paragraph*{Galaxy Sample:} 
The selection criteria for lensed galaxies must be accurately matched
to the reference deep field catalog to ensure an unbiased $\avebeta$
estimate. As an illustrative example,we consider a galaxy size cut
where we only accept objects with sizes 15\% larger than the PSF size
to minimize stellar contamination. To test how well applying an equivalent
cut with respect to the COSMOS PSF size recovers $\avebeta$, we
examined a cluster field, with photometric redshifts available, that
was observed twice, with both 0.4 and 0.6 arcsecond seeing. The
$\avebeta$ values estimated from the two samples, after cutting with
respect to the different PSF sizes but using identical detection and
\photoz catalogs, differ by up to 5\%, depending on the redshift of
the cluster. The seeing for the COSMOS field, at 0.7 arcseconds, is
larger than our average seeing of 0.6 arcseconds, which could
impart up to a 3\% systematic bias for the sample. We have not studied
how other analysis procedures (e.g., filter transformations or lensing
cut approximations) may impact $\avebeta$ estimation.

\paragraph*{Contamination Correction:}
The contamination correction results in an $\approx 8\%$ correction to
each cluster mass using the color-cut method. As demonstrated
previously, this correction is sensitive to details in the derivation,
such as accounting for foreground and cluster galaxies, masking
background galaxies, and the assumed contamination profile. The
correction, as implemented, does not account for ``sheets'' of cluster
galaxies that could exist in filaments and pancakes extending from the
cluster to the edge of the field of view. Currently, no publicly
available image simulations exist to test the accuracy of the
contamination correction procedure. Such simulations will be
challenging to perform robustly, required a detailed understanding of
galaxy evolution and the impact of galaxy-cluster interactions.

Without rigorous quantification of these and other systematic
uncertainties, or without an external calibration from a method with
quantified systematic uncertainties, color-cut style weak-lensing
masses have limited value for calibrating other cluster mass
proxies. The effects discussed here can easily shift the mean cluster
mass by 5-10\%. While the systematic effects highlighted in this
section could in principle be modeled given sufficient computer and
manpower, no effort currently exists, to our knowledge.


\section{Lensing Masses with Photometric Redshift Probability Distributions}
\label{sec:maxlike_method}

A statistical model that includes the redshift for each lensed
source should in principle offer the most reliable mass estimates
from lensing data.  However, photometric redshift estimates are
inherently noisy and are subject to systematic uncertainties. The posterior probability distribution $P(z)$,
returned by standard \photoz codes, gives the relative probability that a given
source is at a particular redshift. This posterior distribution
contains more information than a simple point estimate, in particular
when multiple redshift solutions match the observed galaxy colors. In principle, we
can use the \pz for a given galaxy as a weighting function when
comparing the observed and predicted shear for that galaxy. Here, however, 
we must also model the biases and scatter between the expected
shear g, given a particular redshift and halo model
(Eq.~\ref{eq:reducedshear}), and the measured shear $\hat g$ for each
galaxy. We use a Bayesian approach to incorporate these effects, which
we refer to as the \pz method in this paper.

Given i) the measured tangential\footnote{We suppress the symbol \textit{t} to prevent clutter in the rest of the paper.} shear $\hat g$ of one galaxy in the survey region of a
cluster, ii) a physical model for the expected shear as a function of
source redshift and cluster mass, $g(z,M)$ (and implicitly, other
properties such as cluster redshift and source position), and iii) the
photometric redshift probability distribution $P_i(z)$ for that
galaxy, then the posterior probability for the mass of the cluster is:

\begin{align} 
  \label{eq:onegal_posterior}
  P(M| \hat g_i) & \propto  P(M)P(\hat g_i | M) \nonumber \\
    & =  P(M)\int_0^{\infty}\! P\left(\hat g_i | g(z, M)\right)P_i(z){\rm d}z \quad.
\end{align} %
Here, $P(M)$ is the prior on the cluster mass and
$P(\hat g | g(z, M))$ is the likelihood function for the shape
measurement, specified below. By marginalizing over $P_i(z)$, we consistently
compare the measured reduced shear $\hat g$ with the proper model
value, weighted for the correct relative probability at each redshift. This may be
naturally extended to many galaxies, with

\begin{align}
  P(M | \mathbf{\hat g}) & \propto  P(M)P(\mathbf{\hat g} | M) \nonumber \\
    & =  P(M)\prod_i P(\hat g_i | M) \nonumber \\
    & =  P(M)\prod_i\int_0^{\infty}\! P(\hat g_i | g(z, M))P_i(z){\rm d}z \quad.
\end{align}

\begin{figure*}
\includegraphics[width=0.48\hsize]{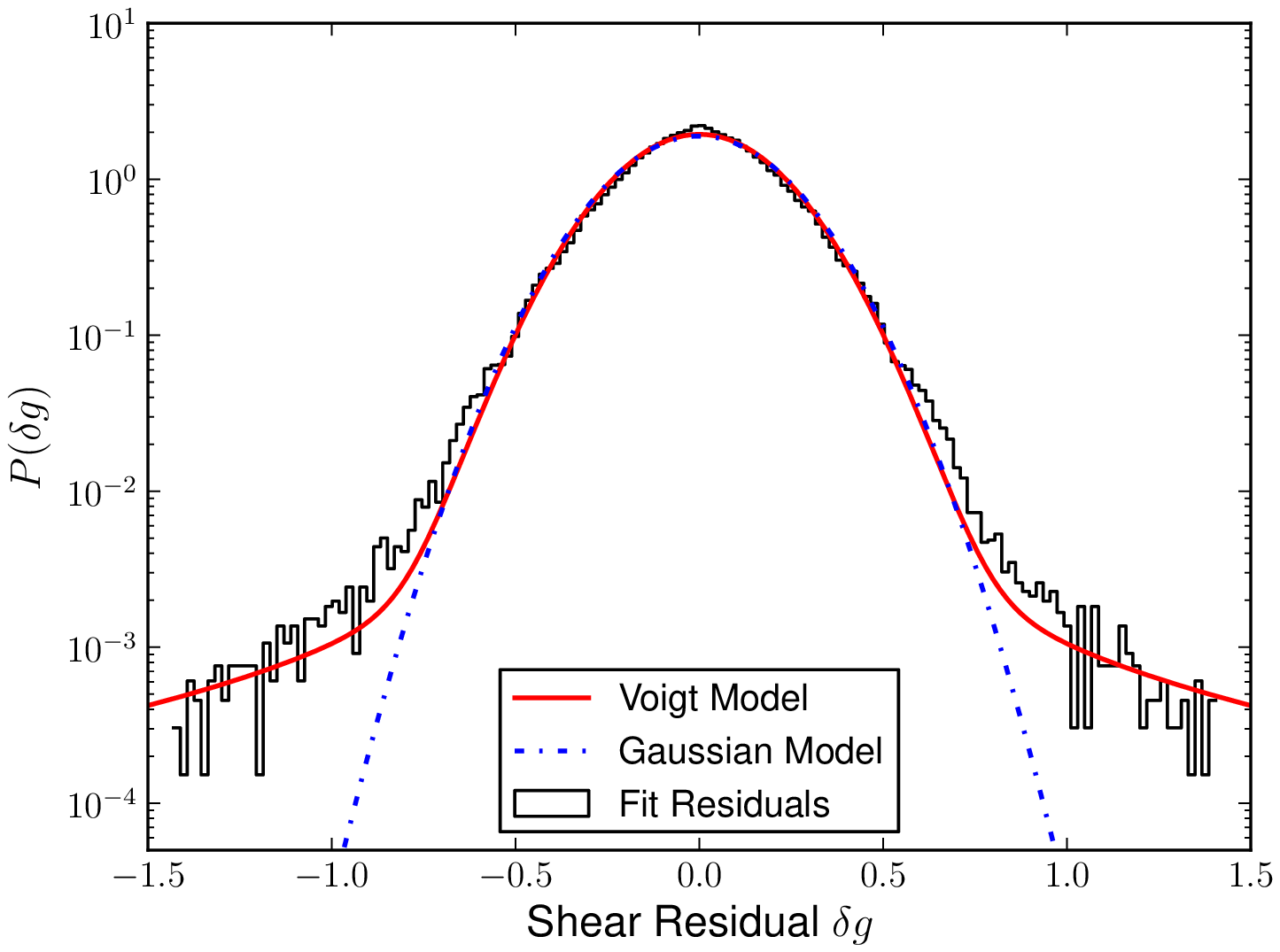}
\includegraphics[width=0.48\hsize]{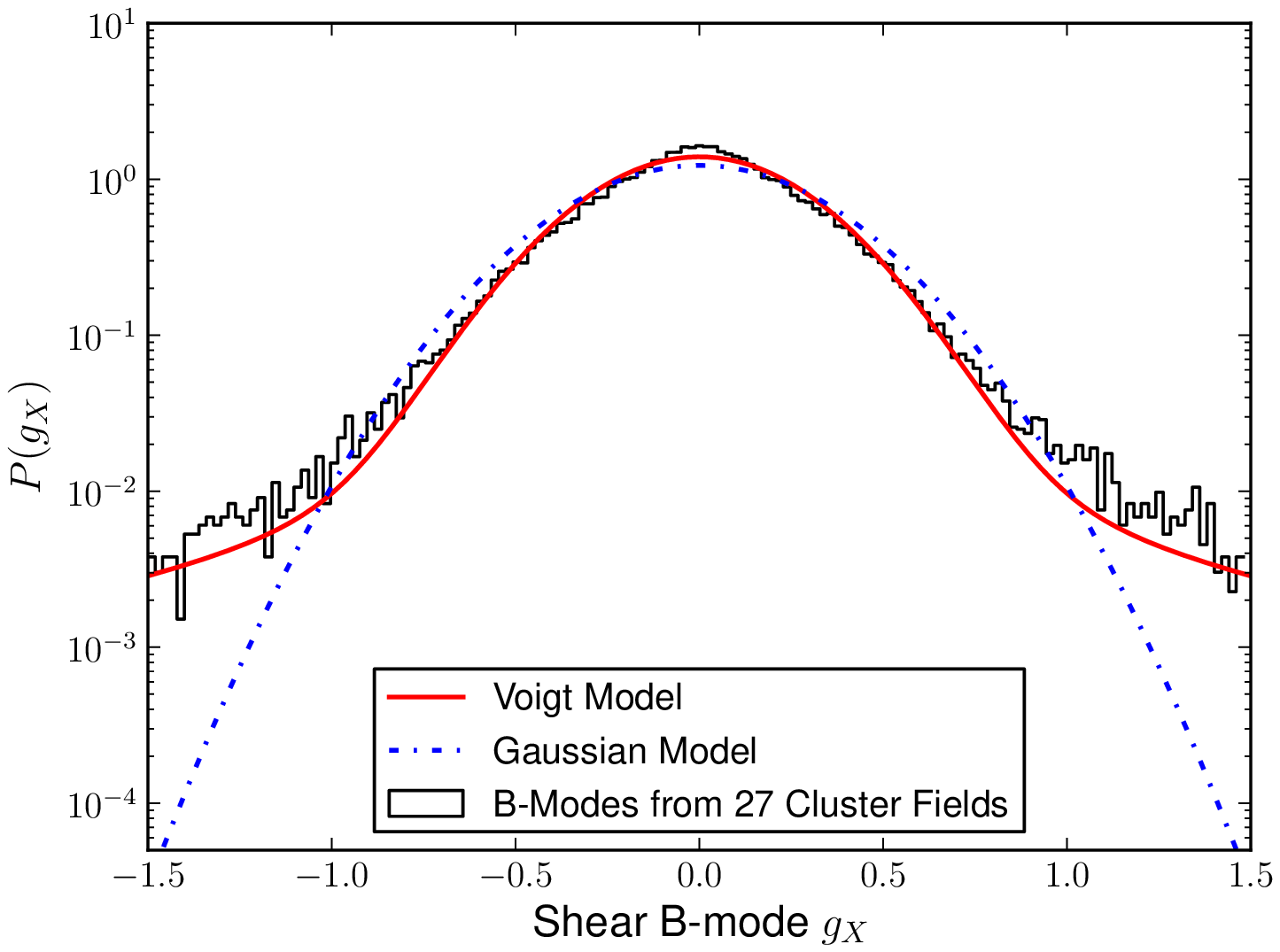}
\caption{Left: Residual scatter measured in STEP2 simulations after
  accounting for the average calibration bias. The shape of the
  residuals is nearly independent of the scatter profile assumed, when
  fitting for the average calibration bias. Right: Shear B-modes measured
  in 27 clusters; the cross terms are expected to be zero on
  average. The overplotted red lines are the best fit Voigt profiles
  for each histogram, while the blue dash-dotted lines are the best
  fit Gaussian profiles. The Gaussian model neglects the significant tails
  present in each distribution, and would therefore place too much
  weight on outlier shear measurements in a maximum-likelihood fit.}
\label{fig:shape_scatter}
\end{figure*}

The likelihood function $P(\hat g | g(z,M))$ encodes the scatter and
bias for the shape measurement $\hat g$. Physically, the bias
is sourced by calibration errors in the shape measurement. The scatter in
the reduced shear is a convolution of the intrinsic galaxy
distribution, after lensing, with the shape measurement error
distribution. Additional scatter is induced by departures from the
assumed mass model. Figure~\ref{fig:shape_scatter} plots the convolved
shape distribution and measurement error, as well as scatter from
assuming a spherical mass model in real systems. We restrict
ourselves to the low shear regime so that we may ignore any shear
dependent, asymmetric scatter \citep{geiger98}. Though not shown in
Fig.~\ref{fig:shape_scatter}, an additional scatter component also
arises from galaxies with redshifts poorly modeled by their \pz
function.

The simplest definition for $P(\hat g | g(z,M))$ is a Gaussian;
however such a model neglects the significant tails in the
distributions seen in Fig.~\ref{fig:shape_scatter}. We apply both the
posterior predictive cross validation and the deviation information
criteria model comparison tests to STEP2 simulated data \citep{step2} to determine the
best fit shape for the scatter. We find that a Voigt profile is a
better description of the scatter than a Gaussian or double Gaussian
model. A Voigt profile (a Gaussian convolved with a Lorentz
distribution) has three parameters, the mean $\mu$, the core width
$\sigma$, and the wing amplitude $\Gamma$.

The mean $\mu$ of the Voigt profile is a function of the predicted shear at
a galaxy's position, given its redshift, and the shear calibration
parameters $m$ and $c$, as defined in the STEP and STEP2 simulations
\citep{step1} for the PSF of the observation:
\begin{align}
  \mu = m({\mathrm size})g(z, M) + c \quad.
  \label{eq:voigt_mean}
\end{align}
The multiplicative bias, $m({\mathrm size})$, is a piecewise-linear function depending on the galaxy
size, and is described by three parameters. We use a multivariate normal prior
with covariances as measured from STEP2 simulations for $m$ and
$c$. See Section~\ref{sec:shape_distribution} for a description of how we
parametrize $m$ and constrain this part of the model.

The Voigt profile scatter parameters $\sigma$ and $\Gamma$ do not
depend on individual galaxy properties as implemented. Instead, we place uninformative, flat
priors on $\sigma$ and $\Gamma$. For simplicity, we refer to the set
of scatter and bias parameters m(size), c, $\sigma$, $\Gamma$ as
$\vec\alpha$, with a joint prior $P(\vec\alpha)$. The values of
$\vec\alpha$ can then be marginalized over for the posterior probability:

\begin{align}
    P(M | \mathbf{\hat g}) & =  P(M)\int_{\forall\vec\alpha} P(\vec\alpha)\Pi_i P(\hat g_i | M, \vec\alpha)d\vec\alpha \nonumber\\
    & =  P(M)\int_{\forall\vec\alpha}P(\vec\alpha)\Pi_i\int_z P(\hat g_i | g(z, M), \vec\alpha)P_i(z){\rm d}z {\rm d}\vec\alpha \quad.
\label{eq:full_posterior}
\end{align}

In Eq.~\ref{eq:full_posterior}, we refer only to the mass of the
spherical NFW halo, M, and suppress the concentration parameter for
clarity. For our analysis, we assume that $c_{200} = 4$ with 0.11 dex
log normal scatter, appropriate for massive halos, $M_{500} >
10^{14}M_{\odot}$ \citep{neto07}. This concentration distribution is also
marginalized over to determine the posterior probability for the
mass. Finally, we set the prior on mass, $P(M)$, to be uniform. Note that making the prior uniform in one measure
of mass typically implies a non-uniform prior in other mass estimates. For
this analysis, we measure masses within an aperture of 1.5 Mpc, and
set the prior accordingly.

In summary, our model has eight parameters: two for the halo
model (mass and concentration); four for the STEP shear correction (three
parameterize the size-dependent multiplicative bias, one for the
additive bias); and two to describe the shear scatter ($\sigma$ and
$\Gamma$ of the Voigt profile). All parameters but the mass are marginalized.

The center of the mass profile and the cluster redshift also enter our
model, but we do not marginalize over these parameters. We anchor our
profiles on the X-ray centroid for each cluster. See
Paper I for how the center is defined in each of the clusters. Cluster
redshifts are determined from spectroscopic follow-up with negligable
uncertainties.

We sample the posterior probability distribution in
Eq.~\ref{eq:full_posterior} using Markov Chain Monte Carlo
with an Adaptive Metropolis step algorithm, as implemented in the PyMC
software package \citep{phf10}. We numerically evaluate the
marginalization over each galaxy's \pz using the Cython extension
\citep{bbs09} for computational optimization.


Other efforts to include photometric redshifts in lensing measurements
exist in the literature. \citet{ss97} relate the measured moments of
an ensemble shape distribution (i.e., $\langle \hat g \rangle$) to an
integral over the redshift distribution. This presupposes that all
galaxies in a sample are statistically identical, independent draws
from a common redshift distribution. One could attempt to apply the
\citet{ss97} method by calculating an expected shear for each galaxy
based on its $P(z)$ and assuming Gaussian scatter. A similar approach is
taken by \citet{dawson11}, where an average critical density is
calculated for each galaxy based on its \pz. The standard deviation of
the average critical density, weighted by the $P(z)$, sets the width of
the assumed Gaussian scatter and serves as a per galaxy weight in that
work. \citet{mandelbaum08} also pursue the approach of an average
expected shear and weight per galaxy, optimized for galaxy-galaxy
lensing. While these approaches rightfully downweight galaxies with
poorly constrained \pz distributions, information is lost, most
notably when a lensed galaxy sits on the steeply rising part of the
$\beta_s$ curve (see Fig.~\ref{fig:beta_s}). This may most easily be
seen by noting that the residuals with respect to the true redshift
and shear will be correlated, which is neglected in these
methods. This correlation is is accounted for with the
marginalization in Eq.~\ref{eq:onegal_posterior}.

The work by \citet{geiger98}, and later \citet{king01}, uses an unbinned,
maximum likelihood approach that marginalizes the \pz in a similar
way to our method. We differ from their work in how the measured shear
relates to the predicted shear. \citet{geiger98} make strong
assumptions about the intrinsic galaxy ellipticity distribution and
ignore measurement uncertainties and biases. However, they explicitly include
the skew induced in the scatter from high shear. We restrict ourselves
to the low shear regime where the skew is negligible, and let the
data determine the proper form of the scatter.

Independent to the work presented in this paper, \citet{kitching11}
employ \pz marginalization for 3D cosmic shear measurements. However,
that work does not introduce the $P(\hat g|g)$ formalism that
generalizes the relationship between measured shear estimator and true
shear.


\section{Calibrating the Shear Likelihood Function}
\label{sec:shape_distribution}

\begin{table*}
  \caption{Best fit values for the Voigt profile parameters $\sigma$ and $\Gamma$ in bins of shape S/N. Quoted uncertainties are the 1$\sigma$ marginalized constraints. Values of $\sigma$ in S/N bins are used to weight shear estimates in the color-cut analysis. Only galaxies with a Lanczos3 S/N $>$ 3 are accepted into the lensing analysis.}
  \begin{tabular}{c | c | c c | c c }
    \hline
    STEP S/N  & Lanczos3 S/N & \multicolumn{2}{c |}{PSF A} & \multicolumn{2}{c}{PSF C} \\
    & & $\sigma$ & $\Gamma$ & $\sigma$ & $\Gamma$  \\
    \hline
    5.00 - 6.00 & 2.17 - 2.61 & 0.24 $\pm$ 1.35e-03 & 2.60e-02 $\pm$ 1.47e-03 & 0.28 $\pm$ 1.90e-03 & 3.23e-02 $\pm$ 2.05e-03 \\ 
6.00 - 8.00 & 2.61 - 3.48 & 0.21 $\pm$ 7.95e-04 & 1.54e-02 $\pm$ 7.87e-04 & 0.23 $\pm$ 1.08e-03 & 2.61e-02 $\pm$ 1.20e-03 \\ 
8.00 - 10.00 & 3.48 - 4.35 & 0.19 $\pm$ 7.95e-04 & 1.38e-02 $\pm$ 7.71e-04 & 0.20 $\pm$ 1.06e-03 & 2.11e-02 $\pm$ 1.13e-03 \\ 
10.00 - 15.00 & 4.35 - 6.52 & 0.19 $\pm$ 5.37e-04 & 3.21e-03 $\pm$ 3.54e-04 & 0.19 $\pm$ 7.38e-04 & 9.50e-03 $\pm$ 6.61e-04 \\ 
15.00 - 20.00 & 6.52 - 8.70 & 0.19 $\pm$ 8.27e-04 & 4.30e-03 $\pm$ 5.75e-04 & 0.19 $\pm$ 9.98e-04 & 8.82e-03 $\pm$ 8.75e-04 \\ 
$>$ 20.00 & $>$ 8.70 & 0.22 $\pm$ 4.88e-04 & 9.34e-04 $\pm$ 2.25e-05 & 0.22 $\pm$ 5.79e-04 & 3.19e-03 $\pm$ 3.01e-04 \\ 
    
   \end{tabular}
  \label{table:step_snratio}
\end{table*}

Various shear estimator algorithms presented in the literature exhibit
biases, with complex dependencies on PSF and galaxy properties. When
all other factors are held constant, KSB+ algorithms (such as the code
{\sc analyseldac} that we employ) show an approximately linear
relationship with true shear in the low shear regime, $g < 0.3$
\citep{erben01, step1}. The STEP simulation studies \citep{step1,
  step2} parameterize this bias as a function of shear using a
multiplicative bias $m$ and an additive bias $c$
(Eq.~\ref{eq:voigt_mean}). \citet{step2} demonstrated explicitly that
the parameters $m$ and $c$ are functions of the point spread function
shape, size, and ellipticity, as well as the lensed galaxy shape S/N
(or magnitude) and size.

Our  \pz method described in Section~\ref{sec:maxlike_method} can correct
for these biases and marginalize out any calibration uncertainty
through the definition of the shear likelihood function $P(\hat g|g)$.  We use
the STEP2 simulations to calibrate the likelihood function, taking
into account PSF and galaxy property dependent biases. 

STEP2 simulated images mimic lensing-quality
SuprimeCam data. Six sets of images were produced (A-F), using five
different PSF's sampled from Subaru images. Five of these sets used
realistic galaxy images derived from shapelet fits to galaxy morphologies in the
COSMOS HST field. Set A and C have PSF sizes of 0.6'' and 0.75'' full
width-half maximum, respectively, spanning the typical seeing range
of our lensing images. The PSF ellipticity in the STEP2 simulations is somewhat smaller than
typical for our images. We quantify the systematic uncertainty from
our use of the STEP2 A and C images in Section~\ref{sec:systematics}.

We detect and measure the size and shape of objects in the STEP2
images using the same algorithms and cuts employed for our data
catalogs. To explore the behavior of the STEP $m$ and $c$ parameters
in bins of various galaxy properties, we perform unbinned maximum
likelihood fits to the STEP2 catalogs. In all fits, we describe the
scatter in measured versus true shear as a Voigt profile where the
mean is given by $\mu = \hat g - (1+m)g - c$. The bias parameters $m$
and $c$, and the Voigt profile scatter parameters $\sigma$ and $\Gamma$,
have uniform priors.  After verifying consistent results, we fit galaxies from the STEP2 original and rotated image sets, as well as
measurements of both shear components simultaneously. Uncertainties
are determined from Markov Chain Monte Carlo exploration of the
parameter space.

\begin{figure*}
\includegraphics[width=0.48\hsize]{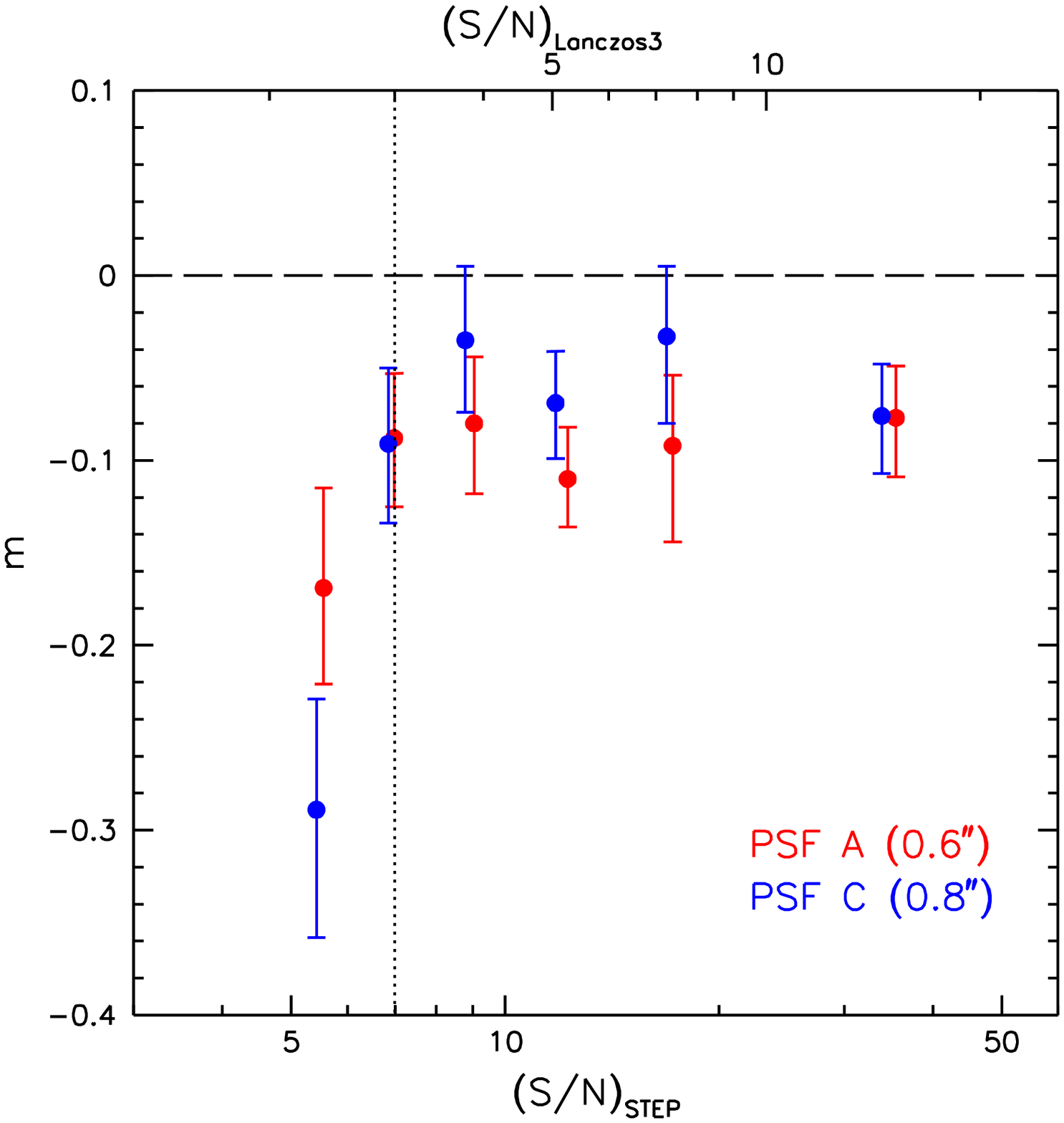}
\includegraphics[width=0.48\hsize]{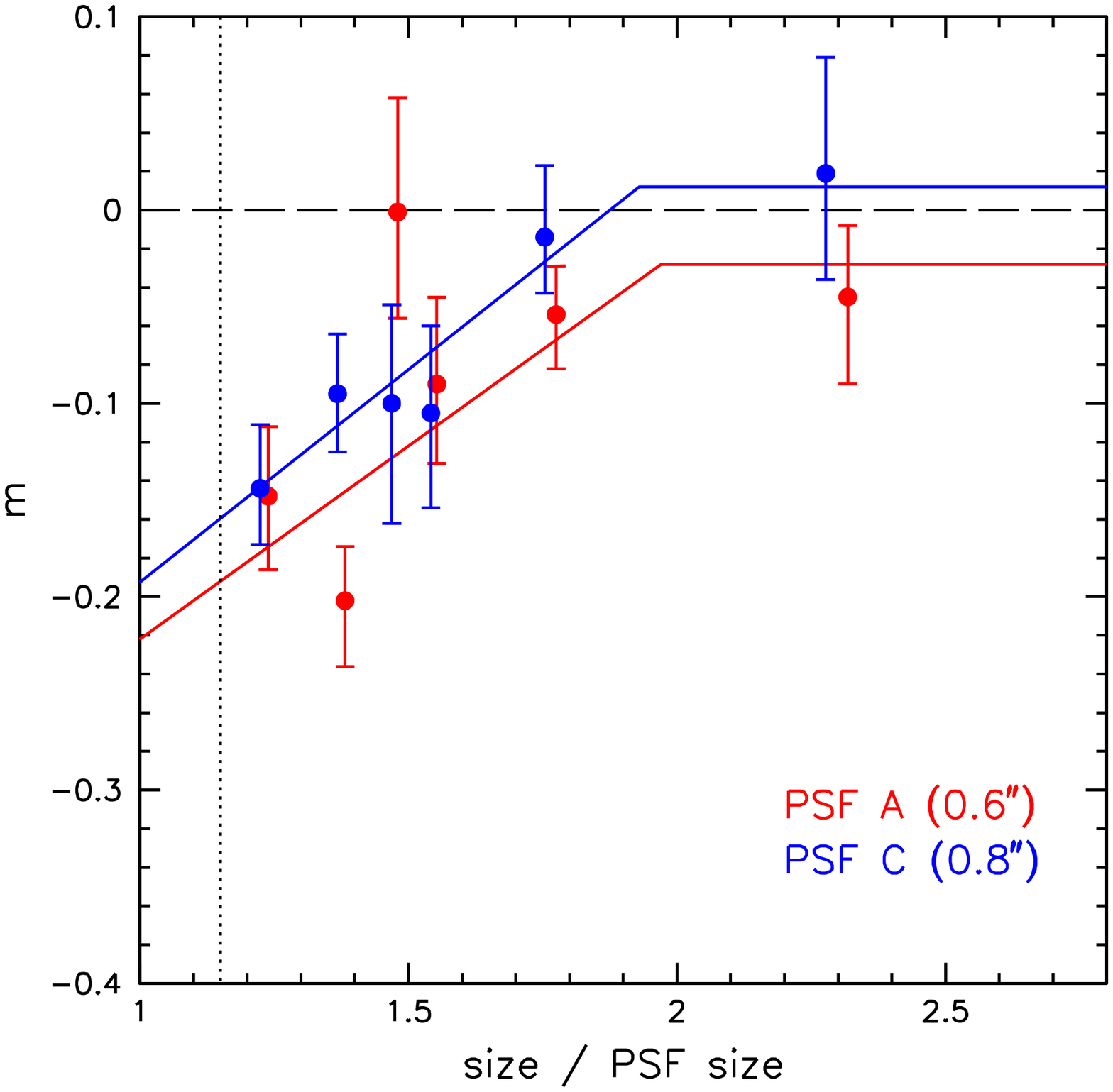}
\caption{The STEP parameter $m$ as a function of shape $S/N$, and
  galaxy size for two different PSFs simulated in STEP2 \citep{step2}
  Left: Galaxies are grouped by $S/N$, as measured by {\sc
    analyseldac}. The parameter $m$ is a sensitive function of $S/N$
  for values $S/N < 6$. The measured $S/N$ values depend on the level
  of correlated noise in the image; STEP2 simulations exhibit much
  stronger correlation (by design) than is present in analysis
  images. Right: Galaxies are grouped by the half-light size, measured
  by {\sc analyseldac} and normalized by the PSF size. Only galaxies
  with shape S/N $> 7$ are used for the right figure. Datapoints show
  the best fit value of $m$ in bins of normalized size.  A dependence
  on the size of the object is clearly seen. We model the size
  dependence of $m$ as linear in size for small objects, and constant
  for large objects, where the break position is a free parameter in
  the model. We use an unbinned fit for the model. (Replicated from
  Paper I.)}
\label{fig:step_size_cor}
\end{figure*}

After our shape S/N $>$ 3 cut to remove false detections
\citep[equivalent to S/N $>$ 12 from {\sc hfindpeaks}, and a STEP2 S/N
$>$ 7][Paper I]{erben01}, we see no S/N dependence in our calibration. The color-cut method uses the best-fit values of
$\sigma$, describing the Voigt profile core width, to weight shear
values when computing the average tangential shear in a radial bin
(Section~\ref{sec:cc_mass_est_proc}). The best-fit results for
$\sigma$ and $\Gamma$, in bins of $S/N$ corresponding to
Fig.~\ref{fig:step_size_cor}a, are shown in
Table~\ref{table:step_snratio}.

We show the results of fitting the STEP $m$ parameter in bins of
galaxy size in Fig.~\ref{fig:step_size_cor}.  We model the clear size dependence
in the STEP2 data using the following fitting function, constrained
in an unbinned analysis. For the definition of the shear likelihood
function $P(\hat g|g)$, we parameterize the parameter $m$ as a
function of the galaxy size relative to the PSF size:

\begin{align}
\label{eq:size_cor}
m & = \begin{cases}
  a\frac{r_{gal}}{r_{PSF}} + b & \text{if} \frac{r_{gal}}{r_{PSF}} < x_p  \notag \\
  b & \text{if} \frac{r_{gal}}{r_{PSF}} \ge x_p   \notag
  \end{cases} \\
c & = const\quad.
\end{align}

We expect the multiplicative bias $m$ to asymptote to zero for galaxies
much larger than the PSF size. We therefore place a Gaussian prior,
with $\mu = 0$ and width $\sigma = 0.03$, on the parameter $b$ in
Eq.~\ref{eq:size_cor} when fitting. We also place a Gaussian prior on
the pivot parameter $x_p$, with $\mu = 0.2$ and width $\sigma =
0.2$. The other parameters ($a$, $b$, $c$) have uniform priors. Due to the small number of galaxies of large size, we
do not consider the
largest objects in the STEP2 images ($\frac{r_{gal}}{r_{PSF}} < 2.5$). 

We
approximate the posterior probability distributions from this model as
a multivariate Gaussian, after marginalizing over the scatter parameters
$\sigma$ and $\Gamma$. We use this multivariate Gaussian as a prior
for the $P(\hat g | g)$ function in the mass modeling described in
Section~\ref{sec:maxlike_method}. Table~\ref{table:step_posterior} shows the best fit values and covariance matrix for the size-dependent shear calibration.

\begin{table}
  \caption{Posterior covariance matrix and best fit values for the size dependent shear calibration, defined in Eq.~\ref{eq:size_cor} and measured from STEP2 images. The covariances of the $c$ parameter are small enough that we set those elements of the matrix to 0. The covariance matrix is approximately the same for both PSFs considered, and is assumed to be constant for mass measurements. }
  \begin{tabular}{r | c c c c}
    \hline
    Covariance ($\times 10^{-3}$) & a & b & $x_p$ & c \\
    \hline
    a & 11.0 & -0.14 & -17.0 & 0.0 \\
    b & -0.14 & 0.6 & 2.9 & 0.0 \\
    $x_p$ & -17.0 & 2.9 & 51.0 & 0.0 \\
    c & 0.0 & 0.0 & 0.0 & 0.4 \\
    \hline
    Best Fit  & & & & \\
    \hline
    PSF A & 0.20 & -0.028 & 1.97 & $-1\times10^{-05}$ \\
    PSF C & 0.22 & 0.012 & 1.93 & $6\times10^{-04}$ \\
    
  \end{tabular}
  \label{table:step_posterior}
\end{table}

For the \pz analysis, we linearly interpolate between the results for
PSFs A \& C to account for the different seeing in each
cluster field. We do not extrapolate the correction to fields with
seeing better than 0.6'' or worse than 0.75'', but instead use the
nearest measured correction. Observations with seeing below 0.5'' are
removed from our analysis because the PSF is undersampled. By implementing a PSF-size dependent STEP
correction, we eliminated what would have been an approximately 8\% systematic uncertainty in the mean
cluster mass, if we had assumed either the STEP set A or set C correction exclusively.



\section{Testing for Mass Biases with \pz Reconstructions}
\label{sec:testing_framework}

Photometric redshifts are inherently more noisy than spectroscopic
redshift measurements. The amount of scatter, and the rate of
outliers, is a strong function of galaxy type and magnitude. Photo-$z$'s
computed with a template based code, such as BPZ, produce a posterior
probability distribution \pz which attempts to characterize the
uncertainty in the measured redshift, within the limits of the assumed
model. In this section, we create simulations based on the COSMOS-30
catalog to test to what extent uncertainties and biases in our \photoz
measurements propagate to mass measurements, given the \photoz cuts we
apply to our catalogs. In addition, we use these simulations to
quantify the effects of cluster galaxy contamination in the catalogs,
since we do not explicitly model the presence of a massive cluster
when we calculate \photoz's. Finally, we investigate the
performance of mass estimator methods that use \photoz point estimates
rather than the full \pz function.


\subsection{Cosmos-30 Based Simulations}
\label{sec:simulations}

For each cluster in the sample, we create a simulated cluster field
with an artificially high density of background galaxies to suppress
shot noise. Galaxy redshifts and photometry are drawn from the
COSMOS-30 catalog, and a measured shear is assigned to galaxies based
on the mass of the simulated cluster. Additional galaxies are added to mimic cluster
contamination. These artificial catalogs are
then passed to \photoz and mass measurement algorithms in the same manner as  real
data.

For our simulations, we assume that the COSMOS-30 redshifts from \citet{ilbert09}
represent the `truth', approximately.  COSMOS-30 used 30 broad and narrow
filter photometry; redshifts were derived using templates optimized
with emission lines to take advantage of the full filter information. \citet{ilbert09} report \photoz accuracy of $\sigma_z /
(1+z) < 0.012$ for $z < 1.25$ and $i_{ab}^+ < 24$. For $24 < i_{ab}^+
< 25,$ performance degrades to $\sigma_z / (1+z) < 0.054$ with a
catastrophic outlier rate of 20\%, where catastrophic is defined as
$\Delta z / (1+z) > 0.15$ \citep[see figure 7 of][]{ilbert09}.

Galaxies from the COSMOS-30 catalog (as identified by the same cuts as
Section~\ref{sec:data_reduction}, when possible), are randomly selected with
replacement to form a blank field with a high number density of galaxies, allowing us to more easily
determine the bias present in our measurements by suppressing shot noise. Galaxies are assigned
a true tangential shear by computing the expected shear appropriate
for a particular NFW halo using Eq.~\ref{eq:reducedshear}, given the
known halo redshift, the COSMOS-30 redshift, and the position of the
galaxy relative to the halo. We then assign noise to create the 'measured'
shear.  In principle, we could assign scatter following the shear
likelihood that we measured in Section~\ref{sec:shape_distribution}. To
reduce computational complexity, we assign Gaussian scatter with
$\sigma = 0.25$. No calibration bias is included. We have checked that using a
Voigt profile as the form of the scatter in our simulations does not
change our conclusions.

When calculating \photoz's for our cluster fields we already know that a
massive cluster is in the field. Ideally, this information should also be
incorporated into the \pz function. Without explicitly modeling the
presence of the cluster, one might expect these cluster galaxies to
bias low the measured mass, as cluster galaxies scatter into the acceptance catalog. We explore the effects of contaminating cluster members in our
simulation by introducing galaxies in the simulated catalogs with zero
net tangential shear (only scatter). We select blue galaxies (COSMOS-30 type $>$ 8) within
$|\Delta z| < 0.05$ of the halo redshift from COSMOS-30 that pass our selection criteria and place them
into the catalogs following an exponential number density profile
centered on the NFW halo:
\begin{equation}
\label{eq:contam_profile}
n(r) = n_{\rm back}f_{500}\exp(1 - r/r_{500})\quad.
\end{equation} %
The parameter $n_{back}$ is the background number density and
$f_{500}$ is the contamination fraction at the halo $r_{500}$. We simulate halos at the redshifts and masses of the clusters in our
sample, with three levels of cluster-galaxy contamination.  We
generate 50 realizations for each redshift, mass, and contamination
set.

We model our \photoz measurements using a subset of the available wide optical
filters in the COSMOS-30 catalog. We follow the same procedure used to
calculate \photoz's as our analysis, including re-calibrating the
photometric zeropoints with stellar locus regression and using BPZ
with modified priors and templates. See Paper II for an exploration of our
\photoz quality.

For our simulation analysis, we accept galaxies to the COSMOS
magnitude limit $i^{+} < 25$. We note, however, that COSMOS-30 galaxies are subject to
multiple \photoz solutions at the faintest magnitudes ($24.5 < i^{+} < 25$;
T. Schrabback, priv. comm), and point estimates adopted by the COSMOS team cannot be taken
as reasonable approximations of the truth. The full \pz
posterior probabilities from the COSMOS-30 study are not published. We have run
alternative simulations with $i^{+} < 24.5$ and do not see any
significant impact on our conclusions.

These simulations discussed in this section model the effects of photometric redshift
uncertainties and deficiencies with respect to cluster
galaxies. We emphasize, however, that these simulations reflect only one
realization of the photometric calibration and \photoz code on one
cosmic variance limited field (although we do not expect much additional
scatter for clusters in the studied redshift range).  We do not simulate the effects of large scale structure,
correlated nearby structure, or triaxiality, which are better handled
with N-body simulations, and can be considered a separable problem.
We also emphasize that these are catalog based
simulations, as we do not simulate the problem of measuring shapes
from images. Mass profile and shear measurements issues are addressed
in Section~\ref{sec:systematics}.


\subsection{\pz Method Performance}
\label{sec:pz_performance}

\begin{figure}
\includegraphics[width=\hsize]{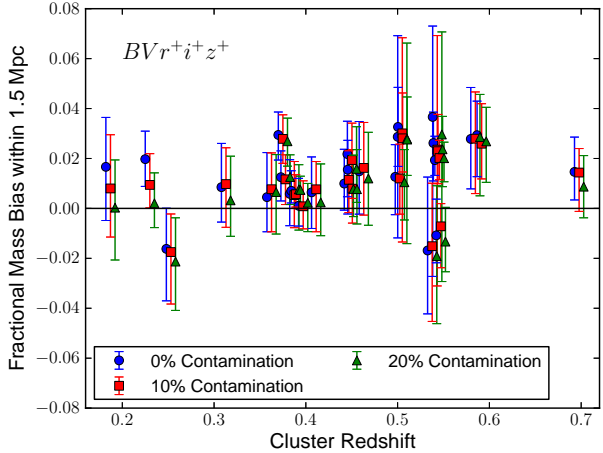}
\caption{Expected fractional bias in the mass within 1.5 Mpc (mean and
  68\% confidence interval on the mean from bootstrapping) for each
  cluster in the \pz sample from high galaxy-density simulations using
  the \BVriz filter set. We show the bias for differing levels of
  contamination from cluster galaxies, offset in redshift for clarity.
  For our redshift range, we detect an overestimate of the mean cluster mass of 1.2\%,
  with a 1\% sensitivity to the assumed cluster galaxy contamination
  level.}
\label{fig:clustersims_bvriz_contam}
\end{figure}

We apply the \pz method to the COSMOS-30 based simulated catalogs
using the \BVriz photometry available in the COSMOS-30 catalog.  In
these results, we mimic the selection cuts that we apply to real
data. This includes applying a size cut, rejecting galaxies with a
photometric redshift point estimate outside the range $z_{cluster} +
0.1 < z < 1.25$ and rejecting galaxies with very wide \pz, $\Delta
z_{95\%} \ge 2.5$. Since we assign shears to the galaxies directly, we
cannot replicate shape measurement quality cuts. Unless otherwise
stated, we assume that we know the form of the shear scatter (which for simplicity in the simulations is a Gaussian
with standard deviation $\sigma_{\rm g} = 0.25$). The results are shown in
Fig.~\ref{fig:clustersims_bvriz_contam}. When using \BVriz photometry,
the expected bias on the mean mass for any single cluster mass and
redshift combination never exceeds $\pm5\%$. Furthermore, as we show
below, the statistical constraints on the mean bias for the sample is
significantly better.

We use these redshift- and mass-matched simulations to derive a
composite bias and systematic uncertainty for our cluster
sample. We fit for a constant ratio between the recovered mass and the
underlying true mass assuming a per-cluster Gaussian
scatter as reported by the fit to each realization.
Table~\ref{table:bvriz_simcal} shows the best fit results for the ratio
and 68\% confidence interval for each of the different contamination
levels. The posterior probability distribution for the ratio is well
described by a Gaussian, and we quote the results accordingly. We also
checked for the presence of a normally distributed intrinsic scatter
component; none was detected. 

For a typical level of cluster galaxy contamination (10\% at
$R_{500}$), the expected multiplicative bias over the sample due to
the effects of \photoz uncertainties and catastrophic outliers is a 1.2\% overestimate of the mean cluster mass.
The statistical
$1\sigma$ uncertainties on this bias are less than 1\%. The presence
of cluster galaxies has a minimal impact on the bias, ranging from 1.4\% to 1.1\% as the contamination fraction varies from 0 to 20\%. 
We can apply this overestimate of 1.2\% as a correction to the \pz masses.
Because we do not know the average contamination
for our cluster sample, but expect it to be $\sim10\%$ (see
Section~\ref{sec:color_cuts}), we quote a systematic uncertainty that
spans the calibration results from 0\% to 20\% contamination. Even
with this caveat, our systematic uncertainty on the bias from \photoz
errors, given the performance in the COSMOS field,  is at most 1\%.

\begin{table}
  \caption{Summary of the sample bias and uncertainty from simulations due to photometric redshift errors while using \BVriz filters, for different fractions of cluster galaxy contamination. (1) The fraction of galaxies that are cluster members at $R_{500}$ with respect to field galaxies (Eq~\ref{eq:contcorr})  (2) The mean fractional bias for measured masses within 1.5 Mpc, for the population of 26 galaxy clusters simulated, and 1$\sigma$ uncertainties. The posterior probability distributions for the mean fractional mass bias are well modeled by a Gaussian.}
  \begin{tabular}{c | c }
    \hline
    Contamination  Fraction  & Mean Fractional Mass Bias    \\ 
    (1) & (2) \\ \hline
    0\% & 1.014 $\pm$ 0.003   \\ 
    10\% & 1.012 $\pm$ 0.003  \\ 
    20\% & 1.011 $\pm$ 0.003  \\ 
  \end{tabular}
\label{table:bvriz_simcal}
\end{table}

We also ran simulations using shear scatter that follows a Voigt
profile instead of a Gaussian. We ran simulations with both one and
two populations of galaxies with different scatter parameters
($\sigma, \Gamma$) and reconstructed the masses using a wide uniform
prior on $\sigma$ and $\Gamma$. We see no significant change in our
results.  In an additional test, we reconstructed masses from
simulations with shear scatter $\sigma_{g} = 0.20$, but fixing the
prior to $\sigma_{g} = 0.16$. Such an error leads to an underestimate
in the recovered mass of $\approx5\%$, emphasizing the need for
flexible priors on scatter parameters.


\subsection{Point Estimator Performance}
\label{sec:point_est_performance}

To emphasize the importance of using the full \pz information, we
have also examined an alternative mass reconstruction method that uses only \photoz point estimators.
A point estimator is usually the redshift at which the \pz is maximum, though it could in principle also be the mean
or the median of the posterior probability. BPZ reports the most likely redshift,
marginalized over all templates, as its point estimator $z_b$. 

\begin{figure}
\includegraphics[width=0.95\columnwidth]{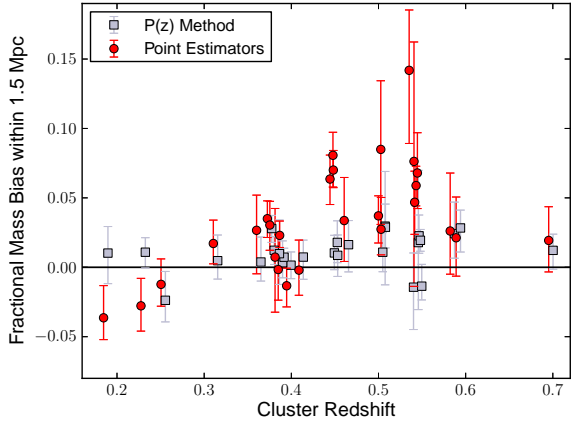}
\caption{Simulated fractional bias in cluster mass reconstructions
  when only photometric redshift point estimators are used, based on \BVriz photometry. The
  simulated input catalogs are identical to those used in
  Fig.~\ref{fig:clustersims_bvriz_contam} with 10\% contamination,
  though different analysis cuts are applied. For reference, the \pz method results are shown in light grey, and are offset slightly in redshift. A significant redshift dependent bias is
  clearly seen when using the point estimators.}
\label{fig:newman_bias}
\end{figure}

We use an alternative set of simulations that include galaxies at all
redshifts, and have our red sequence cuts applied. Assuming Gaussian
scatter for the tangential shears (which is an accurate assumption for
the baseline simulations), we perform a $\chi^2$ fit between measured
and predicted shear using only the \photoz point estimator:
\begin{align}
\chi^2 & = \sum_{i} \frac{\left (\hat g_i - g(z, M)\right )^2}{(f\sigma)^2} \quad.
\end{align} %
We follow the implementation outlined in \citet{newman09, newman11},
which includes a slight inflation of the assumed $\sigma$ between
model and measurement, $f = 1.02$, to account for the uncertainty
in redshift estimates. Galaxies are excluded from the fit if 25\% of
the probability in the \pz is at $z \le z_{cluster}$. Simulation
results using photometric redshifts calculated with both u\BVriz and
\BVriz photometry, with cluster galaxy contamination normalized to
10\% at $R_{500}$, are provided in Figure~\ref{fig:newman_bias}.  This
method and associated cuts display a clear redshift-dependent bias. At higher
redshifts ($z \ge 0.4$), the use of \photoz point estimators leads to a $\approx$7\%
systematic bias.

\section{Mass Measurement Results}
\label{sec:final_masses}

In this section, we report the masses measured for each cluster using
both the color-cut and \pz methods. Table~\ref{table:mass_results}
lists the best fit masses and 68\% statistical confidence intervals for each
method. Mass point estimates are median values, with 68\% confidence
intervals defined as the 16th to the 84th percentile values.  Sections~\ref{sec:color_cuts} and \ref{sec:maxlike_method}
detail how the statistical uncertainties are derived for each
method. The uncertainties on individual cluster masses from
triaxiality, line-of-sight structure, and correlated structure are not
included in the quoted confidence intervals. In combination, these effects can
be expected to add to the uncertainty in individual masses at the 20\%
level \citep{becker11, hhh11, cok07}. The overall systematic uncertainty from these
sources on the mean cluster-sample mass are addressed separately in
Section~\ref{sec:systematics}.

\begin{table*}
  \caption{Lensing masses from the \pz method and the color-cut method. (1) cluster name, (2) cluster redshift; Columns (3) \& (4) report results from the \pz method: (3) median scale radius and 68\% confidence interval after marginalization; (4) median mass within 1.5 Mpc of the cluster center and 68\% confidence interval after marginalization. Columns (5) \& (6) report results from the Color-Cut Method: (5) median fit scale radius and 68\% confidence interval (concentration set to four); (6) median mass within 1.5Mpc of the cluster center and 68\% confidence interval.  Masses are in units of $10^{14}M_{\odot}$. \pz masses do not include the calibration correction from Section~\ref{sec:pz_performance}. }
  \begin{tabular}{ l c || c c || c c }
    \hline
    & & \multicolumn{2}{c ||}{\pz Method} & \multicolumn{2}{c}{Color-Cut Method} \\ \cline{3-6} 
    Cluster & Redshift & $r_s$ & $M(<1.5 {\rm Mpc})$ & $r_s$ & $M(<1.5 {\rm Mpc})$  \\
    & & Mpc & $10^{14}M_{\odot}$ & Mpc & $10^{14}M_{\odot}$ \\
    (1) & (2) & (3) & (4) & (5) & (6)  \\
    \hline
    A2204 & 0.152 &  - & - & $0.65^{+0.04}_{-0.04}$ & $14.5^{+1.8}_{-1.9}$ \\ 
A750 & 0.163 &  - & - & $0.52^{+0.05}_{-0.06}$ & $9.1^{+2.0}_{-2.0}$ \\ 
RXJ1720.1+2638 & 0.164 &  - & - & $0.40^{+0.05}_{-0.06}$ & $5.2^{+1.6}_{-1.6}$ \\ 
A383 & 0.188 &  $0.46^{+0.15}_{-0.12}$ & $7.3^{+1.4}_{-1.4}$ & $0.45^{+0.04}_{-0.04}$ & $7.1^{+1.4}_{-1.4}$ \\ 
A209 & 0.206 &  - & - & $0.56^{+0.03}_{-0.04}$ & $11.3^{+1.5}_{-1.5}$ \\ 
A963 & 0.206 &  - & - & $0.42^{+0.05}_{-0.05}$ & $5.9^{+1.5}_{-1.5}$ \\ 
A2261 & 0.224 &  - & - & $0.63^{+0.03}_{-0.03}$ & $14.4^{+1.5}_{-1.5}$ \\ 
A2219 & 0.228 &  - & - & $0.57^{+0.03}_{-0.03}$ & $12.0^{+1.5}_{-1.5}$ \\ 
A2390 & 0.233 &  $0.41^{+0.19}_{-0.08}$ & $10.2^{+1.8}_{-1.7}$ & $0.56^{+0.04}_{-0.04}$ & $11.6^{+1.8}_{-1.8}$ \\ 
RXJ2129.6+0005 & 0.235 &  - & - & $0.39^{+0.05}_{-0.06}$ & $5.4^{+1.7}_{-1.6}$ \\ 
A521 & 0.247 &  $0.42^{+0.14}_{-0.11}$ & $8.1^{+1.6}_{-1.6}$ & $0.50^{+0.04}_{-0.04}$ & $9.3^{+1.5}_{-1.5}$ \\ 
A1835 & 0.253 &  - & - & $0.62^{+0.06}_{-0.06}$ & $14.3^{+2.8}_{-2.8}$ \\ 
A68 & 0.255 &  - & - & $0.52^{+0.03}_{-0.03}$ & $9.9^{+1.3}_{-1.3}$ \\ 
A2631 & 0.278 &  - & - & $0.56^{+0.03}_{-0.03}$ & $12.2^{+1.4}_{-1.4}$ \\ 
A1758N & 0.279 &  - & - & $0.61^{+0.03}_{-0.03}$ & $14.5^{+1.5}_{-1.5}$ \\ 
RXJ0142.0+2131 & 0.280 &  - & - & $0.42^{+0.05}_{-0.06}$ & $6.6^{+1.7}_{-1.8}$ \\ 
A611 & 0.288 &  - & - & $0.50^{+0.04}_{-0.05}$ & $9.5^{+1.9}_{-1.8}$ \\ 
Zw7215 & 0.290 &  - & - & $0.43^{+0.04}_{-0.04}$ & $6.9^{+1.3}_{-1.3}$ \\ 
A2552 & 0.302 &  - & - & $0.58^{+0.04}_{-0.04}$ & $13.4^{+1.8}_{-1.8}$ \\ 
MS2137.3-2353 & 0.313 &  $0.40^{+0.17}_{-0.09}$ & $8.1^{+1.7}_{-1.7}$ & $0.41^{+0.04}_{-0.05}$ & $6.6^{+1.5}_{-1.6}$ \\ 
MACSJ1115.8+0129 & 0.355 &  - & - & $0.51^{+0.04}_{-0.05}$ & $10.9^{+2.1}_{-2.1}$ \\ 
RXJ1532.8+3021 & 0.363 &  $0.37^{+0.16}_{-0.08}$ & $9.5^{+2.3}_{-2.3}$ & $0.48^{+0.05}_{-0.05}$ & $9.5^{+2.0}_{-2.0}$ \\ 
A370 & 0.375 &  $0.53^{+0.21}_{-0.12}$ & $15.4^{+2.0}_{-2.0}$ & $0.60^{+0.03}_{-0.03}$ & $15.6^{+1.7}_{-1.7}$ \\ 
MACSJ0850.1+3604 & 0.378 &  $0.55^{+0.23}_{-0.13}$ & $15.8^{+2.6}_{-2.6}$ & $0.62^{+0.06}_{-0.06}$ & $16.9^{+3.2}_{-3.2}$ \\ 
MACSJ0949.8+1708 & 0.384 &  $0.41^{+0.13}_{-0.15}$ & $9.4^{+2.9}_{-3.0}$ & $0.50^{+0.07}_{-0.08}$ & $10.6^{+3.6}_{-3.5}$ \\ 
MACSJ1720.2+3536 & 0.387 &  $0.51^{+0.17}_{-0.15}$ & $13.7^{+2.8}_{-2.7}$ & $0.47^{+0.06}_{-0.08}$ & $9.6^{+3.0}_{-3.0}$ \\ 
MACSJ1731.6+2252 & 0.389 &  - & - & $0.67^{+0.04}_{-0.04}$ & $19.7^{+2.2}_{-2.2}$ \\ 
MACSJ2211.7-0349 & 0.397 &  $0.52^{+0.11}_{-0.17}$ & $16.9^{+2.1}_{-2.2}$ & $0.56^{+0.04}_{-0.04}$ & $14.0^{+2.2}_{-2.1}$ \\ 
MACSJ0429.6-0253 & 0.399 &  - & - & $0.51^{+0.05}_{-0.05}$ & $11.3^{+2.5}_{-2.4}$ \\ 
RXJ2228.6+2037 & 0.411 &  $0.44^{+0.19}_{-0.10}$ & $11.6^{+2.7}_{-2.7}$ & $0.51^{+0.04}_{-0.05}$ & $11.3^{+2.1}_{-2.0}$ \\ 
MACSJ0451.9+0006 & 0.429 &  - & - & $0.44^{+0.06}_{-0.07}$ & $8.4^{+2.7}_{-2.8}$ \\ 
MACSJ1206.2-0847 & 0.439 &  - & - & $0.50^{+0.06}_{-0.07}$ & $11.2^{+3.1}_{-3.2}$ \\ 
MACSJ0417.5-1154 & 0.443 &  - & - & $0.64^{+0.04}_{-0.04}$ & $18.9^{+2.6}_{-2.5}$ \\ 
MACSJ2243.3-0935 & 0.447 &  $0.59^{+0.22}_{-0.17}$ & $15.7^{+2.5}_{-2.6}$ & $0.63^{+0.04}_{-0.04}$ & $18.6^{+2.4}_{-2.3}$ \\ 
MACSJ0329.6-0211 & 0.450 &  $0.41^{+0.17}_{-0.09}$ & $12.6^{+2.4}_{-2.4}$ & $0.52^{+0.04}_{-0.04}$ & $12.8^{+2.2}_{-2.1}$ \\ 
RXJ1347.5-1144 & 0.451 &  $0.51^{+0.21}_{-0.14}$ & $14.2^{+3.0}_{-3.0}$ & $0.53^{+0.05}_{-0.05}$ & $13.1^{+2.6}_{-2.6}$ \\ 
MACSJ1621.3+3810 & 0.463 &  $0.39^{+0.14}_{-0.11}$ & $8.5^{+2.3}_{-2.1}$ & $0.44^{+0.05}_{-0.05}$ & $8.8^{+2.2}_{-2.1}$ \\ 
MACSJ1108.8+0906 & 0.466 &  - & - & $0.43^{+0.06}_{-0.08}$ & $8.4^{+2.9}_{-3.0}$ \\ 
MACSJ1427.2+4407 & 0.487 &  - & - & $0.36^{+0.06}_{-0.08}$ & $5.9^{+2.4}_{-2.5}$ \\ 
MACSJ2214.9-1359 & 0.502 &  $0.61^{+0.16}_{-0.21}$ & $14.7^{+2.7}_{-2.6}$ & $0.48^{+0.04}_{-0.04}$ & $11.5^{+2.3}_{-2.1}$ \\ 
MACSJ0911.2+1746 & 0.505 &  $0.35^{+0.13}_{-0.12}$ & $7.1^{+2.7}_{-2.8}$ & $0.50^{+0.05}_{-0.05}$ & $12.2^{+2.6}_{-2.6}$ \\ 
MACSJ0257.1-2325 & 0.505 &  $0.46^{+0.15}_{-0.13}$ & $11.9^{+3.0}_{-3.0}$ & $0.54^{+0.04}_{-0.05}$ & $14.5^{+2.5}_{-2.5}$ \\ 
MS0451.6-0305 & 0.538 &  $0.39^{+0.15}_{-0.11}$ & $8.8^{+3.3}_{-3.2}$ & $0.49^{+0.06}_{-0.07}$ & $12.5^{+3.5}_{-3.7}$ \\ 
MACSJ1423.8+2404 & 0.543 &  $0.25^{+0.11}_{-0.11}$ & $3.7^{+2.8}_{-2.2}$ & $0.42^{+0.07}_{-0.09}$ & $8.8^{+3.6}_{-3.6}$ \\ 
MACSJ1149.5+2223 & 0.544 &  $0.49^{+0.18}_{-0.13}$ & $14.4^{+3.3}_{-3.3}$ & $0.51^{+0.05}_{-0.06}$ & $13.6^{+3.1}_{-3.1}$ \\ 
MACSJ0717.5+3745 & 0.546 &  $0.68^{+0.27}_{-0.18}$ & $25.3^{+4.1}_{-4.2}$ & $0.66^{+0.05}_{-0.06}$ & $23.1^{+3.7}_{-3.8}$ \\ 
CL0016+16 & 0.547 &  $0.54^{+0.18}_{-0.15}$ & $15.0^{+3.7}_{-3.6}$ & $0.58^{+0.05}_{-0.05}$ & $17.5^{+3.2}_{-3.1}$ \\ 
MACSJ0025.4-1222 & 0.585 &  $0.41^{+0.14}_{-0.13}$ & $11.5^{+3.0}_{-3.1}$ & $0.48^{+0.05}_{-0.06}$ & $12.3^{+2.9}_{-2.9}$ \\ 
MACSJ2129.4-0741 & 0.588 &  - & - & $0.52^{+0.05}_{-0.05}$ & $15.1^{+3.2}_{-3.1}$ \\ 
MACSJ0647.7+7015 & 0.592 &  $0.45^{+0.19}_{-0.14}$ & $13.3^{+5.7}_{-5.6}$ & $0.52^{+0.08}_{-0.10}$ & $14.9^{+5.2}_{-5.3}$ \\ 
MACSJ0744.8+3927 & 0.698 &  $0.48^{+0.19}_{-0.12}$ & $20.5^{+5.7}_{-5.7}$ & $0.56^{+0.06}_{-0.06}$ & $20.0^{+4.5}_{-4.4}$ \\ 

  \end{tabular}
\label{table:mass_results}
\end{table*}

\begin{figure}
\includegraphics[width=\hsize]{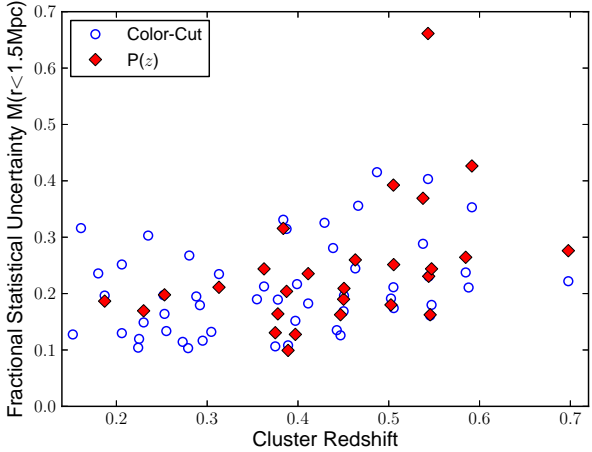}
\caption{A comparison of the precision of the mass measurement for all
  clusters in the sample, measured as the fractional uncertainty in
  the measured mass, for the color-cut (51 clusters) and the \pz
  methods (27 clusters). The asymmetric uncertainties listed in
  Table~\ref{table:mass_results} have been averaged for this
  plot. Both methods achieve a similar level of precision, with a \pz
  to color-cut statistical-error mean ratio of 1.16. The outlier at
  $z=0.54$ is MACS1423+24 \citep[see also][]{lem2010}.}
\label{fig:aveerr_redshift}
\end{figure}

First, we compare the statistical precision with which the color-cut
method and the \pz method constrain cluster
masses. Figure~\ref{fig:aveerr_redshift} shows the fractional
uncertainty measured for each cluster in our sample, as a function of
redshift. Recall that \pz masses are only determined for 27 of the
51 clusters. Both methods measure cluster masses with comparable
statistical precision, despite the lower number of galaxies available for the
\pz method analysis. The mean ratio of \pz to color-cut 68\% statistical
errors is 1.16. The mild loss of precision for both methods at
high redshift is driven primarily by the decrease in the angular size
of the clusters, and thus the smaller number of galaxies accepted into
the shear profile fit.

\begin{figure}
\includegraphics[width=\columnwidth]{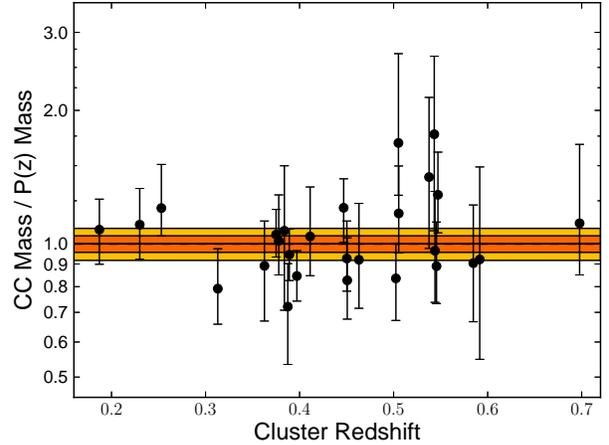}
\caption{A comparison of masses recovered from the color-cut (CC) and the \pz methods. Error bars for each cluster point are determined by
  bootstrapping the input catalog for both methods
  simultaneously. Points are the median ratio and 68\% confidence
  interval for each cluster from the bootstrap realizations. \pz masses do not include the calibration correction from Section~\ref{sec:pz_performance}. The
  dashed line and red shaded region is the best fit ratio between the
  two methods, $\beta = 0.999^{+0.046}_{-0.041}$, for all 27 clusters with \BVRiz photometry.  }
\label{fig:ml_cc_comp_z}
\end{figure}

Next, we compare the mass estimates from the
photometric redshift based \pz and the color-cut methods, to
check for internal consistency. Figure~\ref{fig:ml_cc_comp_z} shows the ratio between the
color-cut and \pz method for each cluster, and for the sample as a
whole. The mass measurements from our two methods are correlated
because the same galaxy catalogs are used as input for each. We bootstrap the last common
galaxy catalog for each cluster to determine the correlated
uncertainties between the two methods. 

For individual cluster masses, scatter between the two methods arises due to the effects of cosmic
variance in the color-cut method and the differences in galaxy selection. Correlation between the two methods should
decrease with increasing cluster redshift as the effects of cosmic variance become more
pronounced and galaxy selection diverges. The uncertainty in the cross-calibration
ratio at $z < 0.4$ is $\sim20\%$, while in contrast, uncertainty on the ratio at $z > 0.4$ grows to
$\sim40\%$ at the highest redshifts.

We use the bootstrapped masses to measure the ratio and
intrinsic scatter between the two methods. Any systematic offset between
the two methods would most likely indicate one or more of the following:
systematic errors in the color-cut method, arising from a mismatch
between the COSMOS field and the average cluster field; the use of the
wrong number density model or an incorrect estimate of the galaxy
density background level for the color-cut method's contamination
correction; or bias in the color-cut masses due to using the single
source plane approximation. Additionally, intrinsic scatter between
the methods may be induced if systematic scatter exists in the
derived field-galaxy redshift distribution (with respect to COSMOS) or
the contamination correction in the color-cut method. The severity of
any systematic bias and scatter (statistical or intrinsic) is expected
to worsen for higher redshift fields (see
Section~\ref{sec:color_cut_sys_errs}), but we see no evidence for a redshift dependence given the noise level present in our data.

To measure an offset, we fit for the ratio $\beta$ between the color-cut and \pz masses,
with an additional intrinsic, log normal scatter with width
$\sigma_{int}$. Assuming uniform priors, the model likelihood is
\begin{equation}
P(\beta, \sigma_{int}) \propto \prod_i\iint\limits_{M_{i,p(z)},M_{i,cc}}\!N\left (\ln \frac{M_{i,cc}}{\beta M_{i,p(z)}}, \sigma_{\rm int}\right )P(M_{i,p(z)}, M_{i,{\rm cc}})\dif \vec M,
\label{eq:method_comp}
\end{equation}
where the correlated uncertainties between the two measurements,
$P(M_{i,p(z)},M_{i,cc})$ are defined by bootstrap sampling on the last
common galaxy catalog.  We numerically integrate the marginalization
integral by converting it to a sum over bootstrap samples. Note that
this formulation breaks down for small values of $\sigma_{int}$ due to
the limited number of available bootstrap samples; we therefore only
consider $\sigma_{int} \ge 0.02$. The best fit value for the offset is $\beta = 0.999^{+0.046}_{-0.041}$, showing no offset between the \pz and
color-cut methods. We do not claim a detection of intrinsic scatter
between the two methods, and instead measure a $2\sigma$ upper bound
at 15\%. We must stress that these calibration results apply to our particular implementation of the color-cut
method \emph{only}.


\section{Overall Systematic Uncertainty  \& Cross-Checks}
\label{sec:systematics}

Systematic biases can enter a lensing analysis from three primary
sources: galaxy shape measurements; the mass model; and the
already discussed uncertainties associated with the redshift
distribution (or, as in the case of the \pz method, propagation of \pz
errors directly into mass measurements).  Accurate quantification of
these uncertainties is particularly important to maintain the power
and integrity of cosmological constraints \citep[e.g.][]{ars08, mae08,
  mantz10a, heidi10, aem11}.  In this section, we estimate the level
of each of these systematic uncertainties. We also present a series of
cross-checks. Since the primary goal of this series of papers is to
calibrate mass proxies for cosmological studies, we will concentrate
our discussion on systematic uncertainties affecting the measured mean-mass
of our sample.  Section~\ref{sec:systematics_summary}
provides a summary of all significant systematic uncertainties in the
analysis.


\subsection{Shape Measurement Uncertainty}
\label{sec:systematics_shape}

The dominant systematic uncertainty for lensed-galaxy shape
measurements is the shear correction derived from STEP2. Both the
precision to which we can measure the shear calibration using STEP2
simulations, and the differences between our measurements and what was
simulated in STEP2, contribute to the systematic uncertainty in the mean
cluster mass.

Our shear calibration model (Eq.~\ref{eq:size_cor}) has 4 free
parameters at a fixed PSF size. We measure the mean values and
covariance of these parameters from STEP2 simulations. The finite size
of the STEP2 simulations places a limit on our ability to
constrain the correction parameters. While the uncertainty on the
shear correction is subdominant to the statistical noise for
\textit{individual clusters} (this is marginalized over in the \pz
method and ignored in the color-cut method), the shear correction
scatters coherently for the cluster sample and will affect the cluster
mean-mass. Most importantly, the multiplicative shear bias, m, will
scale the mean-mass of the sample approximately linearly with $(1+m)$.

We approximate the uncertainty on the mean mass arising from the
multiplicative shear bias by measuring the distribution width of
$(1+m)$ at fixed object size. The distribution of $(1+m)$ is
approximately Gaussian, with a standard deviation no larger than 3\%
of the correction value, for objects at nearly all sizes for STEP
image sets A and C (on which we base our correction). We verified that
this result is insensitive to the prior we place on the shear
correction model (Eq.~\ref{eq:size_cor}). A 3\% uncertainty is
conservative, as it roughly represents a vertical translation to the
curves shown in Fig.~\ref{fig:step_size_cor}. The detailed correlations
between correction parameters, convolved with the object shape
distributions in each cluster, will likely result in smaller mean mass
variations.

In addition to the statistical limits of the STEP2 calibration, we are 
susceptible to the finite sample of PSF's tested in STEP2. Shear
calibration corrections depend on details of the PSF size and
ellipticity \citep{step1, step2}. While the STEP2 simulations are designed to mimic SuprimeCam
observations, the simulations do not span the entire space of observed
PSFs in our observations. We estimate our corrections from two STEP2 image
sets, A (seeing 0.6'') \& C (seeing 0.75''), that represent well
behaved observations. 

\begin{figure*}
\includegraphics[width=0.45\textwidth]{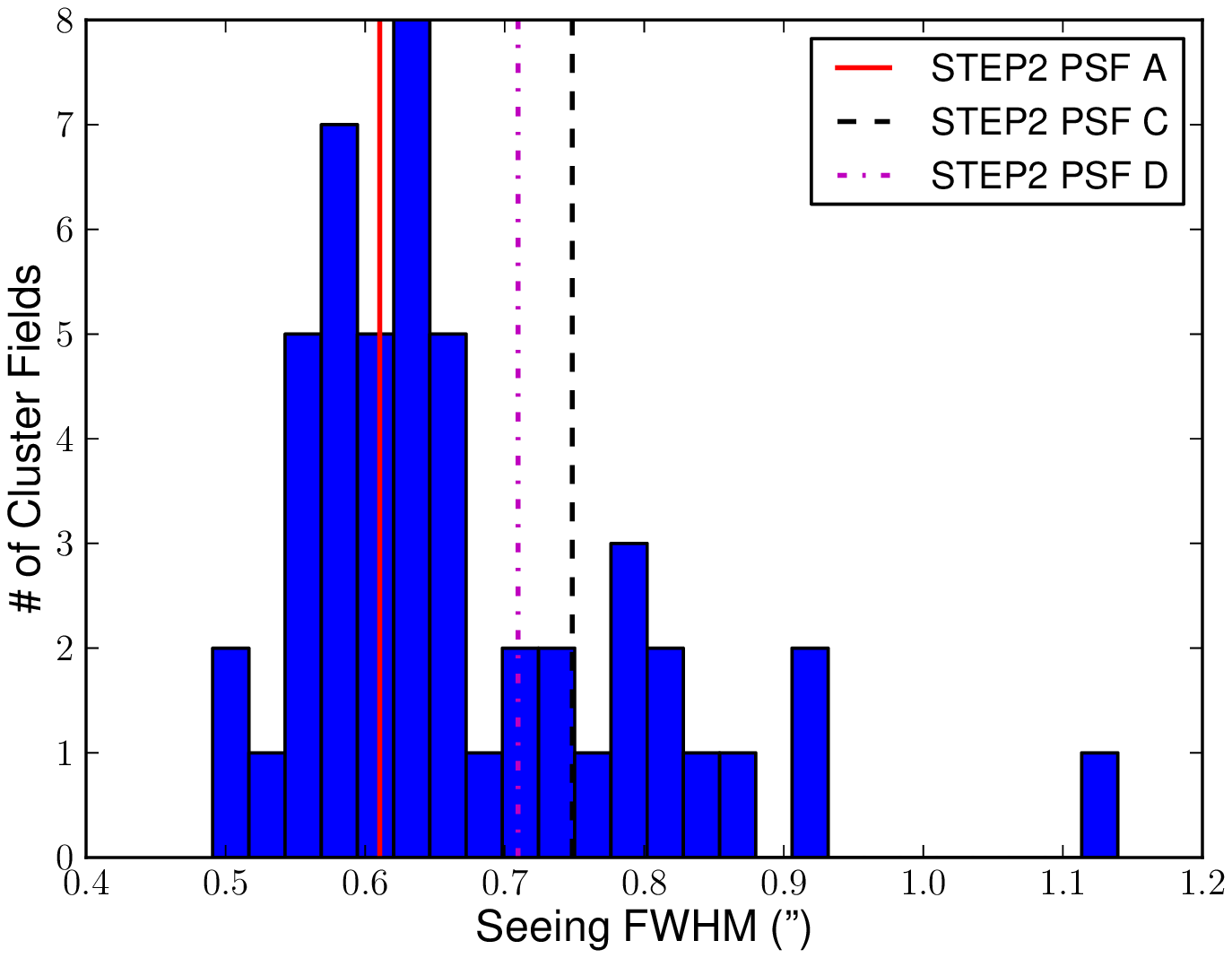}
\includegraphics[width=0.45\textwidth]{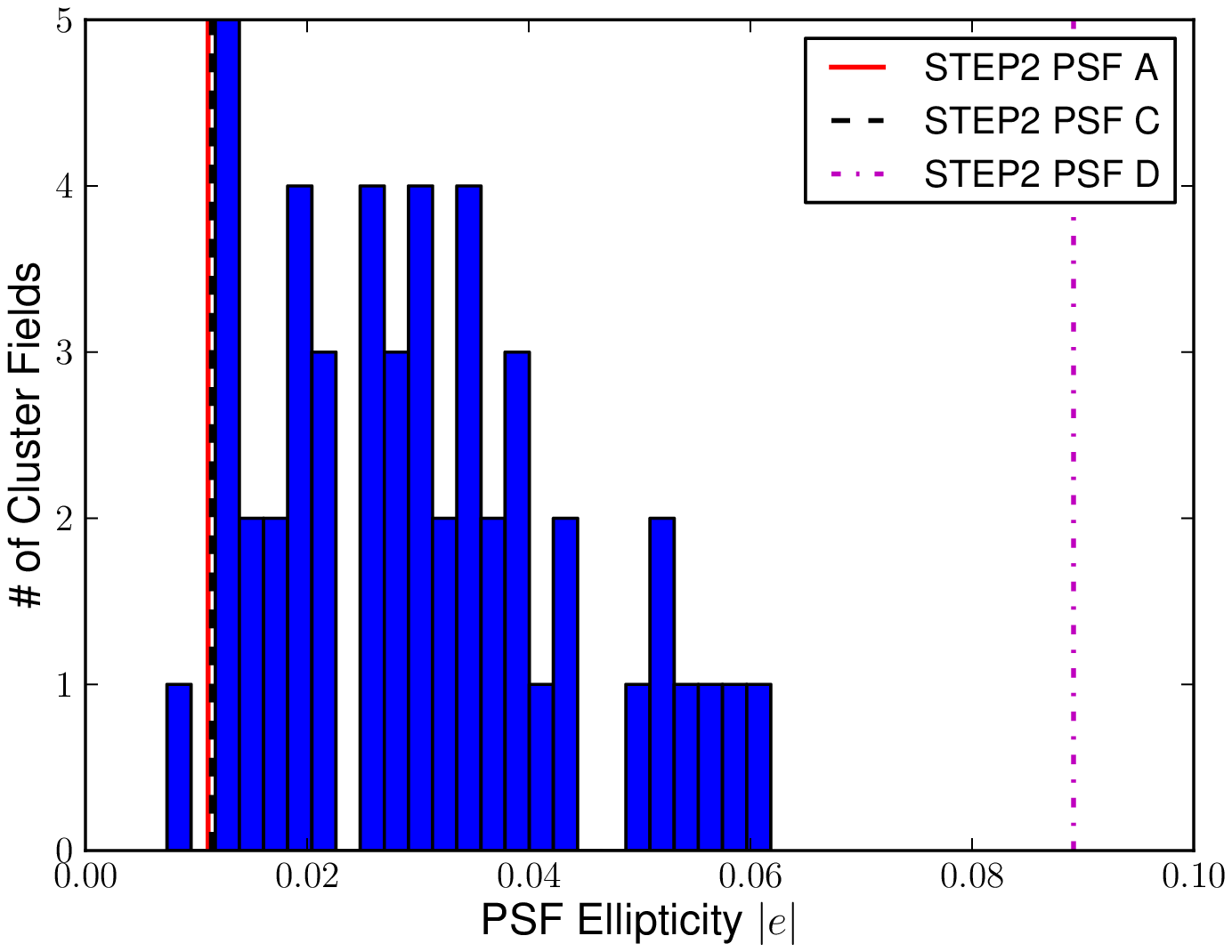}
\caption{Left: The distribution of seeing for cluster observations in
  the comparison set. Right: The ellipticity distribution for those
  same fields. We take the median ellipticity magnitude from stars in
  a field as the PSF ellipticity. Vertical lines mark values for STEP2
  image sets.}
\label{fig:obs_conditions}
\end{figure*}

Figure~\ref{fig:obs_conditions} shows the measured PSF size and
ellipticity distributions for our images, with the STEP2 image sets
marked for comparison. A PSF size of 0.6'' is typical of our
observations, with 0.75'' seeing bracketing the majority of our
observations from above. The mean mass does
not depend sensitively to the details of how we interpolate between the two shear
corrections. If we instead linearly extrapolate the shear correction
beyond the seeing spanned by the STEP2 images (to less than 0.6'' or
larger than 0.75''), the mean mass shifts by no more than 1\% 
(ie, most of our observations are within the interpolation regime).

In addition to the PSF size, STEP2 simulations only coarsely span the
PSF ellipticities observed in the cluster fields. We compare the
distribution of PSF ellipticity magnitudes from the data to available
STEP2 images in Figure~\ref{fig:obs_conditions}. Our observations tend
to have a more elliptical PSF than the baseline image sets A(0.6'',
$\epsilon=0.01$) and C(0.75'', $\epsilon=0.01$), but are significantly
less elliptical than set D (0.7'', $\epsilon=0.09$) .  If we were to
compare the average cluster masses recovered from applying a
correction derived purely from the STEP2 high ellipticity image set,
D, to the baseline correction (interpolating between sets A and C),
our masses would be 2\% lower. Since the shear correction derived from
image set D does not take into account the known size dependence, this
2\% difference is an upper bound on the systematic uncertainty on our
mean cluster mass measurement from unmodeled size and ellipticity
dependencies in our shear calibration.

Because of the STEP2 simulation's use of realistic galaxy shapes, we
do not expect a significant systematic bias from a mismatch in the
galaxy population sampled in the STEP2 images compared to our Subaru
observations.

Additional systematic uncertainties from shape measurements may arise
from either the image coaddition process, or how the PSF model is
interpolated over each coadded field. Section 5.5 and Appendix B from
Paper I detail extensive checks of the PSF and coaddition
procedure. Cluster masses measured from image coadditions for
different camera rotations, and different nights of comparable seeing,
showed no signs of systematic bias. In addition, our measured cluster
masses are insensitive to the particular polynomial order used to
interpolate the PSF across the field of view, above a minimum order
optimized for each observation. From our analysis in Paper I, we
estimate that the systematic uncertainty associated with these sources
is no more than 1\%.

Stellar contamination in the lensing catalogs can cause a systematic
dilution of lensing masses. For our analysis, we have selected
galaxies by only accepting objects at least 15\% larger than the PSF
size. Masses are consistent at the 0.5\% level when we employ a PSF
size cut of 20\%, taking into account the correlations between such
mass measurements, implying that our standard 15\% size cut is
sufficient.  We have also visually inspected the color-color
photometry diagrams for a stack of 30 cluster catalogs for evidence of
a stellar locus; no stellar locus was observed.

In conclusion, the total systematic uncertainty associated with shear
measurements is 4\% for our analysis.


\subsection{Mass Model Uncertainty}
\label{sec:astro_model_uncertainty}

In this section, we consider the uncertainties associated with modeling our observations as spherical NFW halos.
We discuss possible biases from (in order): selection effects, our adopted radial fit range, deviations from an NFW density profile, our adopted mass-concentration relation, and profile centering.

In the limit of perfect shear and \photoz measurements, simulations and
analytical investigations show that the application of spherical NFW models to lensing analyses will suffer a $\sim$20\% system-to-system scatter, although the overall mean sample bias should be small if the observed sample draws fairly from the triaxial distribution of halos \citep[and if one is careful about
the radial range considered:][]{becker11, cok07}.  We expect no
significant systematic bias originating from the selection function for our sample, since our
clusters are X-ray selected and therefore should approximately fairly
sample all possible orientation angles. However, the finite number of clusters in our sample limits
our ability to average over triaxial orientations and we therefore expect some residual scatter in the sample mean mass from this source.  For our full set of 51
clusters, we expect a systematic uncertainty of
$20\%/\sqrt{51} \approx 3\%$.
This is a representative calculation; any comparison with other mass proxies should explicitly fit and account for an intrinsic scatter when measuring a normalization.

While an NFW halo model is the traditional choice for lensing analyses,
some recent works have argued that the NFW model is a poor
description of dark matter halos beyond the virial radius
\citep{ogh11, bahe2011}. Halo density profiles drop off faster than
NFW beyond the virial radius, before a stochastic ``two-halo''
term becomes important at much larger radii. Multiple authors
\citep[e.g.][]{becker11, ogh11, bahe2011} have shown that fitting an NFW
profile to arbitrarily large projected radii can result in significant
mass-estimate biases. However, the same authors also show that a judiciously
chosen outer fit radius essentially eliminates this bias, motivating our outer radius
cut at 3 Mpc. 

To check for an outer fit-radius dependent bias,  we compare our baseline mass
measurements, measured in a radial range of 750~kpc - 3~Mpc, to masses
measured from 750~kpc - 5~Mpc (or until the edge of the available field
for low redshift clusters). Using bootstrapped catalogs to properly
account for the correlations between the two measurements, we see only
a marginal shift in the mean cluster mass. Masses from fits to data
out to 5Mpc are, on average, lower by a factor of
$0.987^{+0.012}_{-0.010}$. 

To test the susceptibility of our measured masses to the specific form of the
model density profile, we also measure masses using
two smoothly truncated NFW-like profiles from \citet{bmo09}, referred to as
`BMO-1' and `BMO-2' in \citet{ogh11}. The BMO profiles
truncate the NFW model at different rates with a truncation parameter
$\tau$. The BMO-2 profile was verified against N-body simulations in
\citet{ogh11}. Given the nature of our data, we do not include the
stochastic ``two-halo'' term when fitting the BMO profiles (we also use a different radial fit range and measure a
different mass than \citealt{ogh11}). We find that the BMO-2 profile returns
a mean cluster mass 2\% smaller than the NFW profile, with mass and
redshift dependence. The BMO-1 profile, returns a mean mass $3-4\%$
smaller than the NFW profile, depending on the assumed truncation
parameter value, $\tau=2.6$ or $\tau = 2$, respectively. 

Note that publicly available simulations do not yet probe cluster populations in the specific mass range ($M_{500} > 10^{15}M_{\odot}$) studied  here  with sufficient
statistics. New, large-volume simulations will offer improved
guidance to the systematic uncertainties for such objects, and will
likely improve these systematic tolerances. Until such simulation guidance becomes available, however, we associate a 3\% systematic uncertainty from our use of NFW models.

Another source of systematic uncertainty in the mass model is the assumed mass-concentration relation.
We minimize our sensitivity to concentration by fitting to shear profiles outside 750~kpc, by measuring masses within an aperture of 1.5~Mpc, and by fitting for and marginalizing out concentration in the \pz method fits.
 We test our sensitivity to concentration by remodelling shear profiles and recalculating masses at 1.5Mpc with concentrations of $\pm50\%$ around our baseline value of $c_{200} = 4.0$.
Cluster masses are higher by a median factor of $1.05$ at $c_{200} = 2.0$, and lower by a median factor of $0.961$ at $c_{200} = 6.0$. 
Such large shifts in the mass-concentration relation are extreme: 
\citet{bhattacharya13} find that the normalization of the mass-concentration relations from the literature \citep[][for example]{neto07} and their own work vary by $\sim10-20\%$.

We revisit the issue of our prior on concentration in Section~\ref{sec:aveconcentration}, in the context of measuring an average concentration for the sample.
In short, when we fit for an average concentration for our sample, we find $c_{200} = 4.1_{-0.9}^{+1.2}$.
This result is within $1\sigma$ of our adopted prior of $c_{200} = 4.0$.

We have verified that our choice of center for the NFW profile fit,
the X-ray centroid, does not lead to a systematic uncertainty in the
cluster mass. We see no change in the mean cluster mass to 1\% if we
instead center the NFW profile fits on the BCG in each system (see
Paper I).

In summary, uncertainties in the mass model contribute a
total systematic uncertainty of $\approx 4\%$.


\subsection{Redshift Distribution Uncertainty}
\label{sec:redshift_uncertainty}

In Section~\ref{sec:testing_framework} we estimated the uncertainty in
the \pz mass measurements due to a range of systematic uncertainties
associated with \photoz performance. For our sample redshift range,
and using \BVRiz photometry, we showed the \pz method typically
overestimates the mass by 1.2\%, with an uncertainty of $\approx1\%$.
This uncertainty is dominated by the unknown fraction of contaminating
cluster galaxies present in each cluster.

Several additional sources of potential systematic scatter, associated with photometric calibration (Paper II), are not captured by the
simulations discussed in Section~\ref{sec:testing_framework}. The photometric calibration could use other versions of
the stellar locus, and/or employ zeropoint training in the
$B_{\rm J}$ filter. Overall, our study of alternative prescriptions for the photometric calibration result in shifts in the mean cluster sample mass
of up to $\approx3\%$. Nonetheless, residual calibration and \photoz systematic uncertainties are
subdominant to other sources of error in the analysis.

COSMOS uses the $r^{+}$ and $i^{+}$ filters while our observations
typically use the SuprimeCam $\mathrm{R_c}$ and $\mathrm{I_c}$
filters. The SuprimeCam filter set resembles the SDSS $r^{+}$ and
$i^{+}$ filters more so than tradition Johnson-Cousin
filters\footnote{http://www.naoj.org/Observing/Instruments/SCam/sensitivity.htm}. While
we cannot explicitly verify the biases discussed above using
$\mathrm{R_c}$ and $\mathrm{I_c}$, we do not anticipate significant
differences from the values quoted. We also note that the depths of
observations used for this study vary, and do not exactly match the
depths to which the COSMOS field was observed. However, we again
expect that these issues are secondary to the direct uncertainties
from \photoz's discussed, and are not dominant in the analysis. To be
conservative, we allocate an additional 1\% uncertainty for these
effects.

Adding these uncertainties in quadrature, we estimate a total
systematic uncertainty associated with the redshift distribution of
$\approx 3\%$.


\subsection{Data-Driven Systematic Cross-Checks}

A number of additional systematic cross-checks were performed to verify the
accuracy of the \pz method. These checks consisted of splitting the
galaxies in each field into subsamples and determining
cluster masses for each subsample. These checks provide reassurance
that the statistical model assumed in the \pz method is adequate, but
do not provide sufficient statistical power to quantify additional
residual systematic uncertainties in the analysis.  We know from the
COSMOS-30 simulations in Section~\ref{sec:testing_framework} that the
mean masses should be accurate. Therefore any offsets detected in
these tests represent internal tensions in the analysis and avenues
for improvement in the precision of individual cluster masses. 

\begin{figure*}
  \centering
  \begin{minipage}{0.48\linewidth}
    \centering
    \includegraphics[width=\columnwidth]{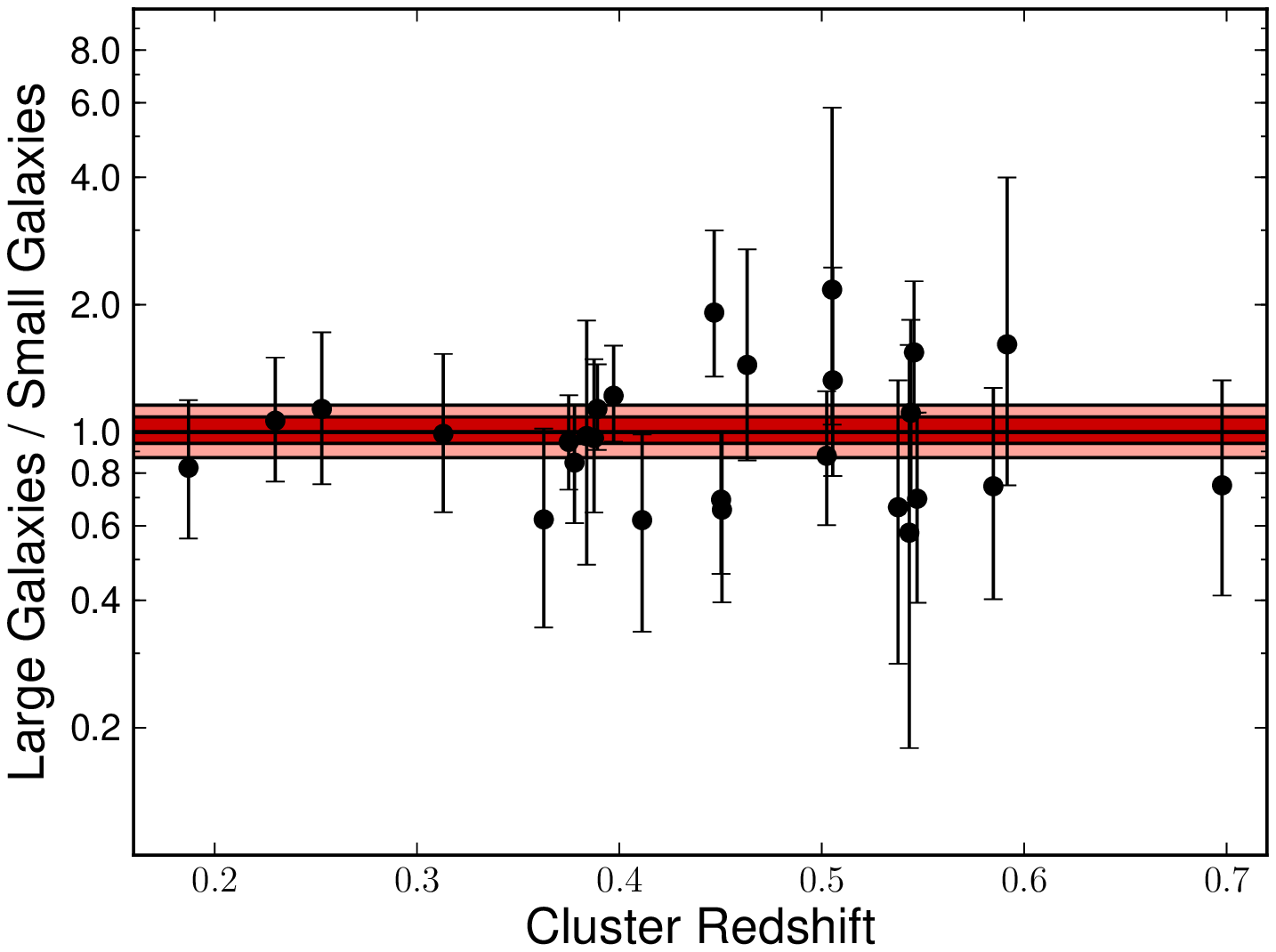}
    (a) Galaxy Size
    \label{fig:rh_split}
  \end{minipage}
  \begin{minipage}{0.48\linewidth}
    \centering
    \includegraphics[width=\columnwidth]{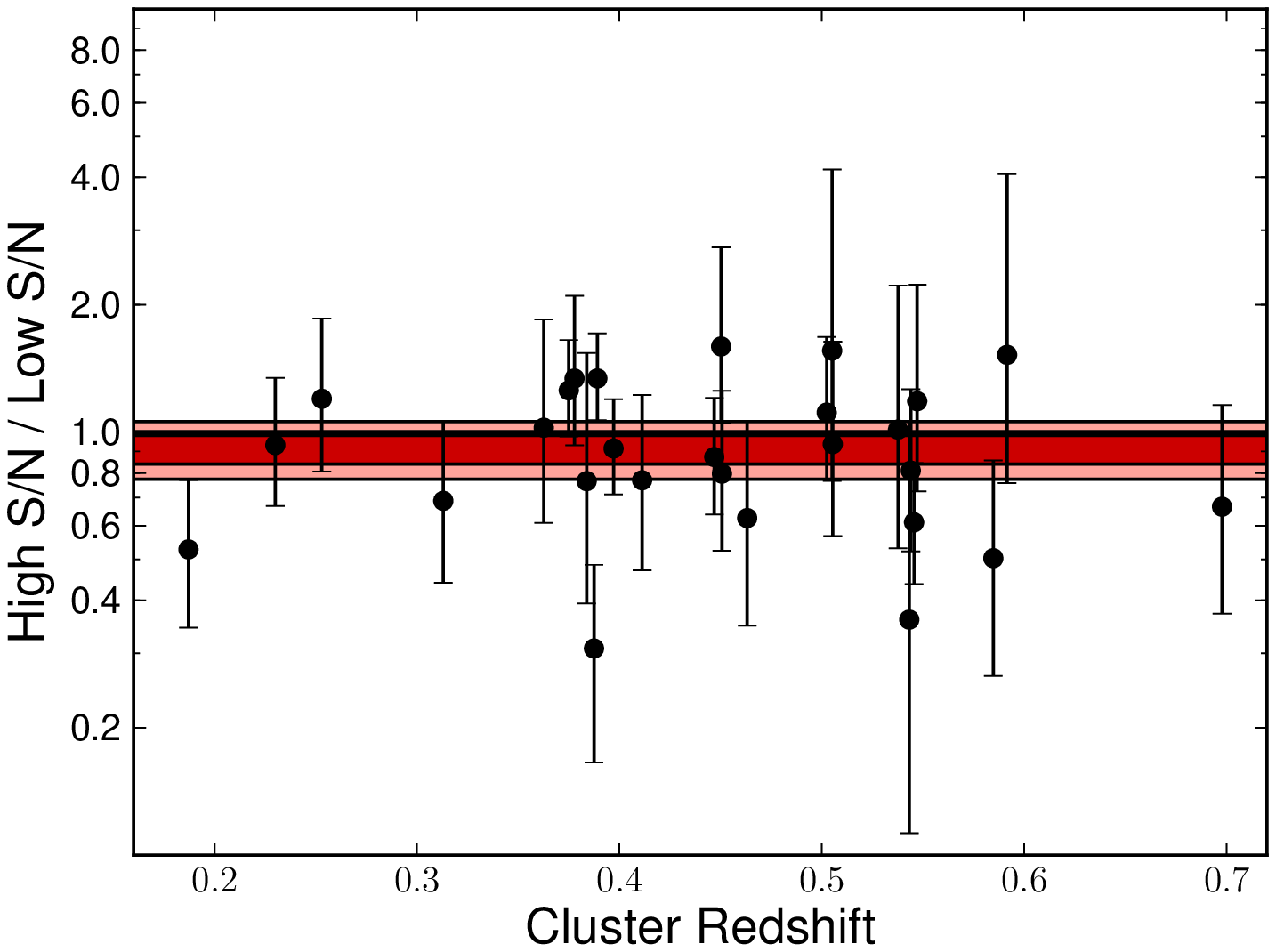}
    (b) Shape S/N
    \label{fig:sn_split}
  \end{minipage}
  \\
  \begin{minipage}{0.48\linewidth}
    \centering
    \includegraphics[width=\columnwidth]{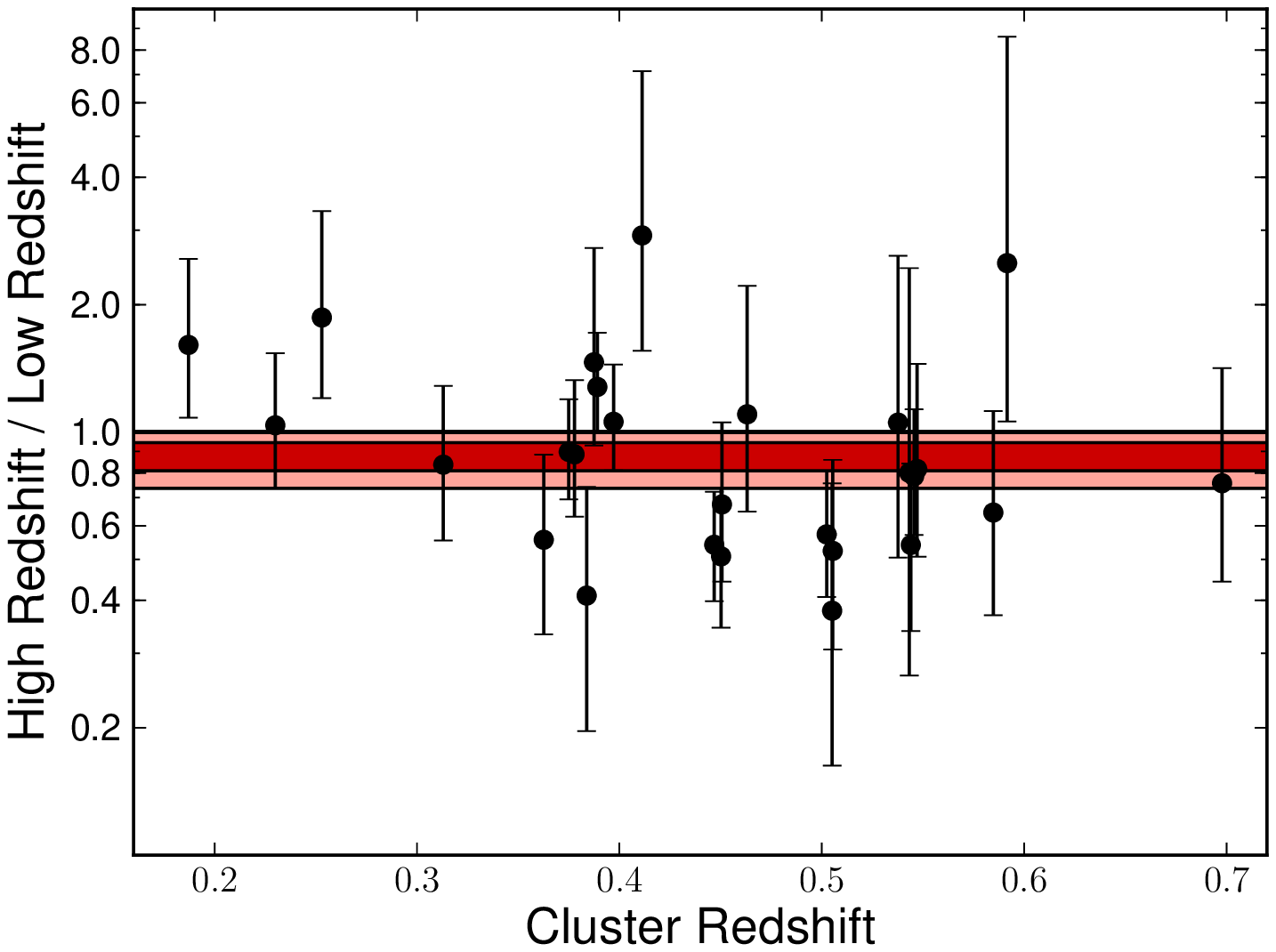}
    (c) Galaxy Redshift
    \label{fig:zsplit}
  \end{minipage}
  \begin{minipage}{0.48\linewidth}
    \centering
    \includegraphics[width=\columnwidth]{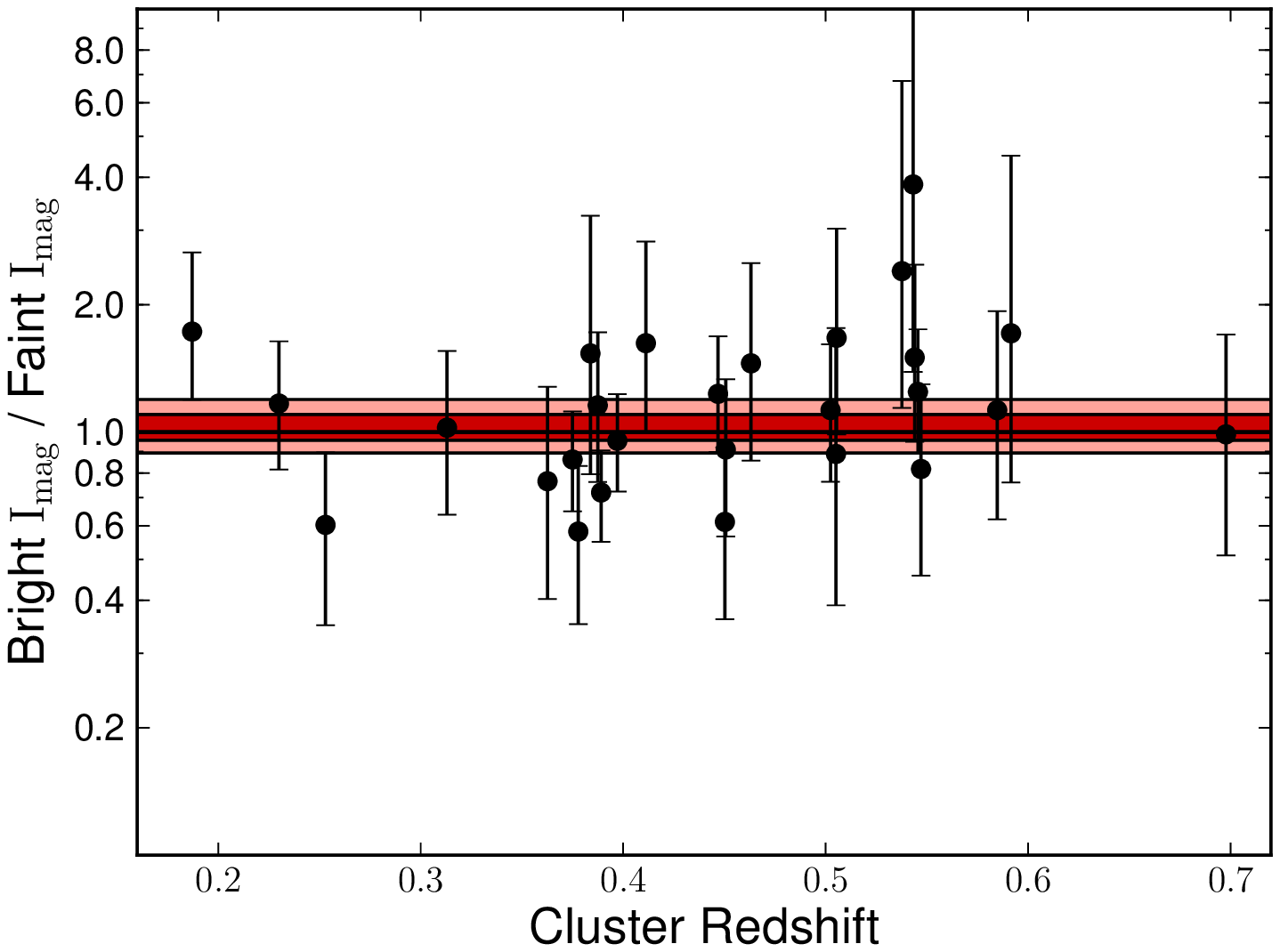}
    (d) Galaxy I magnitude
    \label{fig:imag_split}
  \end{minipage}
  \caption{For each galaxy cluster with five filter photometry, we
    reconstruct masses independently for different subsamples of equal
    statistical weight.  For each cluster, we split galaxies at the
    median value of (a) background galaxy size, (b) background galaxy
    shape S/N as reported by {\sc analyseldac}, (c) background galaxy
    redshift, (d) and I-band magnitude. The black points show the ratios and
    $1\sigma$ uncertainties for each cluster in the sample. The dark and light red
    bands are the $1\sigma$ and $2\sigma$ uncertainties
    on the ratio for the sample. Out of the four tests, we detect a
    ratio deviating from one by more than $1\sigma$ level in one test
    (splitting by shape $S/N$), and one test deviating at $2\sigma$
    (splitting by galaxy redshift). This internal tension will
    increase the cluster-to-cluster scatter and will increase the per
    cluster error; however it should not impact the mean sample
    mass. }
  \label{fig:crosschecks}
\end{figure*}

As a cross-check on the suitability of the size dependent shear
calibration, we independently measure cluster masses with galaxies
above and below the median galaxy size in each
field. Figure~\ref{fig:rh_split} shows the ratio between the two masses
for each cluster in the \pz sample, as a function of redshift. The
mean offset between the two reconstructions is
$1.01^{+0.08}_{-0.07}$, consistent with zero offset within the
$1\sigma$ uncertainties. We repeat the exercise in
Fig.~\ref{fig:sn_split}, now splitting galaxies into
samples at the median shape S/N value in each field. The mean offset
between the two reconstructions is $0.91^{+0.07}_{-0.07}$, a 1$\sigma$ offset. For both cross-checks, galaxies were
accepted into the fit from a larger fit range, 600 kpc $<$ R $<$ 5
Mpc, to increase the available galaxy statistics.

To check for consistencies in the shear signal between background galaxies at different redshifts, we
fit masses to independent samples below and above the median lensed
galaxy redshift in each field. This is equivalent to a simplified
``shear-ratio'' consistency check \citep{tkb07}.  The median redshift
is $z\approx0.8$, depending on the redshift of the cluster. The
results are shown in Fig.~\ref{fig:zsplit}. From the bootstrap
analysis, we see that masses reconstructed with high redshift galaxies
tend to be $0.88^{+0.06}_{-0.08}\%$ less massive than masses
reconstructed with low redshift galaxies. The $2\sigma$ confidence region is 0.74 to 1.00.  We also split lensed galaxies by their $i^+$
magnitude, either measured or reconstructed, and independently
measured the cluster masses (Fig.~\ref{fig:imag_split}). The ratio of
masses from bright objects to faint objects is
$1.02^{+0.08}_{-0.06}$. 

We see that internal tension exists in the data, particularly when
galaxies are divided by their shape $S/N$ or by their redshift. We do
not interpret this tension as a sign of systematic error. Instead,
this tension represents a promising avenue to increase the precision
of individual cluster masses in future work.


\subsection{Summary of Systematic Uncertainties}
\label{sec:systematics_summary}

Systematic uncertainties in our mass calibration arise from the shear
measurements, the assumed mass model, and from uncertainties
in the lensed-galaxy redshifts. We have investigated each in turn with
checks from simulations and data. Table~\ref{table:syserr_summary}
summarizes our estimates of the systematic uncertainties in the
analysis.

The systematic uncertainties associated with the shear measurements and
the mass model apply \emph{both} to the color-cut and the \pz
methods. The uncertainties associated with redshift measurements
listed in the table apply only to the \pz method, and are constrained
to be less than 2\%.  The systematic uncertainties that only apply to
the color-cut method are more difficult to quantify
(Section~\ref{sec:color_cut_sys_errs}). To gauge the uncertainties in
those measurements, we therefore pursue a strategy of
cross-calibration. By scaling the color-cut masses by the average
ratio between the two methods (as measured in
Section~\ref{sec:final_masses}), we calibrate out the unknown
systematic uncertainties in the color-cut analysis for the price of
adding the statistical uncertainty in the average ratio. For the 27
clusters with \BVRiz photometry, the uncertainty in the
cross-calibration between the color-cut and \pz methods is $\approx
4\%$.

\begin{table*}
  \caption{Summary of the sources and levels of systematic uncertainty in the analysis. Shear measurement and mass model uncertainties apply equally to the color-cut and \pz methods. The quoted redshift measurement uncertainties apply only to the \pz method; the color-cut method is subject to other, more difficult to quantify systematic uncertainties discussed in the text when not cross-calibrated from the \pz method. Additionally, systematic uncertainties are quoted for all 51 clusters studied with the color-cut method, and 27 clusters that use the \pz masses directly.  Values in the table are reported to single-digit precision.}
\begin{tabular}{l | c c c  }
  \hline
  Uncertainty Source & \multicolumn{3}{c}{\% of Mean Cluster Mass} \\
  \hline
  & Color-Cut Method & & P(z) Method \\
  \hline
  \textbf{Shear Measurements} & &  &\\
  Multiplicative Shear Bias Cor & & 3\% & \\
  STEP PSF Mismatch & & 2\% & \\
  Coaddition \& PSF Interpolation & & 1\% & \\
  \hline
  \textbf{Mass Model} &  & & \\
  Profile Uncertainty & & 3\% & \\ 
  \hline
  \textbf{Photo-$z$ Measurements} & &  & \\
  Residual Photometry Systematics & & 3\% & \\
  Simulated Photo-$z$ Bias & &1\% & \\
  Depth \& Filter Mismatch & &1\% & \\
  \hline
  \textbf{Method Cross-Calibration}& 4\% & & -  \\
  \hline
  \hline
  \textbf{Method Systematic Uncertainty} & 7\% &  & 6\%  \\
  \hline
  Triaxiality \& LOS Structure & 3\% &  & 4\% \\
  \hline
  \hline
  \textbf{Total Systematic Uncertainty} & 8\% & & 7\% \\

\end{tabular}
\label{table:syserr_summary}
\end{table*}

We expect each source of systematic uncertainty to be independent, and
have approximated each source as a Gaussian. Our total systematic
uncertainty on the mean cluster mass, for 51 clusters, is therefore
7\%. Results are comparable when only masses measured with the \pz
method are used.



\section{Average Sample Concentration}
\label{sec:aveconcentration}

In addition to asking, as we do above, at what level our mass measurements are sensitive to our choice of priors on $c_{200}$ (Section~\ref{sec:astro_model_uncertainty}), we can also ask how well we can measure the average concentration of the sample.
For this, we use only our \pz measurements, in order to avoid the clear degeneracy between concentration and the contamination correction in the color-cut analysis (Section~\ref{sec:contamination_correction}).
We use the same radial range for this analysis as in our standard mass modeling, from 750 kpc to 3 Mpc.
We assume that cluster concentrations are drawn from a log-normal distribution with mean $\mu_c$ and width $\sigma_c$.
We simultaneously fit for the mass and concentration of each cluster, as well as the sample mean and width.
The width $\sigma_c$ has a uniform prior from 0.286 to 0.318 \citep[][this range accounts for the expected scatter in concentration due to triaxiality and large-scale structure]{bahe2011}.
As with our mass measurements, we include priors on the seeing-dependent mean STEP correction, but instead lock the parameters of the shape dispersion to the best fit values measured in our fits for cluster masses (see Section~\ref{sec:shape_distribution} for more details).

\begin{figure}
\includegraphics[width=\columnwidth]{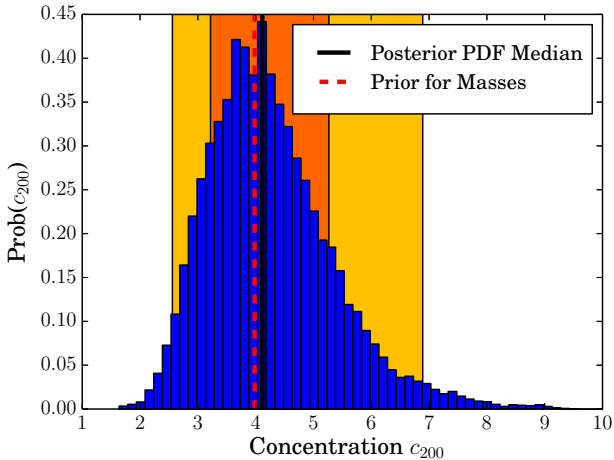}
\caption{Posterior probability distribution for the mean concentration of the 27-cluster \pz sample, where cluster concentrations are assumed to be log-normally distributed.
The black solid line and orange contours mark the median, 68\%, and 95\% median-centered confidence intervals for the average concentration of the sample.
The 68\% interval is $c_{200} = 4.1_{-0.9}^{+1.2}$.
The dashed red line shows our prior on concentration for the mass measurements reported in Section~\ref{sec:final_masses}, $c_{200} = 4.0$.}
\label{fig:aveconcentration}
\end{figure}

Figure~\ref{fig:aveconcentration} shows the posterior probability distribution for $\mu_c$.
All other parameters, including the masses and concentrations of individual clusters, have been marginalized out.
The 68\% confidence interval on the mean concentration $\mu_c$ is $[3.2, 5.3]$, and is centered at $c_{200} = 4.1.$
This is consistent with the standard prior used in our mass modeling at the 68\% confidence level, confirming the robustness of the results in Section~\ref{sec:final_masses}).

\section{Comparison to the Literature}
\label{sec:lit_comp}

\begin{figure*}
\includegraphics[width=0.46\hsize]{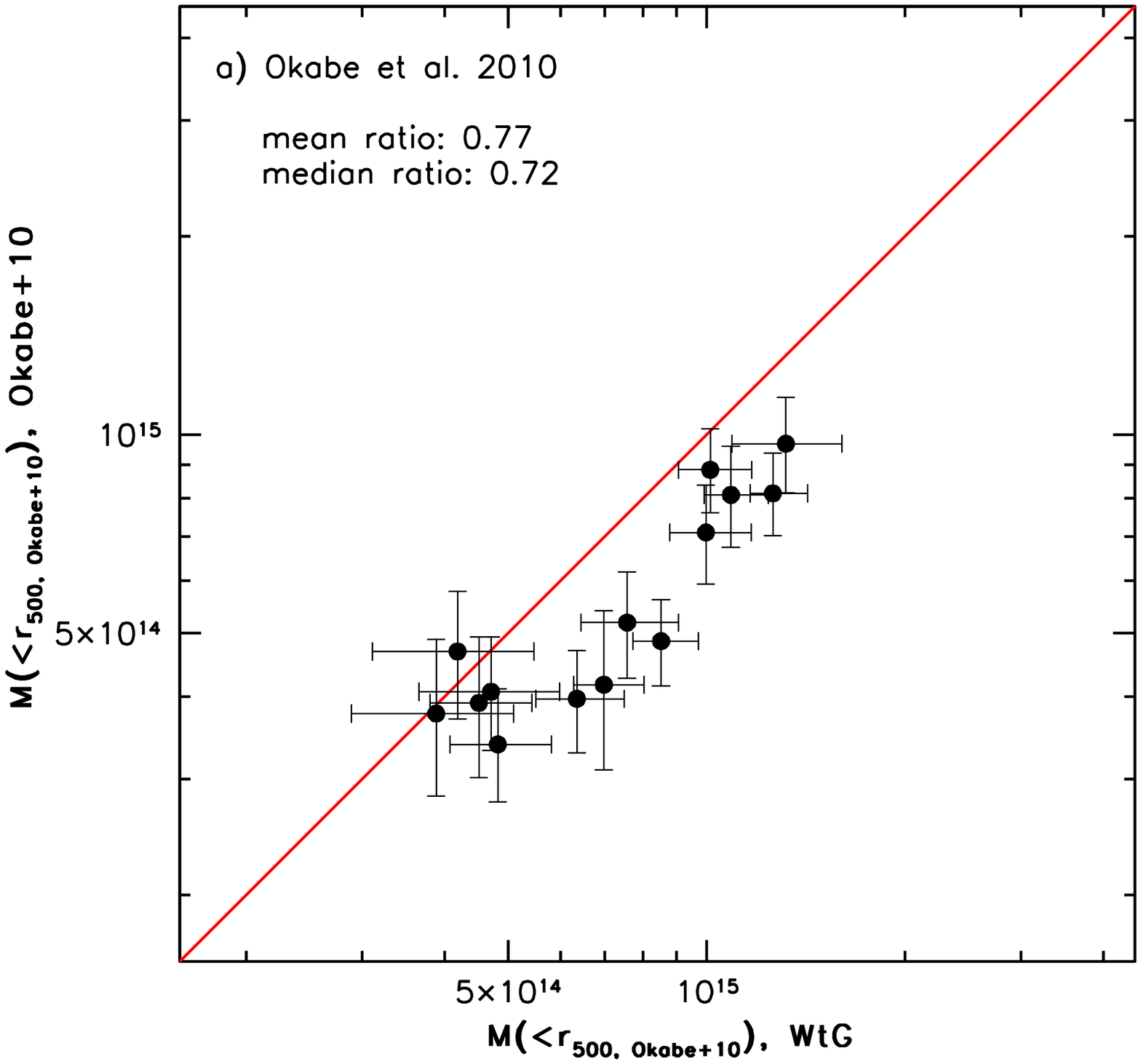}\hspace{0.04\hsize}%
\includegraphics[width=0.46\hsize]{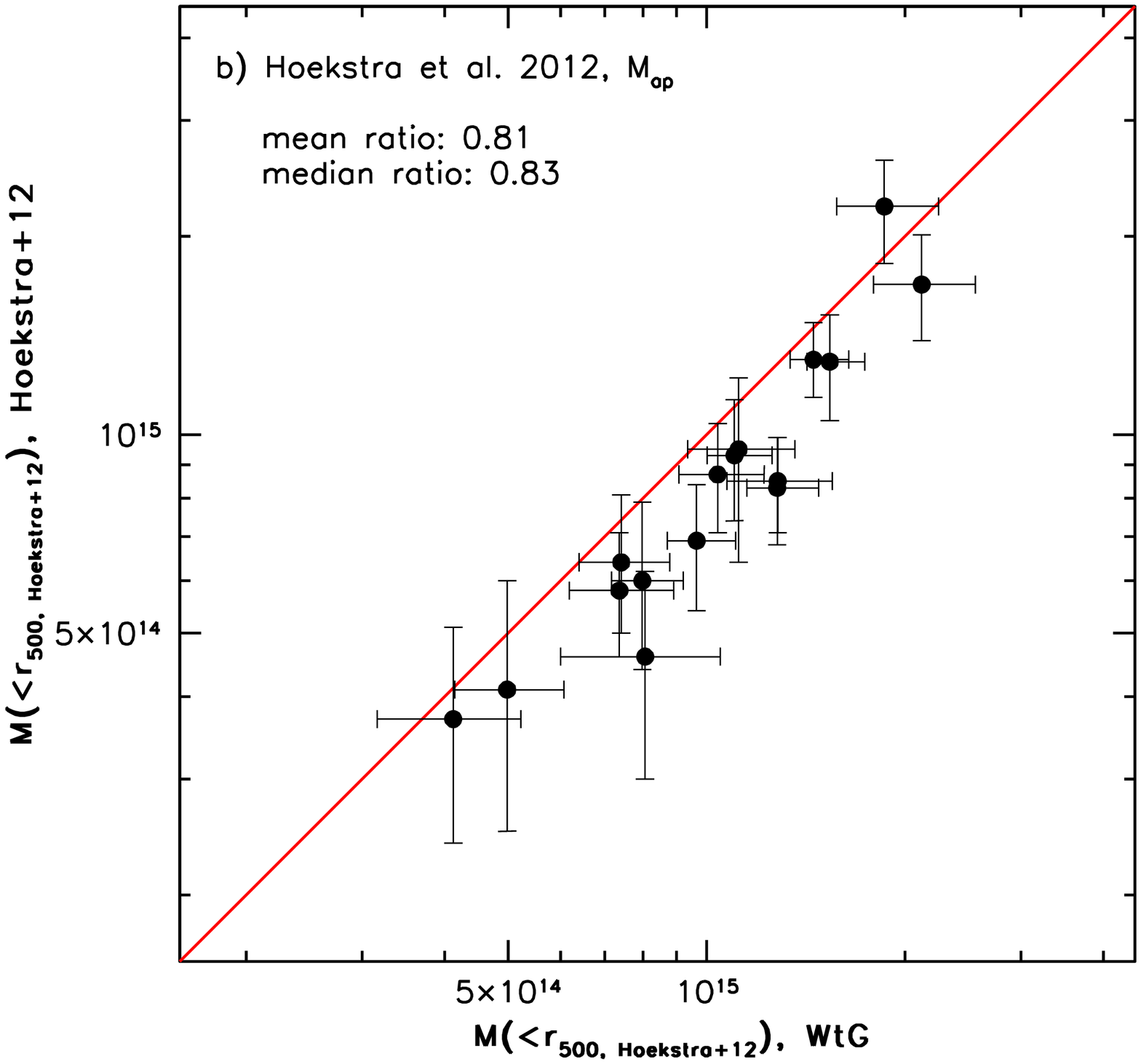}
\includegraphics[width=0.46\hsize]{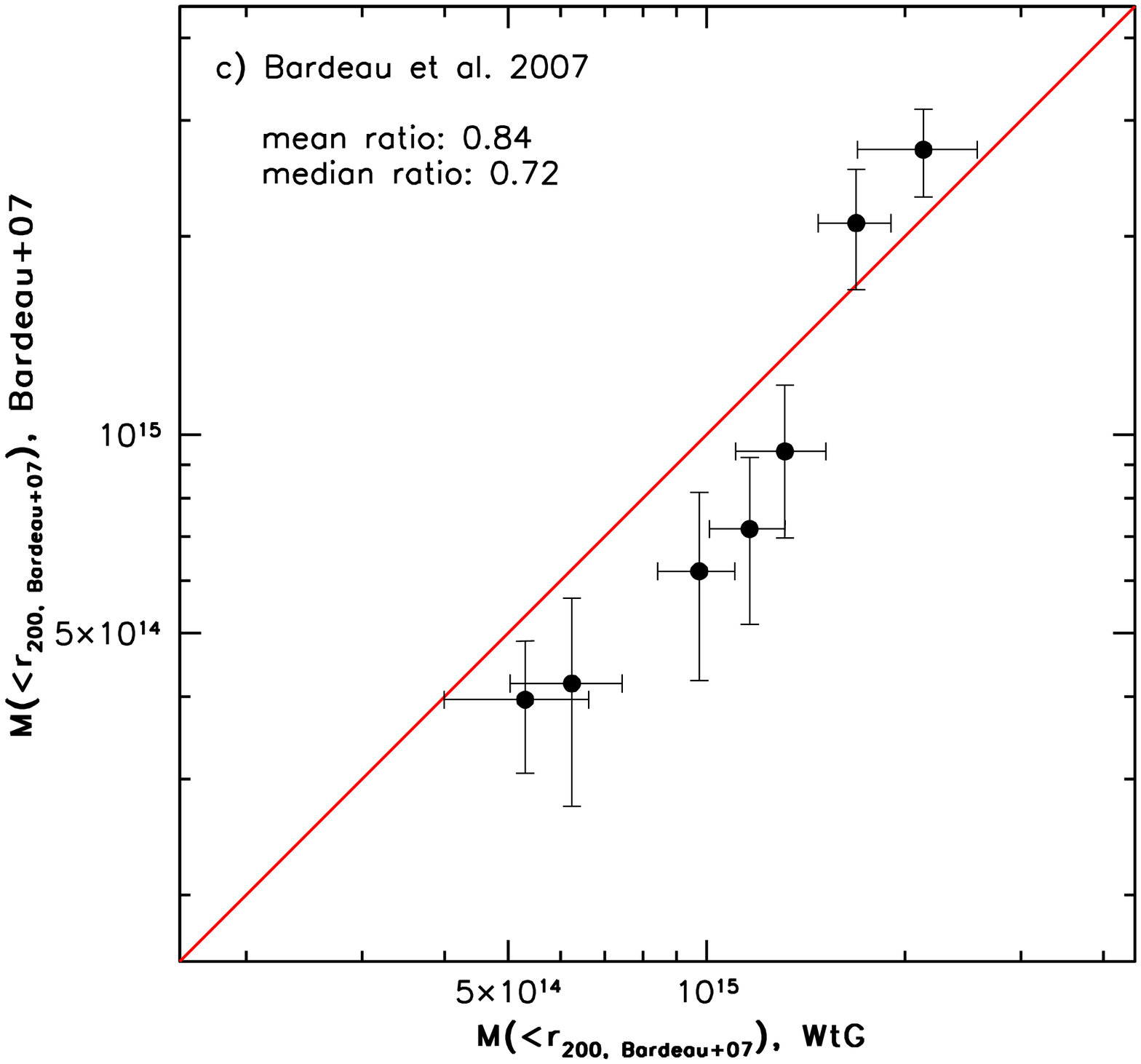}\hspace{0.04\hsize}%
\includegraphics[width=0.46\hsize]{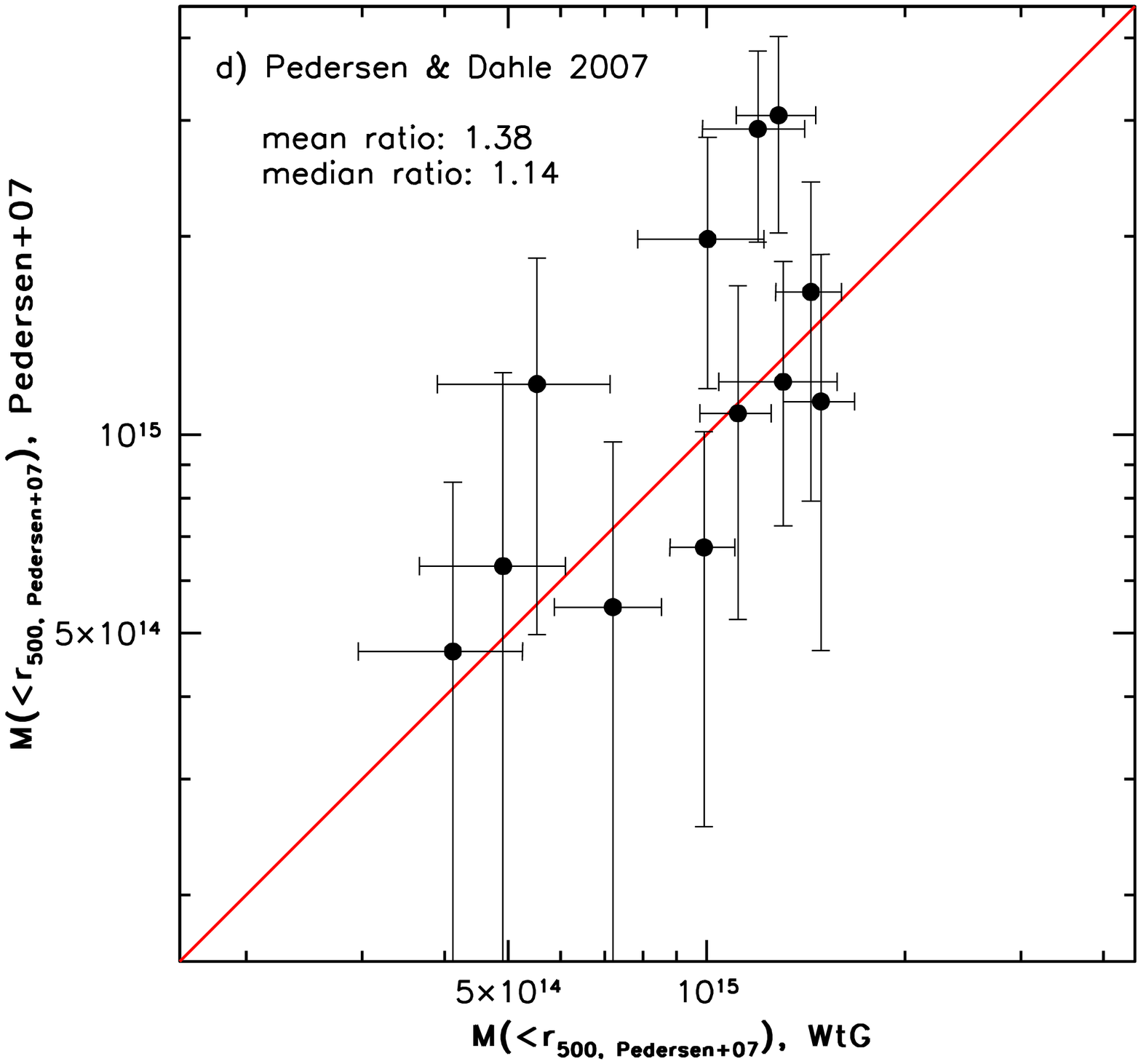}
\caption{Comparison of results in the literature to our mass measurements (WtG). Panel a) shows the comparison to \citet{okabe_masses}, panel b) to \citet{hoekstra12}, panel c) to \citet{bsk07}, and panel d) to \citet{ped07}. For each comparison, we measure the mass within the
  same overdensity radius $r_{\Delta}$ and with the same cosmology as used in the respective work. The solid line indicates a one-to-one ratio. The value of the mean ratio shown on each plot is calculated as the unweighted average of the ratios of the measured mass from the literature to our mass; the ratios are not weighted in the average because the measurements in each ratio are correlated due to overlap in the sample of source galaxies.}
\label{fig:lit_comp}
\end{figure*}

For a number of clusters considered here, previous weak-lensing mass
measurements have been reported in the literature. In this section, we
compare our mass measurements with those works, in cases
where those studies have employed a homogeneous weak-lensing
methodology, have at least five clusters in common with the present
study, and quote mass measurements at a suitably large density
contrast ($M_{500}$, $M_{200}$, etc.). All of the previously reported
mass measurements considered here are based on variations of the color-cut
method. Almost all of the clusters overlapping the present study are
at relatively low redshifts, $z\sim0.2$; in this redshift regime, the
color-cut method can in principle provide robust mass measurements,
although significant care needs to be taken in calibrating the shear
measurements, correcting for contamination from cluster galaxies, and
estimating the redshift distribution of background galaxies
(Section~\ref{sec:color_cuts}).

To facilitate the most direct comparison of mass measurements, we
redetermine masses within the same radius used in each work in the literature. In cases where the measurement radius is not explicitly reported, we
calculate it from the cited overdensity mass, $r_{\Delta} = \left( 3
  M_{\Delta} / 4\pi \Delta \rho_c(z) \right)^{1/3}$, adopting the
cosmology used in the literature work. In each case, we fit the shear
profile over the radial range 750~kpc -- 3~Mpc, and
measure the mass within the measurement radius using the calibrated color-cut method and NFW-model fits.

Weak-lensing calibration of X-ray measured masses is a major goal for X-ray cluster cosmology \citep{aem11}.
Not surprisingly, a number of the weak-lensing studies we examine in this section have previously been compared to X-ray hydrostatic-equilibrium measurements \citep[\textit{e.g.,~}][]{hoekstra07,mahdavi08,vbe09,zof10,mahdavi13}. 
However, we defer a detailed discussion of the implications of these comparisons to \citet{applegate_nontherm}.

\subsection{Comparison to \citet{okabe_masses}}

The study of \citet[][30 clusters in total]{okabe_masses} has the largest
overlap with our sample, with 14 clusters in common. Their mass
measurements are based on two-filter imaging with SuprimeCam, and some
of the raw data are in common with our study.

We find a significant offset between the \citet{okabe_masses} mass
measurements and ours, with the former being lower on average by
$\sim$ 25\% (Fig.~\ref{fig:lit_comp}a). 
Such a large offset is likely due to a combination of effects. One possible cause for a sizable
bias lies in the different depths of the galaxy catalogs used for the
lensing analysis. \citet{okabe_masses} use galaxies to typically
$i^{+}<26$, at least a magnitude fainter than our completeness limits
(Section~\ref{sec:color_cuts}). The completeness limits in our study
were set such that the signal-to-noise of each galaxy's shape
measurement is large enough to ensure that the shear measurement bias
can be robustly calibrated (see Paper I and
Section~\ref{sec:shape_distribution} in this paper). The significant
shear bias that we find for fainter objects (15--30\%,
Fig.~\ref{fig:step_size_cor}) is typical for KSB-based algorithms
\citep{step2}, and thus is likely to affect the study of
\citet{okabe_masses}, since a large fraction of the total galaxy
sample is fainter than our signal-to-noise criterion
(Fig.~\ref{fig:mag_histo}).

Another difference between the two studies is the radial range over
which the profile is fit. \citet{okabe_masses} fit the shear profile
from the core (1~arcmin, corresponding to 200~kpc at $z=0.2$) over the
entire field of view, i.e. out to $\sim$20~arcmin. We fit out to only 3~Mpc, which corresponds to $\sim
11 - 15$~arcmin for most of the clusters in the comparison
set. Numerical simulations find that mass estimates based on fitting
NFW profiles to radii as large as those used in the
\citet{okabe_masses} study are likely to be biased low at the 5--10\%
level \citep{becker11,ogh11,bahe2011}. The bias can be reduced or
essentially eliminated if the fit range is restricted to within $\sim
2 \times r_{500}$ (about 3~Mpc for the most massive clusters here) as
in this work.  Note, however, that we did not find a significant mass
offset between fitting to 3~Mpc and fitting to 5~Mpc in our own data
(Section~\ref{sec:astro_model_uncertainty}).

Possible biases arising due to the smaller inner radial cut-off chosen
by \citet{okabe_masses} are more difficult to estimate. Simulations do
not indicate a significant bias introduced by fitting to small scales
\citep{bahe2011}, but observational biases could play a much larger
role. For example, any residual contamination by cluster members would
be more pronounced, and most detrimental to the measured shear, at the
cluster center. Furthermore, shear testing programs have not yet
investigated the calibration bias for shear values typically found at
cluster center ($g\gtrsim0.1$). We also note that \citet{okabe_masses}
fit for the concentration, rather than assuming a specific value of
mass-concentration relation. Shear values and mass measurements are
more sensitive to the concentration at small radii; hence this may
also cause an offset in the mass measurements.

Following the submission and preprinting of the present study, the authors of \citet{okabe_masses} published a major revision of their analysis in \citet{okabe13}.
The newer work presents a stacked analysis of 50 galaxy clusters, 21 of which are in common with \citet{okabe_masses}.
Individual cluster masses are not reported.
The authors significantly revise their color-cut technique used to select background galaxies and modify their shear estimation algorithm.
In particular, \citet{okabe13} now use high-S/N galaxies to calibrate the isotropic correction step during shear estimation.
The authors report a significant shift in cluster masses due to this new calibration, but do not state the magnitude of the correction.
The authors report that individual cluster masses are now higher by 15\%-20\% due to all modifications, and are compatible with our study.

\subsection{Comparison to \citet{hoekstra12} and \citet{mahdavi08}}

\citet{hoekstra12} present weak-lensing mass measurements for 50 clusters, 16 of which are in common with the present study.
These clusters were observed in two groups, using different cameras and filter sets, as well as different observing and data reduction strategies.
The mass-measurement methodology presented in \citet{hoekstra12} tracks closely to that presented in \citet{hoekstra07}.
Important updates include using the CFHT Legacy Survey Deep Fields \citep{cfhtlsdeep} as the reference redshift distribution instead of the relatively small Hubble Deep
Field \citep{hdf}, as well as using a newer mass-concentration relation reported in \citet{duffy08} instead of \cite{bullock01}.

The aperture-mass measurements from \citet{hoekstra12} are 15--20\% lower than ours (Fig.~\ref{fig:lit_comp}b), but correlate reasonably well.
In general, our color-cut methodology
follows closely that of \citet{hoekstra12}; the most
significant difference is that \citet{hoekstra12} use the aperture-mass
method to determine cluster masses, rather than fitting an NFW
profile.\footnote{\citet{hoekstra07,hoekstra12} both present masses measured from NFW-profile fits. However, the authors adopt mass-aperture measurements for their multiwavelength comparisons in \citet{hoekstra07, mahdavi08, hoekstra12, mahdavi13}.}  In the aperture-mass method, the mass within a specific radius
is determined only from galaxies at larger projected
radii. Nevertheless, there is considerable overlap between the source
galaxies used in the two studies, since our measurements are based on
the radial range of 0.75--3~Mpc, and the $r_{500}$ measurements of
\citet{hoekstra12} use a radial range starting from $\approx$ 1-1.5~Mpc to 10-15 arcminutes.

The cause for the offset in the \textbf{\citet{hoekstra12}} masses is not
clear.  For example, the redshift distributions of COSMOS and
CFHT-LS have been shown to be in excellent agreement with each other
\citep{ilbert09}.  \citet{hoekstra12} set a similar limiting magnitude
for the galaxies in their CFH12K source sample as ours; since their exposure
times for these clusters are typically about four times longer than
our observations, but using a telescope of half the diameter, this
should yield a similar minimum signal-to-noise ratio, and avoid the
larger shear bias of lower-SNR objects. 
However, clusters observed with Megacam in \citet{hoekstra12} use similar or less total exposure time than for clusters in the CFH12K sample \citep[originally studied in][]{hoekstra07,mahdavi08} at comparable redshift.
Furthermore, \citet{hoekstra12} detect objects and measure shapes in coadded images of half the total exposure time.
This may increase the influence of lower-SNR objects, where the amplitude of the required shear calibration correction is larger and SNR dependent.
For reference, the masses reported in \citet{mahdavi08} \citep[updating][masses]{hoekstra07} are 10--15\% lower than our work.

Another possible (partial) cause for the difference between the \citet{hoekstra12} results and ours is the use in
\citet{hoekstra12} of an NFW profile to correct for the mass-sheet
degeneracy, as well as in converting the measured 2D mass to a 3D mass
estimate. \citet{mrm10} show that the use of aperture-mass method with such
NFW corrections can lead to a similar bias as directly fitting an NFW
profile to a large radial range.

\subsection{Comparison to \citet{bsk07}}

\citet{bsk07} measured weak-lensing masses for eleven clusters using
three-band imaging with the CFH12K camera at the CFHT (some of these data were
also used in the \citealt{hoekstra07} and \citealt{mahdavi08} analyses). Seven
of these clusters are also in our sample. The comparison to our mass
measurements is somewhat inconclusive: while for five of the seven
clusters, the \citet{bsk07} masses are lower by $\sim 30$\%, their
masses for the two most massive clusters in common with our sample are
$\sim 25$\% higher (Fig.~\ref{fig:lit_comp}c).

\subsection{Comparison to \citet{ped07}, \citet{dah06} and \citet{dki02}}

The first large sample of weak-lensing mass measurements for galaxy clusters was
compiled by \citet{dki02}, with a total sample size of 38 clusters.
These measurements were used by \citet{dah06} to study the cluster
mass function, and by \citet{ped07} to calibrate the scaling relation
between X-ray temperature and total mass.  Their weak lensing
observations were obtained with the 2.5-m Nordic Optical Telescope and
the 2.2-m University of Hawaii Telescope. The smaller aperture, but
similar exposure times to the present work, means that the \citet{dki02} data are
significantly shallower than ours (and the other works
considered here). Furthermore, most of the data were taken with
single-CCD cameras, restricting the available field of view and radial
fit range.

The weak lensing methodologies of \citet{dki02}, \citet{dah06}, and
\citet{ped07} are identical; here we compare to the measurements of
$M_{500}$ presented in \citet{ped07}, since our own work focuses in
particular on measuring $M_{500}$. There are twelve clusters in common
across the two studies. The scatter between the two sets of
measurements is significant (Fig.~\ref{fig:lit_comp}d), although this
is partly due to the large statistical uncertainties of the shallower
data.  Furthermore, the overlap in the source galaxy catalogs between
the two studies is small, since for all but two of the clusters,
\citet{ped07} fit the profile from imaging with a small field of view,
spanning radii of only $\sim$0.9--3~arcmin. In comparison, our inner
radial cut-off corresponds to 2.5--4~arcmin for these clusters -- the
measurements are thus nearly independent.

On average, the masses of \citet{ped07} are 10--40\% higher than
ours; it is the only literature sample considered here that
overestimates the cluster masses compared to our measurements. It is
important to keep in mind that the \citet{ped07} mass measurements are
largely derived from the inner cluster regions, which we have
explicitly excluded in our analysis. As discussed in
Sect.~\ref{sec:data_reduction}, the correction for cluster galaxy
contamination is large in these regions, and the shear measurement
bias for such large shear values remains uncalibrated by simulations, leading to
larger systematic uncertainties. \footnote{We emphasize that weak lensing cluster mass measurement efforts
must be accompanied by simulations well matched to the observations, in order to calibrate
biases specific to the observational methodology.}

In part, the apparent mass over-estimates of \citet{ped07} may also
reflect the tendency of the shape measurement algorithm used in
\citet{dki02} to overestimate the true shear \citep{step1}. However,
over the range of simulated shear values, $0<|\gamma|<0.1$, the bias
is no more than 5\%, and indeed \citet{ped07} find that accounting for
it changes the mass estimates by only a few percent.


\subsection{Minimizing Observer's Bias}

The next stage of our project will be to calibrate current X-ray mass
proxies. We emphasize that we have deliberately avoided any
comparisons between lensing and X-ray derived masses in this study to
date. With respect to the X-ray to lensing-mass ratio, our analysis is `blind' (see the
discussion in \citealt{aem11}).  To this end, all previous lensing
efforts by teams affiliated with the authors, using the same datasets
as this paper, were ignored -- including raw data
reduction. Development of the color-cut and \pz methods proceeded in
parallel, and key parts of the algorithm and test simulations were
independently coded and cross-checked. Only once both algorithms were
finalized, and all cross-calibration and systematic uncertainty
analysis was complete, did we compare our measurements to lensing
masses in the literature, as reported above. The X-ray analysis team
has as yet had no access to the final lensing masses reported here
(and vice versa), while they independently update the X-ray masses with improved instrument
calibrations and analysis packages. All draft copies of this paper
have had the lensing masses redacted for all coauthors, excluding
Applegate and von der Linden. The two independent efforts will be
combined in subsequent papers.

After we compared our lensing masses to literature measurements, it
was noted that the mass for MACS1731+22 was measured using a coadded
image containing some frames with seeing smaller than 0.45''. We
report updated values in the text after fixing this oversight. The
results did not change appreciably and no conclusions were altered.
For reference, the mass for MACS1731+22 used in the initial literature
comparison was $23.9^{+2.3}_{-2.4} 10^{14} M_{\odot}$ and the
color-cut
to \pz calibration was $\beta = 0.998^{+0.044}_{-0.042}$.


\section{Conclusions \& Outlook}
\label{sec:conclusions}

We have developed and employed two separate weak-lensing mass
measurement algorithms to derive accurate lensing masses for a sample
of 51 massive, X-ray-selected clusters. We have used a traditional,
but improved, ``color-cut'' analysis to derive masses for the entire
sample, and a new method incorporating the full individual \photoz
posterior probability distributions for galaxies in each cluster field
for the 27 clusters observed in at least five filters. We have arrived
at the following conclusions:

\begin{itemize}
\item The color-cut method, while requiring the least observing time,
  does not easily allow for systematic uncertainties to be
  quantified without cross-calibration by other techniques, such as our \pz method. Systematic uncertainties associated with obtaining
  appropriately matched deep fields and other details in the analysis
  can easily shift the mean cluster mass by 5-10\%, depending on the
  cluster-sample redshift range. It is currently unclear whether the
  color-cut method can be successfully extended to high redshifts
  ($z\ge0.7$) while simultaneously achieving the 2\% systematic
  uncertainty goal required for upcoming surveys. `Stacking'
  analyses, where many cluster catalogs are combined for joint analysis,
  does not evade these sources of systematic error, though
  additional systematics (e.g., miscentering) may dominate in such cases \citep{rrk11}.

\item The \pz method, which uses full \photoz posterior probability
  distributions, is shown to accurately recover the mean cluster mass of the
  present sample. Considering only uncertainties from \pz distributions, we
  recover the mean cluster mass to better than 2\% accuracy when using \BVriz photometry with current \photoz codes. This systematic
  uncertainty is subdominant to current uncertainties associated with the shear
  calibration, sample size, and assumed mass model in the present study. Though requiring
  observations made in more filters than the traditional color-cut
  approach, the benefits of the \pz method in both accuracy and the straightforward
  quantification of systematic uncertainties are evident.

\item Current photo-$z$ point estimators cannot be used to accurately measure
  mean cluster masses over the redshift range examined in this study. We find systematic biases in excess of 5\%
  using algorithms currently in the literature, with significant
  redshift dependencies. In contrast, our method, which uses the full
  \pz information, is accurate to better than 2\% for clusters at $0.15 < z < 0.7$. 

\item We have used the subsample of clusters with \pz mass
  measurements to calibrate the systematic uncertainties in our
  color-cut method. The agreement in the mean mass for the color-cut
  and \pz analysis codes is excellent, with a mean ratio of $1.00 \pm 0.04$.
We know the total systematic uncertainty in our color-cut method only because we have cross-calibrated it against our \pz method. This cross-calibration measures the systematic uncertainties for only this implementation of the color-cut method and only for the filters and cluster redshifts used in this study.
 The statistical uncertainty in the mean ratio
  is comparable to the systematic uncertainties associated
  with the shear measurements and the assumed mass model. The
  overall systematic uncertainty on the mean-mass for the
  entire sample of 51 clusters is $\approx 7\%$. 

\item Currently, the dominant systematic uncertainties associated with both the shear calibration and
  the mass model are limited by available simulations, not data. New STEP-like simulations, using the
  exact PSFs observed in lensing studies, as well as larger-volume
  dark matter simulations probing the appropriate cluster mass range,
  should reduce these systematic uncertainties to below the 2\%
  level. We are working with collaborators to realize these
  improvements.

\item The excellent performance of the \pz method offers significant
  promise for extracting cosmological constraints with galaxy clusters
  in future wide, deep imaging surveys. For clusters in the redshift
  range of interest to these surveys ($z < 0.6$), our results show
  that it is possible to determine the redshift distributions of
  lensed galaxies without relying on deep spectroscopic
  surveys for calibration \citep[similar to][]{carlos11}. Surveys such as DES and LSST will offer precise five- or six-
  filter photometry similar to that used here, enabling a
  straightforward application of the \pz technique. However, we
  caution that the particular filter sets employed by these surveys
  must be calibrated for bias against suitably deep fields with
  precise and accurate photo-$z$'s calculated from many, e.g. 30+,
  filters. As the systematic tolerances push below 2\%, a larger deep
  field with many-filter coverage and spectroscopic redshift
  validation will be needed to verify \photoz performance. The
  performance of the \pz method also suggests that a similar
  statistical approach to cosmic shear may prove fruitful, while
  deep-field tests should provide insight into possible systematic
  biases.

\item Targeted follow-up of high redshift ($z > 0.5$) clusters
  (e.g. SPT or eRosita follow-up) should benefit from the addition of
  near infrared filters, which should in principle allow the extension
  of the \pz method to higher redshifts. Simulations similar to those
  pursued in this study, again utilizing a suitably large and well
  studied deep field, should also be undertaken to determine the bias
  in the cluster masses obtained in this regime.

\end{itemize}

The weak-lensing masses reported in this paper have sufficient
accuracy to realize most of the potential of current X-ray derived cluster
samples. We emphasize that all measurements reported here were derived
blindly with respect to X-ray mass measurements, and other
lensing analyses in the literature. Improved measurements of X-ray
scaling relations and cosmological parameters using these results will
be reported in forthcoming papers.


\section*{Acknowledgments}

We thank David Donovan for his efforts in collecting much of
the weak lensing data used here. We also thank Henk Hoekstra,
as well as members of the LoCuSS team, for insightful discussions
about the results presented in Sect.~\ref{sec:lit_comp}. In addition,
we thank Phil Marshall for discussions on testing statistical
models, and PM \& David Hogg for discussions on model selection with
the posterior predictive cross-validation technique.

This work is supported in part by the U.S. Department of Energy under
contract number DE-AC02-76SF00515. This work was also supported by the
National Science Foundation under Grant No. AST-0807458. MTA and PRB
acknowledge the support of NSF grant PHY-0969487. AM acknowledges the
support of NSF grants AST-0838187 and AST-1140019.  The authors
acknowledge support from programs HST-AR-12654.01-A,
HST-GO-12009.02-A, and HST-GO-11100.02-A provided by NASA through a
grant from the Space Telescope Science Institute, which is operated by
the Association of Universities for Research in Astronomy, Inc., under
NASA contract NAS 5-26555. This work is also supported by the National
Aeronautics and Space Administration through Chandra Award Numbers
TM1-12010X, GO0-11149X, GO9-0141X , and GO8-9119X issued by the
Chandra X-ray Observatory Center, which is operated by the Smithsonian
Astrophysical Observatory for and on behalf of the National
Aeronautics Space Administration under contract NAS8-03060.  The Dark
Cosmology Centre (DARK) is funded by the Danish National Research
Foundation. DEA recognizes the support of a Hewlett Foundation
Stanford Graduate Fellowship and of the German Federal Ministry of Economics and Technology (BMWi) under project 50 OR 1210.

Based in part on data collected at Subaru Telescope (University of
Tokyo) and obtained from the SMOKA, which is operated by the Astronomy
Data Center, National Astronomical Observatory of Japan.  Based on
observations obtained with MegaPrime/MegaCam, a joint project of CFHT
and CEA/DAPNIA, at the Canada-France-Hawaii Telescope (CFHT) which is
operated by the National Research Council (NRC) of Canada, the
Institute National des Sciences de l'Univers of the Centre National de
la Recherche Scientifique of France, and the University of Hawaii.
This research used the facilities of the Canadian Astronomy Data
Centre operated by the National Research Council of Canada with the
support of the Canadian Space Agency.  This research has made use of
the VizieR catalogue access tool, CDS, Strasbourg, France. Funding for
SDSS-III has been provided by the Alfred P. Sloan Foundation, the
Participating Institutions, the National Science Foundation, and the
U.S. Department of Energy Office of Science. The SDSS-III web site is
http://www.sdss3.org/. This research has made use of the NASA/IPAC
Extragalactic Database (NED), which is operated by the Jet Propulsion
Laboratory, Caltech, under contract with NASA.

\bibliography{paper3.bib}

\appendix

\label{lastpage}

\end{document}